\documentclass[%
 aip,
 amsmath,amssymb,
preprint,%
]{revtex4-1}

\usepackage{graphicx}
\usepackage{dcolumn}
\usepackage{bm}

\usepackage[utf8]{inputenc}
\usepackage[T1]{fontenc}
\usepackage{mathptmx}
\usepackage{etoolbox}

\usepackage[english]{babel}     
\usepackage[table,xcdraw]{xcolor}
\usepackage{xcolor}
\usepackage{longtable}
\usepackage{graphicx}
\usepackage{amssymb}
\usepackage[hidelinks]{hyperref}
\hypersetup{colorlinks=false}

\usepackage{xr}
\makeatletter
\def\@email#1#2{%
 \endgroup
 \patchcmd{\titleblock@produce}
  {\frontmatter@RRAPformat}
  {\frontmatter@RRAPformat{\produce@RRAP{*#1\href{mailto:#2}{#2}}}\frontmatter@RRAPformat}
  {}{}
}%
\makeatother
\begin{document}

\preprint{AIP/123-QED}

\title[ZORA Vibrational Corrections to Molecular Properties]{Vibrational corrections to molecular properties including relativistic corrections at the level of the Zeroth-Order Regular Approximation}
\author{Louise M{\o}ller Jessen}
\author{Ronan Gleeson}
\altaffiliation[Now at ]{Department of Energy Conversion and Storage, Technical University of Denmark, Anker Engelunds Vej, 
Building 301, DK-2800 Kongens Lyngby, Denmark.}

\author{Lars Hemmingsen}

\author{Stephan P. A. Sauer}
\email{sauer@chem.ku.dk}
\affiliation{%
Department of Chemistry, University of Copenhagen, DK-2100 Copenhagen Ø, Denmark.
}%

\date{\today}

\begin{abstract}
The vibrational averaging module of the Dalton Project was extended to work also with the Amsterdam Density Functional (ADF) program, making it possible to calculate vibrational corrections to properties and at the same time include a treatment of relativistic effects for heavier atoms at the level of the Zeroth-Order Regular Approximation (ZORA). 
To illustrate the importance of the relativistic contributions, zero-point vibrational corrections were calculated for the electric field gradient tensor and the two NMR parameters, the isotropic shielding and the spin-spin coupling constants (SSCC), of selected mercury compounds.  
For all three properties, the vibrational corrected values performed closest to experimental values, and the magnitudes of the corrections depended on the level of relativity and the basis set in the calculation. 

\end{abstract}

\maketitle

\section{Introduction}

In recent years, computational chemistry has become well-equipped to assist in interpreting $^{199}$Hg NMR and $^{199m}$Hg PAC spectroscopy parameters.
\cite{Schenberg2024_NMR,Erlemeier2025} 
For complexes with heavy atoms, inclusion of relativistic effects is important to obtain reliable results. 
There are different ways to include relativity, e.g. the fully relativistic four-component linear response methods \cite{jod970801vso, Visscher1999_4comp-relativity, Komorovsk2010a, JCP-152-184101-2020-Repisky, JPCA-124-5157-2020-Komorovsky, Saue2020}  or less computationally expensive but approximate two-component methods such as linear-response elimination of the small component (LR-ESC) \cite{Melo.LRESC1.JCP.2003, Melo.LRESC3.MolPhys.2003, Melo.LRESC2.JCP.2004, Melo.EFG1.IJQC.2019, Melo.EFG2.IJQC.2021, Melo.EFG3.JCP.2022, Melo.EFG3.JPCA.2024} or the zeroth-order regular approximation (ZORA). \cite{Chang1986, Lenthe1993_ZORA_hamiltonian, Lenthe1996_ZORA_hamiltonians, Lenthe1996_ZORA_effects, Lenthe1994_ZORA_energy, Lenthe1999_ZORA_geometry}

Arcisauskaite et al. \cite{Vaida2012_EFG,Vaida2012_relativistic,Vaida2011_NMR} investigated \textit{e.g.} the relativistic effects on the EFG tensor and the chemical shifts by comparing ZORA to the 4-component relativistic approach for mercury halides, and found ZORA to be adequate. 
In general, calculating the NMR parameters for complexes containing heavy atoms with ZORA relativistic effects has been used in many studies.\cite{Buhl2004_other-ZORA-calc, 
Kuapp2004_other-ZORA-calc, Autshbach2004_other-ZORA-calc, Autschbach2004_other-ZORA-calc, AUTSCHBACH2009_other-ZORA-calc, AUTSCHBACH2014_other-ZORA-calc, Repisky2016_other-ZORA-calc, Jankowska2016_other-ZORA-calc, Lino2020_other-ZORA-calc, Vı́cha2020_other-ZORA-calc, Glent-Madsen2021_other-ZORA-calc, Lino2022_other-ZORA-calc, Wu2023_Hg-geometry, Jessen2024} 

Mercury has two NMR active nuclei, $^{199}$Hg and $^{201}$Hg which have the spins $\frac{1}{2}$ and $\frac{3}{2}$, respectively.
Due to its accessibility with a broadband NMR probe, the $^{199}$Hg nucleus is ideally suited for NMR experiments \cite{WRACKMEYER1992_Hg-nmr}. 
The $^{199}$Hg NMR shielding constants or chemical shifts and indirect nuclear spin-spin coupling constants (SSCC) are sensitive to changes in the first coordination sphere of Hg(II),\cite{Utschig1995_MerR-NMR, Iranzo2007_Hg-NMR-first-coordinations-sphere, Luczkowski2008_Hg-NMR-first_coordinations_sphere, Wu2023_Hg-geometry, Jessen2024} and therefore $^{199}$Hg NMR spectroscopy has become a powerful tool to investigate the coordination chemistry of mercury(II) complexes in both smaller complexes and in proteins.\cite{Wagner1993_NMR-protein,Srivatsan2010_NMR-protein,Danielson1996_NMR_proteins,Duus2000_NMR-carbohydrates_proteins,Tosato2024}

The electric field gradient (EFG) is another sensitive probe of the local charge distribution, and it is affected by the chemical surroundings such as the number and type of ligands bound to the atom of interest.\cite{spas051, Vaida2012_EFG,Vaida2012_relativistic,OShea2023,Nagy2024}
It is a property that cannot be directly measured experimentally, but it can be indirectly determined from the nuclear quadruple interaction (NQI), which can be measured with $^{199m}$Hg Perturbed Angular Correlation (PAC) spectroscopy \cite{Hemmingsen2004_PAC}. PAC spectroscopy has been measured on various molecules with different nuclei including biomolecules. \cite{Iranzo2007_Hg-NMR-first-coordinations-sphere, Troger1999_MerR-EFG, Troger2001_PAC-biomolecules, Butz1992_PAC_biomolecules,Ctortecka1999_PAC_biomolecules, Faller2000_PAC_biomolecules, Haas_2017}

In the Born-Oppenheimer approximation, the motion of the electrons and nuclei is separated, and the electronic Schrödinger equation is solved for fixed nuclear geometries. 
However typically when carrying out property calculations, only the electronic Schrödinger equation is employed, which means the properties are only evaluated at a fixed geometry e.g. at the equilibrium geometry for the given molecule.
As a result, any movement of the molecule is ignored, as it is solely accounted for in the nuclear Schrödinger equation.
However, it is known that molecules vibrate even at 0K with $3N-6$ degree of freedom ($3N-5$ for linear molecules), where $N$ is the number of atoms.
The effect of vibrational motion on the spectroscopic properties is therefore not \textit{a priori} included, and isotope effects or the effect of temperature on calculated properties can thus not be accounted for with such calculations.
Therefore for proper comparison with experimental data, one has to include the effect of the nuclear motions, \textit{i.e.} average the calculated electronic properties with a vibrational wavefunction. In most cases this is done at the level of vibrational perturbation theory to second order (VPT2).\cite{Kern1968_vib_avg_original}

For the present work, we have therefore, extended the vibrational averaging module\cite{Gleeson_vib_avg_dalton-project} of the Dalton Project\cite{spas191} to perform also calculations of vibrational corrections to EFG tensors, the NMR shielding constants and spin-spin coupling constants at the ZORA level by  interfacing it to the Amsterdam Density Functional (ADF) program.\cite{ADF2023authors,Velde2001_ADF,Guerra1998-ADF} This makes it possible to include  relativity effects at the ZORA level in the calculation of vibrational corrections. One should mention that vibrational corrections to spin-spin coupling constants including relativistic effects based on 4-component relativistic calculations have recently been presented.\cite{Jakubowska2022_vib-avg_relativity}

\section{Theory}
Normally, when calculating different molecular properties within the Born-Oppenheimer approximation, the calculation relies solely on a fixed geometry (typically the equilibrium geometry P$_{eq}$) and does not include any movement of the nuclei in the molecule e.g. vibration and rotation. 
To include the vibrations, it is necessary to obtain the Boltzmann average of the property values of all the relevant vibrational states for the molecule.\cite{Kern1968_vib_avg_original,Jakubowska2022_vib-avg_relativity,Gleeson_vib_avg_dalton-project,Faber2016_vib-avg_book}
This can be derived from the expectation value of the property of interest P dependent on nuclear coordinates Q with a vibrational wavefunction $\Psi$ of the appropriate vibrational state
\begin{equation}
    \frac{\langle \Psi | P(Q) | \Psi\rangle}{\langle \Psi|\Psi\rangle} .
\end{equation}

Therefore, a correction $\Delta P$ is defined as the difference between the expectation values of the property in a given vibrational state and the value of the property at the equilibrium geometry
\begin{equation}
    \Delta P = \frac{\langle \Psi | P(Q) | \Psi\rangle}{\langle \Psi|\Psi\rangle} - P_{eq} .
\end{equation}

This has been derived first by Kern et al.\cite{Kern1968_vib_avg_original} by using second-order vibrational perturbation theory. They used an equilibrium geometry approach, which means the perturbation expands around the minimum of the potential-energy surface although alternative expansion points have also been suggested.\cite{pro000208rat} 
The formula for the zero-point vibrational correction to the property is given as
\begin{equation}
    \Delta ^{VPT2}P= -\frac{1}{4} \sum\limits_{i} \frac{1}{\omega_i} \frac{\partial P}{\partial q_i}\Bigg|_{q=0} \sum\limits_{j}k_{ijj}+\frac{1}{4}\sum\limits_{i}\frac{\partial^2 P}{\partial q_i^2}\Bigg|_{q=0}
    \label{equation:correction}
\end{equation}
where $q_i$, $k_{ijj}$, and $\omega_i$ are the reduced normal coordinates, the off-diagonal reduced cubic force constant (cm$^{-1}$), and the harmonic vibrational frequency (cm$^{-1}$), respectively. 
The reduced normal coordinates are defined as: 
\begin{equation}
    q_i = \sqrt{\frac{2\pi c \omega_i}{\hbar}}Q_i
\end{equation}
where c is the speed of light (cm s$^{-1}$) and $\hbar$ is the reduced Planck constant; the harmonic vibrational frequency is defined as: 
\begin{equation}
    \omega_i = \sqrt{\frac{\partial^2 E_0^{(0)}}{\partial Q_i^2}}
\end{equation}
and the elements of the reduced cubic force constant tensor are defined as:
\begin{equation}
    k_{ijk} = \frac{\partial^3 E_0^{(0)}}{\partial q_i \partial q_j \partial q_k}
\end{equation}
which represent the anharmonicity of the wavefunction. They are calculated numerically from the Hessian by generating different distorted geometries for each vibrational mode in the molecule.\cite{schneider1989anharmonic, barone2005anharmonic}
The number of distorted geometries varies and depends on which stencil is used (equation \ref{equation:geometry_distortion}). The program has both a 3-point stencil and a 5-point stencil implemented where the 5-point stencil increases the accuracy. 
\begin{equation}
    \Delta x_i = x_0 + t(Q_ih) \quad \quad
    t=\begin{cases}
    \pm 1 & \text{if 3-point stencil}\\
    \pm 1, \pm 2 & \text{if 5-point stencil}
    \end{cases}
    \label{equation:geometry_distortion}
\end{equation}

Likewise are the first and second derivatives of the property calculated numerically.
%
For the 5-point stencil, the first and second derivatives are computed as 
\begin{equation}
    f'(x_0) \approx \frac{-f(x_0+2h) + 8f(x_0+h)-8f(x_0-h)+f(x_0-2h)}{12h} 
\end{equation}
\begin{equation}
    f''(x_0) \approx \frac{-f(x_0+2h)+16f(x_0+h)-30f(x_0)+16f(x_0-h)-f(x_0-2h)}{12h^2}
\end{equation}
with $x_0$ being the equilibrium point and \textit{h} being the step length. 
The first derivative excludes reference to the equilibrium point, whereas the second derivative includes the functional form around the equilibrium point, i.e. $f(x_0)$.
Further details are described in the previous work.\cite{Gleeson_vib_avg_dalton-project}
Since numerical differentiation is used, it is important to choose an appropriate step length to ensure accurate results. If the chosen step length is too small, numerical errors will dominate, and if the step length is too large, higher-order terms will contaminate the derivatives.\cite{Jakubowska2022_vib-avg_relativity} 

\section{Implementation}
The Dalton project\cite{spas191} is a set of Python scripts that among other things includes a vibrational averaging module,\cite{Gleeson_vib_avg_dalton-project}  which was originally implemented in the Dalton Project by Gleeson et al. \cite{Gleeson_vib_avg_dalton-project} using the Dalton program\cite{Dalton2014} for the electronic structure calculations. Since then it has been extended to include an interface to Gaussian. Multiple properties have already been implemented in the module such as polarizability, hyperfine couplings, NMR shielding, and spin-spin coupling constants.

In present study, the vibrational averaging module has been extended to interface to the ADF program for the properties: NMR shielding constants, NMR indirect nuclear spin-spin coupling constants, and electric field gradient tensors, where the latter had previously not been implemented at all.
Including ADF in the module opens the possibility of carrying out the property calculations with a ZORA treatment of relativity, which has been added as an option for the vibrational corrections in \fbox{\textcolor{blue}{qcmethod.py}}. 
Thereby, the corrections can be carried out non-relativistic, scalar relativistic, or spin-orbit relativistic, which can influence the magnitude of the correction, especially for the heavier atoms. 


\subsection{NMR parameters}
As previously mentioned the properties, NMR shielding and SSCC, were already implemented in the module. To include these in vibrational averaging for ADF all the keywords for the input files for the properties were added to the \fbox{\textcolor{blue}{program.py}} script (special for ADF), and the code to read the output files was added to the \fbox{\textcolor{blue}{output\_parser.py}} script (special for ADF). 
For NMR shielding, one function was needed in the \fbox{\textcolor{blue}{output\_parser.py}} script, \fbox{\textcolor{blue}{nmr\_shieldings}} which finds the isotropic shielding tensor for all the distorted geometries. 
For SSCC, two functions were needed in the \fbox{\textcolor{blue}{output\_parser.py}} script, the \fbox{\textcolor{blue}{spin\_spin\_couplings}} that finds the four terms of the SSCC for all the distorted geometries, and the \fbox{\textcolor{blue}{spin\_spin\_labels}} that finds which atoms that are coupled. 
For these two properties, the keyword \textit{DEPENDENCY} is added automatically to the input file, since some molecules have problems with convergence in the property calculation. 
Likewise, to avoid problems in the SSCC calculation, the max number of iterations is changed from the default 25 to 200, just in case some molecules/distorted geometries need it.

\subsection{Electric Field Gradient}
The electric field gradient is a property where typically all the eigenvalues of the diagonalized tensor are of interest and not only the isotropic value. 
Therefore, the vibrational corrections are calculated for all the elements in the 
3x3 EFG tensor individually and only the vibrationally corrected tensor is diagonalized in order to obtain the vibrationally corrected eigenvalues. 
In principle, since the tensor is symmetric it would only necessary to obtain the corrections for six of the nine elements. 
The vibrational averaging module produces corrections as a new 3x3 tensor (\textbf{V$_{\text{Corr}}$}), which is then added to the undiagonalized tensor of the equilibrium geometry (\textbf{V$_{\text{eq}}$}) that is then diagonalized. 

\begin{equation}
    \textbf{V$_{\text{eq}}$} + \textbf{V$_{\text{Corr}}$}=
    \begin{pmatrix}
    V_{xx} & V_{xy} & V_{xz}\\
    V_{yx} & V_{yy} & V_{yz} \\
    V_{zx} & V_{zy} & V_{zz} \\
    \end{pmatrix}
    + 
    \begin{pmatrix}
    \Delta V_{xx} & \Delta V_{xy} & \Delta V_{xz}\\
    \Delta V_{yx} & \Delta V_{yy} & \Delta V_{yz} \\
    \Delta V_{zx} & \Delta V_{zy} & \Delta V_{zz} \\
    \end{pmatrix}
\end{equation}

When running a vibrational correction calculation for the EFG property, the program automatically adds the two tensors and diagonalizes them in the new function \fbox{\textcolor{blue}{diagonalize\_EFG\_tensor}} which was added to the \fbox{\textcolor{blue}{ComputeVibAvCorrection}} class in the \fbox{\textcolor{blue}{vibrational\_averaging.py}} script. 
It saves the vibrational corrected eigenvalues in a new file called \fbox{\textcolor{blue}{EFG\_diagonalized\_tensor.txt}}.  

If the vibrational correction for the EFG tensor is added to another interface, three functions in the \fbox{\textcolor{blue}{output\_parser.py}} are needed to extract all the information. 
The function \fbox{\textcolor{blue}{efg}} extracts the EFG tensor from the output files for all the distorted geometries that the program uses to calculate the corrections.  
The function \fbox{\textcolor{blue}{efg\_labels}} extracts 
the element and the isotope of the element. 
The function \fbox{\textcolor{blue}{efg\_std\_prop}} extracts the EFG tensor for the output file of the equilibrium geometry, which is used in \fbox{\textcolor{blue}{diagonalize\_EFG\_tensor}} to add the corrections to and then diagonalize. 

\section{Computational details}
\subsection{Step length analysis}
A step length analysis was performed on the molecule HgCl$_2$ for the properties NMR shielding, SSCC, and EFG. 
To ensure the step length analysis would be as representative as possible, the analysis was carried out with the spin-orbit ZORA method in ADF with the functional BHandHLYP \cite{Becke_BHandH_BHandHLYP} and basis set QZ4P for NMR shielding and EFG calculations, and the basis sets QZ4P and QZ4P-J \cite{Bryce2009_basis-set_TZ2P-J-QZ4P-J} for SSCC calculations (QZ4P-J only in the SSCC property calculation). 
To ensure that the geometry stayed the same in the analysis, the geometry obtained from one geometry optimization was reused for all the calculations. 
Therefore, the same frequencies and cubic force constants were reused for all the properties at each step length.

\subsection{Testing the dependence on basis set}
To see how much the basis set influences the vibrational corrections, multiple vibrational averaging calculations were carried out for HgCl$_2$ with the spin-orbit ZORA method and the functional BHandHLYP. 
To see how the corrections compare to the one run with QZ4P, the calculations were carried out with DZ, TZP and TZ2P(-J).  

\subsection{Testing the dependence on relativistic effects}
The effect of the relativistic effects on the vibrational corrections were investigated for the mercury-halides (HgCl$_2$, HgBr$_2$, and HgI$_2$) and Hg(SH)$_2$ electric field gradients, for the methylmercury-halides (H$_3$CHgCl, H$_3$CHgBr, and H$_3$CHgI) isotropic shielding constants and for the methylmercury-halides SSCCs. 
For all molecules, a non-relativistic, a scalar ZORA, and a spin-orbit ZORA calculation were carried out using the functional BHandHLYP for the electric field gradient and the functional PBE0\cite{Ernzerhof1999_PBE0} for the isotropic shielding constants and the SSCCs.
All the vibrational averaging calculations began with a geometry optimization with the basis set QZ4P, and the same treatment of relativity as the following property calculations. 
The basis set used in the property calculations was QZ4P for all calculations except for SSCC which used the QZ4P-J basis set. 
Due to problems with convergence in the NMR shielding calculation, the keyword \textit{DEPENDENCY} was added for all the molecules. 

\subsection{Comparison to experimental values}
In order to analyze, if a vibrational correction improves the results relative to experiment, experimental values were found for the three properties: EFG, NMR chemical shift, and SSCC. 
The molecules are HgCl$_2$, HgBr$_2$, HgI$_2$, CdCl$_2$, CdBr$_2$ and CdI$_2$ for the EFG; Hg(CH$_3$)$_2$, the mercury-halides (HgCl$_2$, HgBr$_2$, and HgI$_2$) and the methylmercury-halides (H$_3$CHgCl, H$_3$CHgBr, and H$_3$CHgI) for the chemical shifts and H$_3$CHgCl, H$_3$CHgBr, and H$_3$CHgI for the SSCCs.
The electronic structure calculations of the molecular properties and Hessians at the equilibrium and distorted geometries as well as the geometry optimizations, were carried out with the ADF program with spin-orbit ZORA treatment of relativity included. 
Two different functionals were used, PBE0 for NMR shielding and SSCC calculations with the basis set QZ4P and QZ4P-J, respectively, and BHandHLYP for the EFG calculations with the basis set QZ4P, which was found to be optimal for the respective properties in previous studies.\cite{Vaida2012_relativistic,Wu2023_Hg-geometry} 
Before the property calculations were carried out, a geometry optimization was performed with the same functional as the proceeding property calculation and the QZ4P basis set. 
To see the full effect of the vibrational correction, solvent 
effects are excluded in present study. 

\section{Results and discussion}
With the vibrational averaging module of the Dalton Project extended to ADF, the performance has been investigated 1) with different basis sets, 2) with different levels of treatment of relativistic effects, and 3) when compared to experimental values. 

\subsection{Step length analysis}
To carry out vibrational averaging calculations with numerical derivatives, an appropriate step length has to be chosen to ensure numerically accurate and stable results. 
If the step length is too large, the derivatives could potentially be influenced by higher-order terms, and if it is too small, numerical errors will dominate due to the approximate solution.\cite{Jakubowska2022_vib-avg_relativity}

In previous studies \cite{Faber2012_vib_avg_SSCC, spas154, spas160, Gleeson_vib_avg_dalton-project} the vibrational corrections for SSCC were performed with a step length of 0.05 (in reduced normal coordinates) as this is the default value also in other quantum chemistry programs. However, it should be noted that in the previous studies the atoms in the molecules of investigation were all from the second period of the periodic table and hydrogen.
Furthermore, in earlier work on individual molecules, more extended property surfaces were generated.\cite{spas002, spas011, spas018, spas020, spas022, spas026, spas028, spas032, spas034, spas036, spas049, spas051, spas093}

\begin{figure*}[h!bpt]
    \centering
    \includegraphics[width=1.\textwidth]{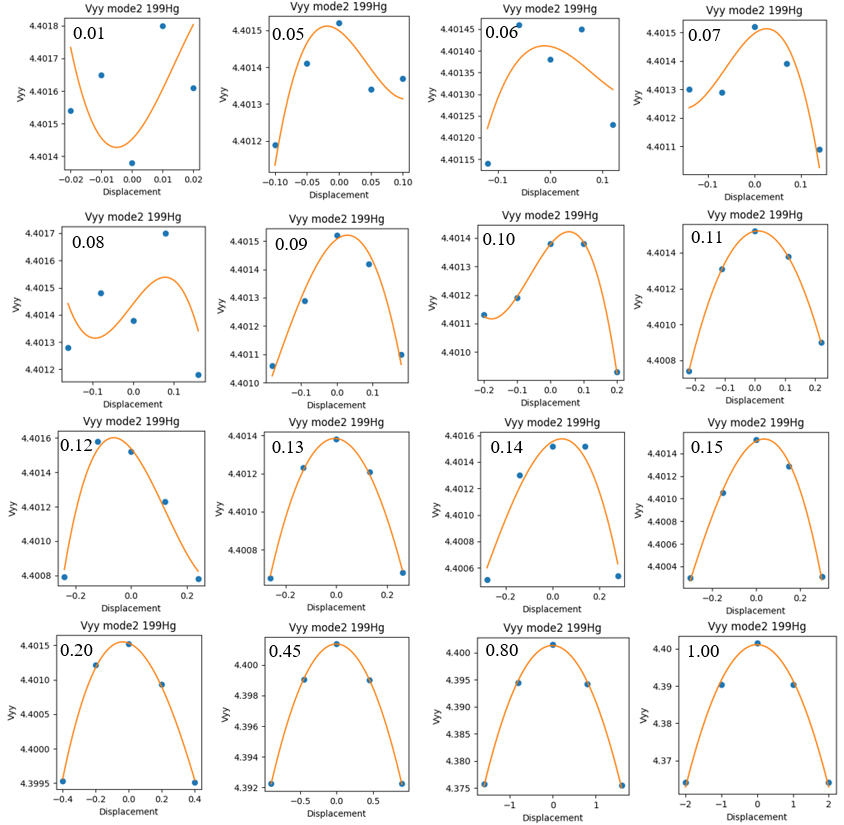}
    \caption{An example for the variations in the fitting of the changes in V$_{\text{yy}}$ (in au) of $^{199}$Hg in HgCl$_2$ along the normal mode 2 with increasing step length. The step lengths can be found in the upper left corner of each graph. }
    \label{fig:vib-avg_step-length_all_small}
\end{figure*}
In the present study, an preliminary investigation concluded a step length of 0.05 was not adequate for mercury complexes in ADF. It was simply too small. 
Therefore, a thorough investigation for an appropriate step length was performed on HgCl$_2$ (at ZORA/BHandHLYP/QZ4P level). 
The vibrational averaging was performed for 17 different step lengths in the range of 0.01 and 1.20 with a five-point stencil for EFG, isotropic shielding, and SSCC for HgCl$_2$. 
Initially it was assumed that the step length would be found in the range between 0.05 and 0.20, and therefore most of the 17 investigated step lengths were within this range. 
Figure \ref{fig:vib-avg_step-length_all_small}, for V$_{\text{yy}}$ of $^{199}$Hg in HgCl$_2$ along the normal mode 2, illustrates how the fit of the variation in a molecular property along a vibrational mode can change with different step lengths, where a step length between 0.45 and 1.00 obtained the best fit. 

\begin{figure}[h!]
    \centering
    \includegraphics[width=0.5\textwidth]{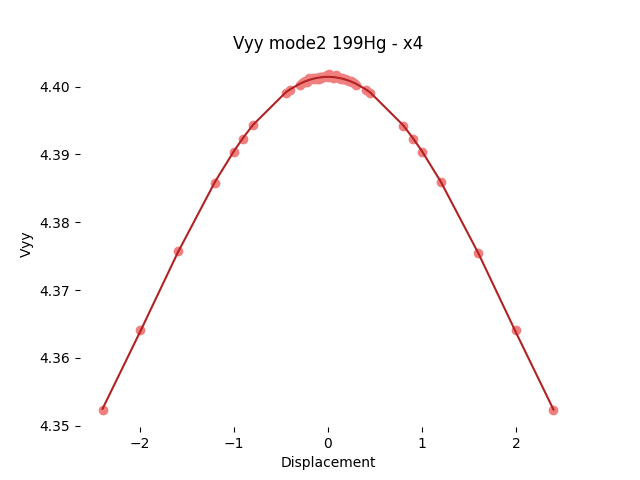}
    \caption{An example of the step length analysis with 61 points and a fourth-degree polynomial of the variation of  V$_{\text{yy}}$ (in au) of $^{199}$Hg in HgCl$_2$ with respect to mode 2.}
    \label{fig:vib-avg_step-length}
\end{figure}

To determine the most appropriate step length for each property, all the values of the properties and the displacement obtained from all the distorted geometries and vibrational modes were plotted together in one graph for each mode of each atom. 
This produced a graph with 61 points (illustrated in Figure \ref{fig:vib-avg_step-length} for the same property), from which the first and second derivatives were extracted from a polynomial regression. 
The polynomial regressions were performed with multiple orders (third-, fourth-, fifth-, sixth-, and seventh-order), to find the accuracy of the number of decimals for each derivative. 
The derivatives can be found in Tables \ref{SI-app:vib-avg_step-length_EFG}, \ref{SI-app:vib-avg_step-length_shielding}, and \ref{SI-app:vib-avg_step-length_sscc} in the supplementary material for the EFG, the isotropic shielding and the SSCCs.  

The derivatives of the 61-points were then compared with the derivatives (Tables \ref{SI-app:vib-avg_EFG_derivatives_of_all_step-length}, \ref{SI-app:vib-avg_shielding_derivatives_of_all_step-length}, and \ref{SI-app:vib-avg_SSCC_derivatives_of_all_step-length} in the supplementary material for the EFG, the isotropic shielding and the SSCCs) of all the separate calculations with only one step length, and a step length of 0.50 was found adequate for all three properties, which is 10 times larger than the value used in previous studies. 

This analysis was performed on HgCl$_2$ (at ZORA/BHandHLYP/QZ4P level) due to its small size, thus keeping the computational cost low. 
For another complexs, the step length may need to be adjusted.
Furthermore, while the isotropic shielding is mainly sensitive to changes in the first coordination sphere,\cite{Wu2023_Hg-geometry} the the EFG tensor is also sensitive to changes in the second coordination sphere.\cite{Hemmingsen2004_PAC} 
Therefore, a change in the bond length between mercury and another atom would affect these properties significantly, and the step length would need some adjustment, which also can explain the large difference in the appropriate step length between the present and previous studies.
Performing vibrational averaging of properties with numerical derivatives, one should thus not blindly rely on the standard settings but investigate the effect of different of step lengths. Furthermore, it would be interesting to investigate how the size of a suitable step length correlates with the magnitude of the bond length.

\subsection{Dependence on basis sets}
In the previous section, all the vibrational averaging calculations have been carried out with the basis set QZ4P. 
However, due to the possibility of reducing computational cost, it was investigated if the choice of basis sets affected the corrections. 

\begin{figure*}[hbpt]
    \centering
    \includegraphics[width=1.0\textwidth]{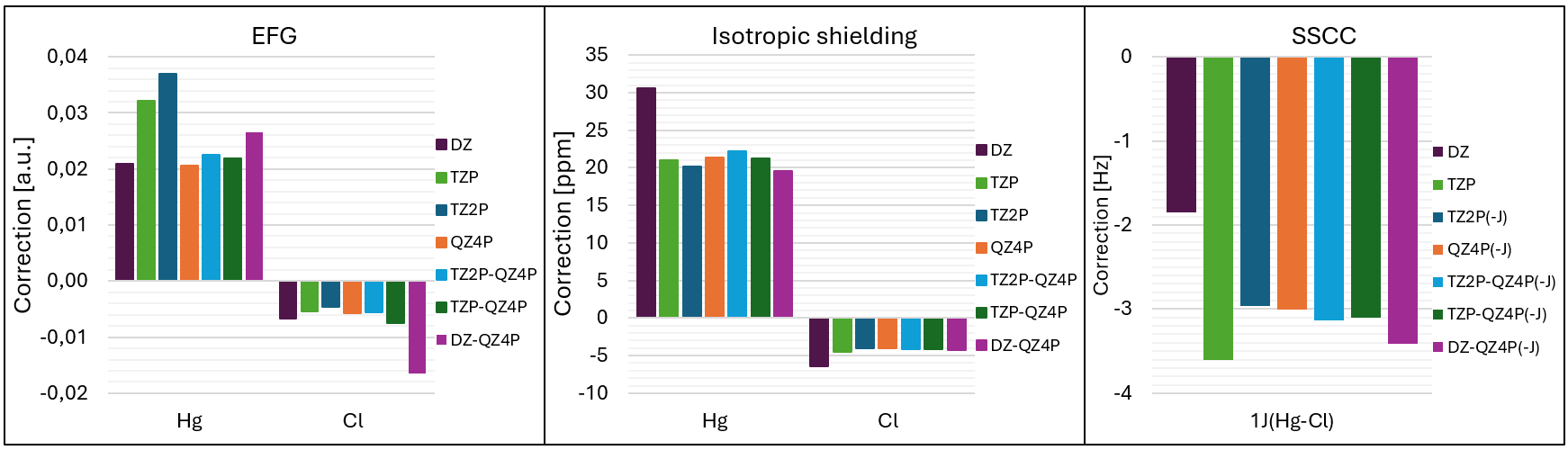}
    \caption{The vibrational corrections for HgCl$_2$ calculated at the spin-orbit ZORA level with different basis sets and the BHandHLYP functional. To the left is the correction to the V$_{zz}$, in the middle is the correction to the isotropic shielding, and to the right is the correction to the SSCC. The values can be found in section \ref{SI-app:sec_vib-avg_basis-set} of the supplementary material.}
    \label{fig:vib-avg_basis-set}
\end{figure*}

Therefore, vibrational averaging calculations were carried out for the three properties and HgCl$_2$ as test molecule. 
In the first series of calculations we employed the DZ, TZP, TZ2P(-J) and QZ4P(-J) basis sets in both the geometry optimization and all the property calculations. In the second series we used a set of mixed basis set with QZ4P(-J) in the geometry optimization and all the property calculations and a smaller basis set in the calculation of the cubic force constants, meaning in the calculation of the Hessian matrix at the distorted geometries. 
The J-version of the basis sets were only employed in the property calculation for SSCC. The results are illustrated in Figure \ref{fig:vib-avg_basis-set} with the values given in Tables \ref{SI-app:vib-avg_basis-set_EFG}, \ref{SI-app:vib-avg_basis-set_shielding} and \ref{SI-app:vib-avg_basis-set_sscc} of the supplementary material.

For the vibrational corrections for V$_{zz}$, the corrections are generally small with differences from QZ4P being below 0.02 a.u. (Table \ref{SI-app:vib-avg_basis-set_EFG}). 
For mercury, the correction for DZ is the same as QZ4P, and the mixed basis set combinations differ by 0.001 a.u. (TZP-QZ4P) and 0.006 a.u. (DZ-QZ4P). 
The largest differences from QZ4P are 0.011 a.u. and 0.016 a.u. for TZP and TZ2P, respectively.  
For chlorine, the corrections are all within 0.001 a.u. of QZ4P except DZ-QZ4P which differs by 0.010 a.u.. 

For the correction of the isotropic shieldings (Table \ref{SI-app:vib-avg_basis-set_shielding}), the basis set that performs closest to QZ4P for mercury is TZP-QZ4P (difference 0.082 ppm) followed by TZP (difference 0.295 ppm). 
For chlorine, the correction for TZ2P is closest to QZ4P by a difference of 0.013 ppm. The corrections for DZ are furthest away from QZ4P which differ by 9.198 ppm and 2.331 ppm for mercury and chlorine, respectively. 

For the SSCC (Table \ref{SI-app:vib-avg_basis-set_sscc}), the correction for TZ2P is closest to QZ4P with a difference of 0.044 Hz, and the one furthest away is the correction to DZ (difference 1.159 Hz). 

Overall, when looking at all the properties, the corrections closest to QZ4P are from the basis set combination TZP-QZ4P, 
followed by the combination TZ2P-QZ4P. 
It should be noted that the calculations with mixed basis sets are, however, not a consistent method, since the frequencies and cubic force constants are then calculated from a non-equilibrium geometry for that method, since the geometry optimizations were performed with QZ4P and the calculations of the Hessian at distorted geometries were performed with TZ2P. 
For the not-mixed basis set, TZ2P gives the results closest to the QZ4P results. 
The basis set that overall gives results furthest away from QZ4P is DZ. 

\subsection{Dependence on relativistic effects}
\begin{figure*}[hbpt]
    \centering
    \includegraphics[width=1.0\textwidth]{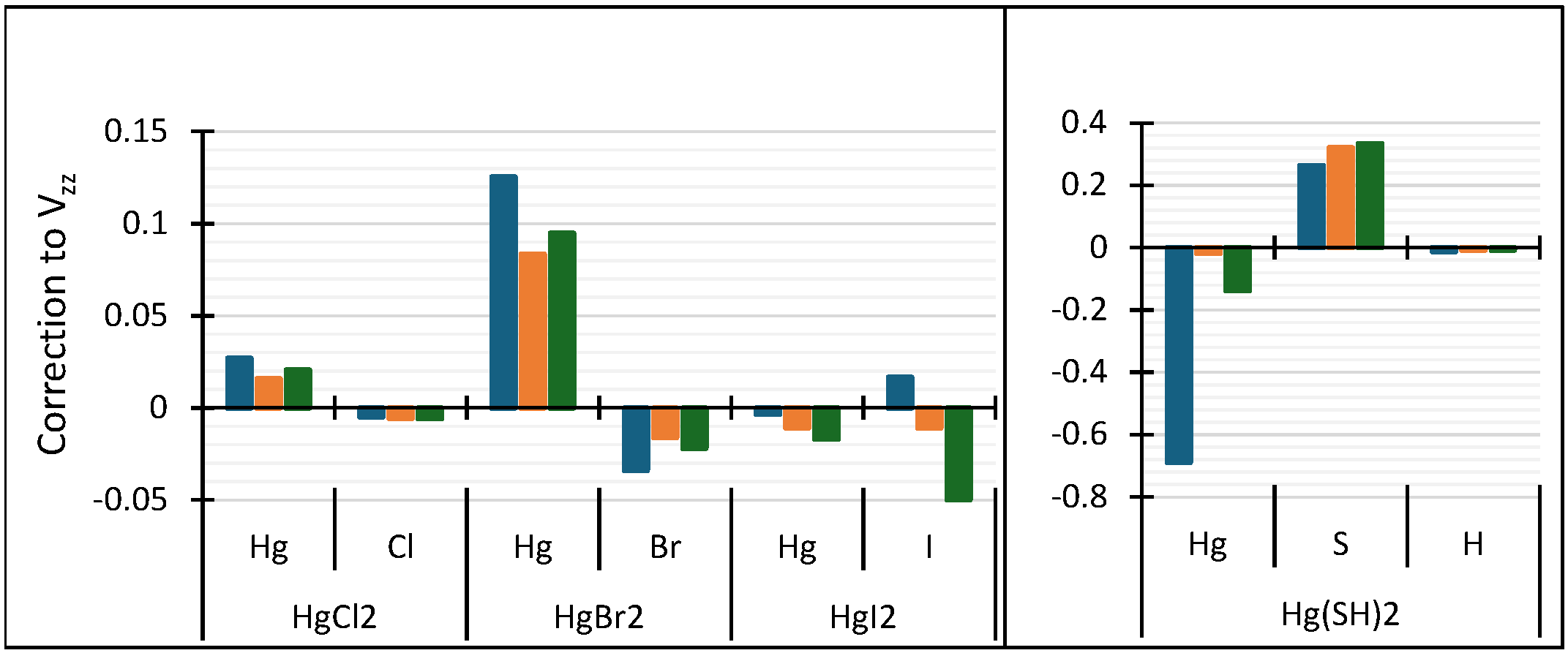}
    \caption{The correction for the electric field gradients $V_{zz}$ calculated with BHandHLYP/QZ4P at different levels of relativity. 
    Without relativistic corrections (none) is shown in blue, with scalar ZORA relativistic corrections (scalar) is shown in orange, and with also spin-orbit ZORA relativistic corrections (spin-orbit) is shown in green. 
    The values and corrections can be found in section \ref{SI-app:sec_vib-avg_relativity} of the supplementary material.}
    \label{fig:vib-avg_relativity-efg}
\end{figure*}

\begin{figure*}[hbpt]
    \centering
    \includegraphics[width=1.0\textwidth]{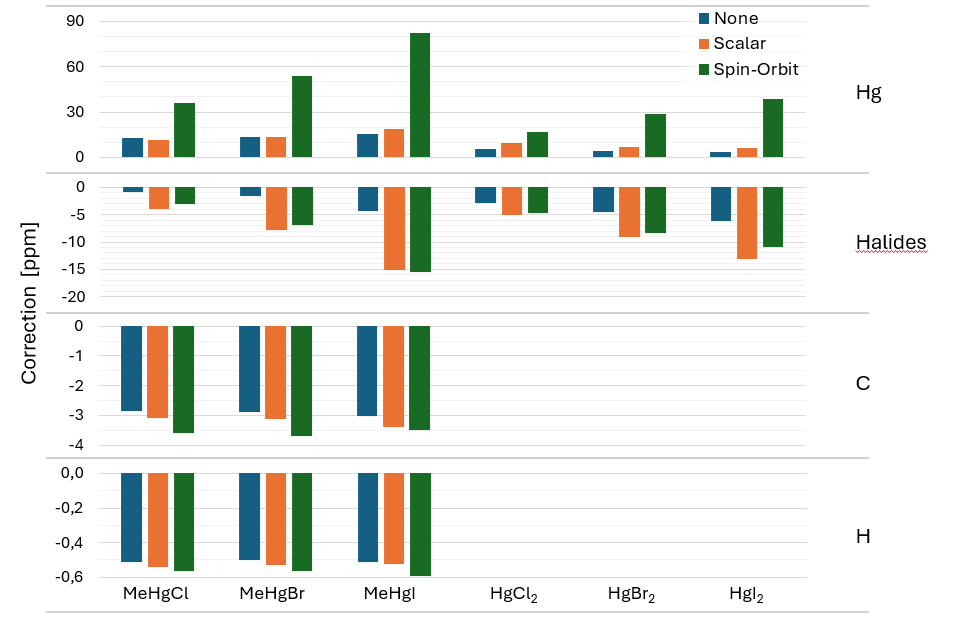}
    \caption{The correction for the isotropic shielding constants [in ppm] calculated with PBE0/QZ4P at different levels of relativity. 
    Without relativistic corrections (none) is shown in blue, with scalar ZORA relativistic corrections (scalar) is shown in orange, and with also spin-orbit ZORA relativistic corrections (spin-orbit) is shown in green. 
    The values and corrections can be found in section \ref{SI-app:sec_vib-avg_relativity} of the supplementary material.}
    \label{fig:vib-avg_relativity-shield}
\end{figure*}

\begin{figure*}[hbpt]
    \centering
    \includegraphics[width=1.0\textwidth]{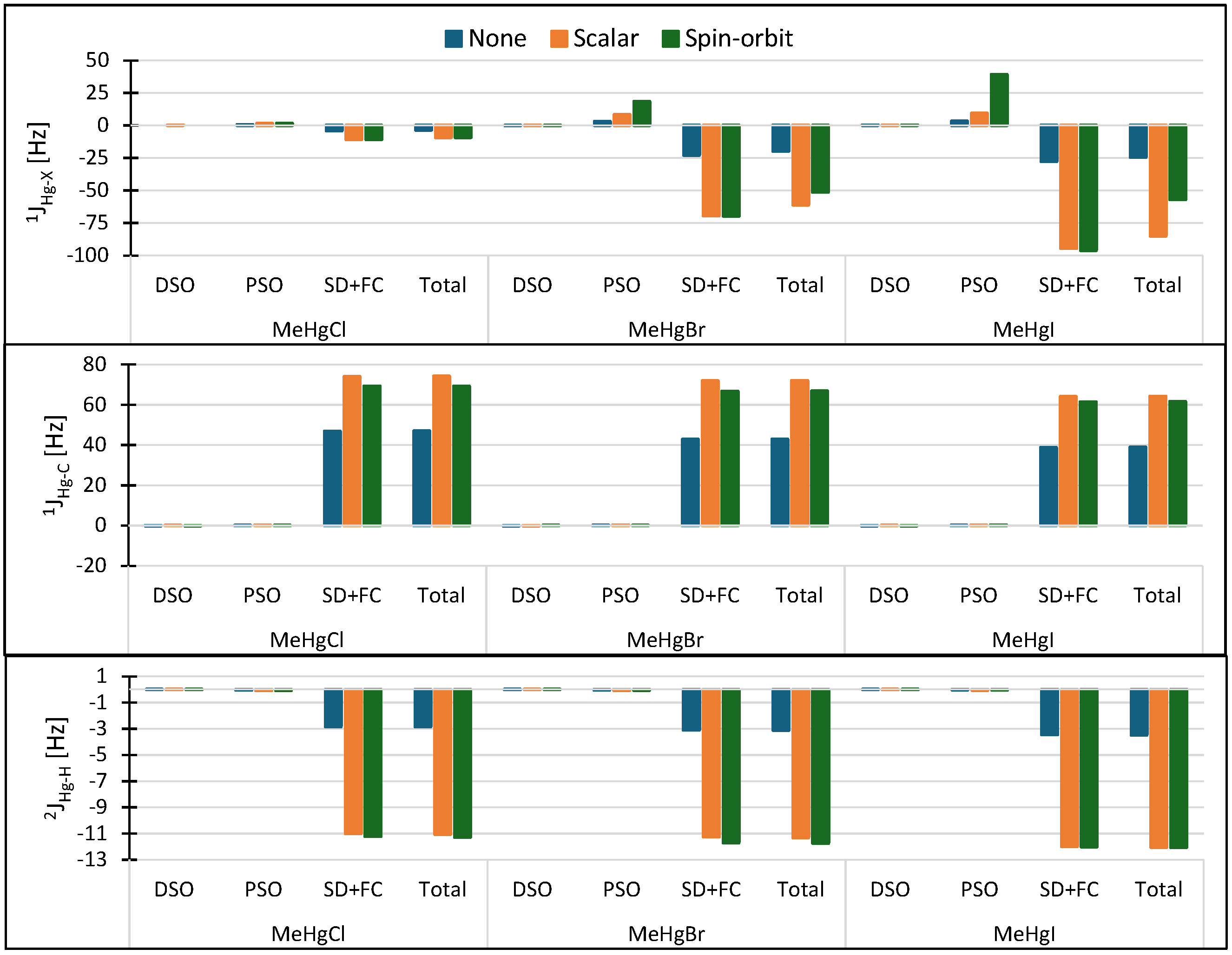}
    \caption{The correction for the $^1$J$_{\text{Hg-X}}$, $^1$J$_{\text{Hg-C}}$ and $^2$J$_{\text{Hg-H}}$ SSCCs in the methylmercury-halides calculated with PBE0/QZ4P-J at different levels of relativity. 
    Without relativistic corrections (none) is shown in blue, with scalar ZORA relativistic corrections (scalar) is shown in orange, and with also spin-orbit ZORA relativistic corrections (spin-orbit) is shown in green. 
    The values and corrections can be found in section \ref{SI-app:sec_vib-avg_relativity} of the supplementary material.}
    \label{fig:vib-avg_relativity}
\end{figure*}

Inclusion of relativistic effects increases the computational cost and therefore it was investigated, if the inclusion of relativistic effects influences the magnitude of the vibrational correction. Otherwise the vibrational correction could be calculated without relativistic effects and then added to an equilibrium geometry value calculated with relativistic effects included. 
The vibrational corrections from this investigation are illustrated in Figures \ref{fig:vib-avg_relativity-efg} - \ref{fig:vib-avg_relativity}.

In general, the corrections for mercury differ much more than the corrections for the other atoms for both EFG and isotropic shielding, which is not surprising, since the values for mercury also differ significantly when calculated at different levels of relativity. 

With respect to the electric field gradients V$_{zz}$ in HgCl$_2$, HgBr$_2$, HgI$_2$ and Hg(SH)$_2$ in Figure \ref{fig:vib-avg_relativity-efg} and Table \ref{SI-app:vib-avg_relativity_EFG}, there is a clear difference between the lighter mercury-halides  (HgCl$_2$ and HgBr$_2$) and Hg(SH)$_2$ on one side and HgI$_2$ on the other.
For the compounds with the lighter ligands, Cl, Br and SH, the largest absolute vibrational corrections of V$_{zz}$ of mercury arises from the calculation without relativitic effects included.
The next largest vibrational corrections are with scalar and spin-orbit coupling included, and the smallest corrections are at the scalar level of relativity.  
For HgI$_2$ with two heavy elements, the largest vibrational correction is obtained with scalar and spin-orbit coupling included and the smallest one in the calculation without relativistic effects included. 
HgI$_2$ differs also from the lighter halide compounds, because the vibrational corrections to the EFG of Hg are negative in contrast to Hg in HgCl$_2$ and HgBr$_2$. 
The same is actually also the case for Hg in Hg(SH)$_2$.
For the corrections for chlorine and sulfur, the calculations at scalar and spin-orbit levels of relativity are similar, which indicates, that if one of these atoms is of interest, relativistic effects may be included (depending on the desired precision). 
For bromine in HgBr$_2$ the same trend as for the Hg EFG is observed, the vibrational correction without relativistic treatment is largest followed by the correction at the spin-orbit level.
Finally, for hydrogen in Hg(SH)$_2$ all the corrections are alike, which illustrates that it is unnecessary to include relativistic effects in the calculation of the vibrational corrections for the EFG of hydrogen. 

The vibrational corrections to the isotropic shielding constants in the mercury-halides and methylmercury-halides are shown in Figure \ref{fig:vib-avg_relativity-shield} and Table \ref{SI-app:vib-avg_relativity_shielding}.
The vibrational corrections to the isotropic shielding constants of mercury increase when the level of relativistic treatment increases. 
By including spin-orbit coupling, the corrections more than triple, whereas the differences between scalar and no relativistic corrections are not larger than 4 ppm.
For the halides, the corrections at scalar and spin-orbit levels of relativity are similar with the largest difference for iodine in HgI$_2$, where the vibrational correction at the scalar relativistic level is by 2 ppm more negative.
In addition, both for Hg and the halides the importance of a relativistic to the vibrational corrections increases strongly from the chlorine to the iodine compounds. For Hg it is the spin-orbit correction, which is very important and increases strongly, whereas for the halide atoms it is mostly the scalar relativistic effects which dominate. 
For carbon and the hydrogens in the methyl groups, the vibrational corrections are similar but increase slightly with the level of relativistic treatment. 

Finally, for SSCCs of the methylmercury-halides (Figure \ref{fig:vib-avg_relativity} and Tables \ref{SI-app:vib-avg_relativity_sscc} and \ref{SI-app:vib-avg_relativity_sscc-1}), the corrections were calculated for each of the terms DSO (diamagnetic spin–orbit), PSO (paramagnetic spin–orbit), and FC+SD (Fermi-contact and spin-dipole (only given as the sum in ADF)). The total correction is the sum of the corrections for each term. 
In general, the vibrational corrections of the DSO term are very small with a maximum of 0.01 Hz. 
The vibrational corrections to the PSO contribution to the mercury-halogen one-bond couplings increase with an increase in relativity, both in respect to the relativistic treatment and in the series from chlorine to iodine.
For the other couplings, the vibrational corrections to the PSO term are not larger than 0.2 Hz and largest, when spin-orbit coupling is included.
The vibrational corrections to FC+SD terms are the most relevant for all three types of couplings, and the total corrections mostly depend on this term. 
Scalar relativistic contributions are very important for all three types of couplings but the changes due to the inclusion of also spin-orbit coupling are small. They are actually largest for the one-bond mercury-carbon couplings.
For the one-bond mercury halogen couplings and to a lesser extent also for the two-bond mercury hydrogen coupling, one observes that both the vibrational corrections and the importance of including relativistic effects increase from the chlorine to the iodine compounds.  
For the one-bond mercury carbon coupling on the other hand, the opposite trend is observed.

Overall for all the properties of the heavier atoms, the corrections vary for the different levels of relativity. 
This indicates, that it is necessary to include spin-orbit relativistic effects in the vibrational averaging. 
It is important to remember, that even though the absolute magnitude for the corrections calculated at the non-relativistic or scalar relativistic level were sometimes larger, it does not mean that they describe the vibrational corrections better. 

\subsection{Comparison to experimental values}
To establish whether the vibrational corrections improve the values or not, the value at equilibrium geometry and the vibrational corrected value are compared to experimental values in this section. 

\subsubsection{Electric Field Gradient} 
For the EFG tensor, there is not a direct experimental value. However, V$_{zz}^{exp}$ can be derived from the measurement of the quadrupole coupling constant, $\nu_Q$
\begin{equation}
    \nu_Q = \frac{eQ^{e}V_{zz}}{h}
\end{equation}
where $h$ is Planck's constant and $e$ the elementary charge. $Q^{e}$ is the nuclear quadruple moment for the nucleus which also needs to be known, and the standard error on this experimentally determined property will propagate to $\nu_Q$. 
To only compare the calculated V$_{zz}$ to the experimental quadrupole coupling constant, the ratio for complexes with the same atom of interest was calculated, which made the nuclear quadruple moment redundant. 
\begin{equation}
    \frac{\nu_Q^{mol1}}{\nu_Q^{mol2}} = \frac{V_{zz}^{mol1}}{V_{zz}^{mol2}}
\end{equation}

\begin{figure*}[hbpt]
    \centering
    \includegraphics[width=1.0\textwidth]{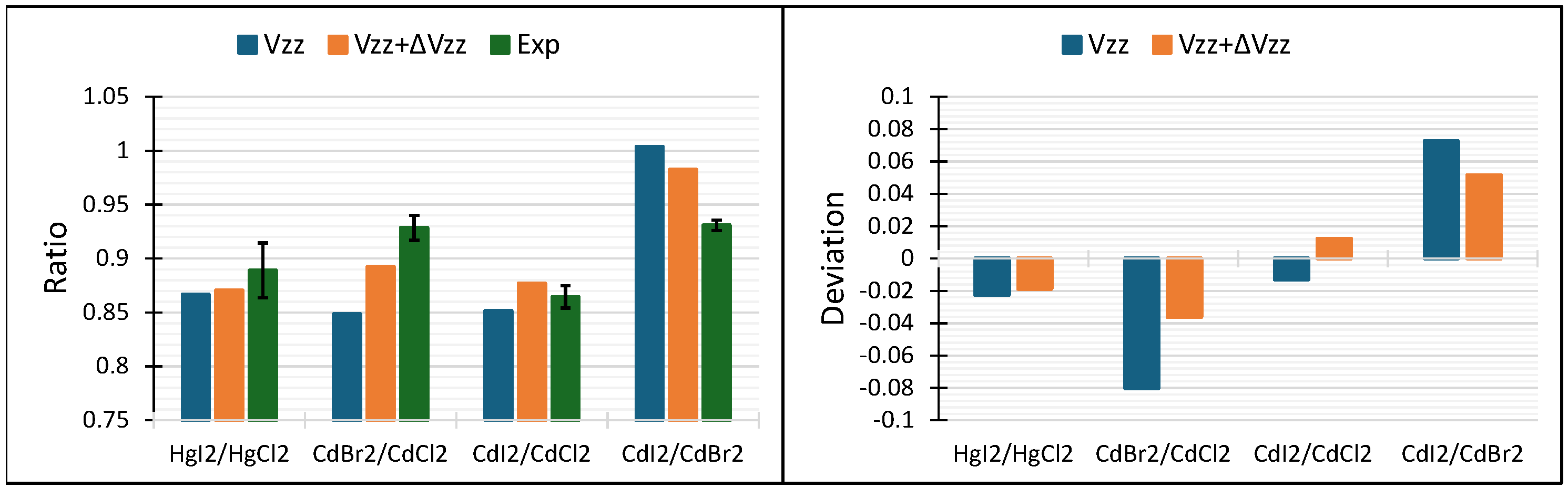}
    \caption{The ratio between V$_{zz}$ for two mercury-halides and between the three cadmium-halides calculated with ZORA(spin-orbit)/BHandHLYP/QZ4P. 
    To the left are the values for the ratio calculated at the equilibrium geometry (blue), the ratio with vibrational correction (orange), and the ratio of the experimental values (green), where the uncertainties are the dotted green line. 
    To the right are the deviations from the ratio of the experimental values. }
    \label{fig:vib-avg_EFG_calc_vs_exp}
\end{figure*}

The atoms of interest in this investigation are $^{199}$Hg and $^{111}$Cd in the compounds HgI$_2$, HgCl$_2$, CdI$_2$, CdBr$_2$ and CdCl$_2$. 
The calculated vibrational corrections and the experimental values \cite{Haas2021_EFG_exp-vibavg_Q} can be found in Table \ref{SI-app:vib-avg_calc-vs-exp_EFG} in the supplementary material.
The ratios of V$_{zz}$ for two of these compounds are illustrated in Figure \ref{fig:vib-avg_EFG_calc_vs_exp}, where the uncertainty of the experimental measurements are shown as error bars on the experimental ratios (Exp). The numerical values of the ratios can be found in Table \ref{SI-app:vib-avg_calc-vs-exp_EFG_ratio} in the supplementary material. 

It can be seen that adding the vibrational correction overall improves the ratio with a mean absolute deviation of 0.030, whereas it was 0.047 for the ratio at equilibrium geometry, 

It is observed that for the HgI$_2$/HgCl$_2$ ratios, both the vibrational corrected value and the value at equilibrium geometry are within the range of the uncertainty of the experimental values. 
For the CdI$_2$/CdCl$_2$ ratios, both values are just outside the uncertainty by 0.002 and 0.003. 

It should be noted that the vibrational averagings were performed at 0K, whereas the experimental values are measured at higher temperatures. 
Possibly, calculating the vibrational correction at higher temperatures could further improve the values as in the higher vibrational states the average bond lengths would further be increased.

\subsubsection{Isotropic shielding}
\begin{figure*}[hbpt]
    \centering
    \includegraphics[width=1.0\textwidth]{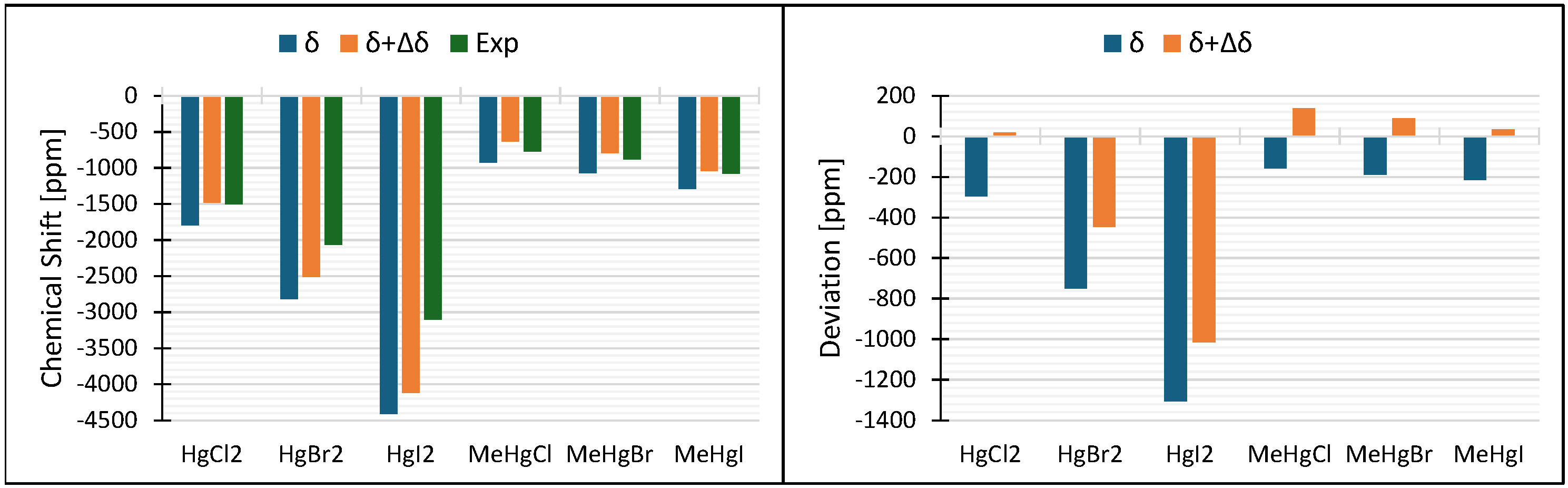}
    \caption{The $^{199}$Hg chemical shift for mercury-halides (HgX$_2$) and the methylmercury-halides (MeHgX) with respect to Hg(CH$_3$)$_2$ calculated with ZORA(spin-orbit)/PBE0/QZ4P. 
    To the left are the values for the shift (blue), the shift with correction (orange), and the experimental values (green). To the right are the deviations from the experimental values for the shift (blue) and the shift with correction (orange). 
    The chemical shift has been calculated with Hg(CH$_3$)$_2$ as the reference.}
    \label{fig:vib-avg_Shift_calc_vs_exp}
\end{figure*}

The magnitude of the corrections for the $^{199}$Hg isotropic shielding in HgCl$_2$, HgBr$_2$, HgI$_2$, H$_3$CHgCl, H$_3$CHgBr, and H$_3$CHgI is found to be between 16 ppm (or 0.15\%) in HgCl$_2$ and 83 ppm (or 0.77\%) in MeHgI (Table \ref{SI-app:vib-avg_calc-vs-exp_shielding}). 
The chemical shifts with respect to Hg(CH$_3$)$_2$ for the equilibrium geometry ($\delta$), the vibrational corrected ($\delta+\Delta\delta$), and the experimental values\cite{TAYLOR2011_mercury-halides_exp, Jokisaari2002_shieling_Methylmercury_Halides_exp} are illustrated in Figure \ref{fig:vib-avg_Shift_calc_vs_exp}. 

The vibrational corrected values ($\delta+\Delta\delta$) are closer to experimental values with a mean absolute deviation of 290 ppm, whereas it is 486 ppm for the values at the equilibrium geometry ($\delta$). 
It should be noted, that a vibrational correction was also added to the shielding of the reference molecule Hg(CH$_3$)$_2$ when calculating the vibrational corrected chemical shift values.
Typically, the vibrational corrections increase the $^{199}$Hg isotropic shielding, and therefore, if using the reference shielding of Hg(CH$_3$)$_2$ without vibrational correction, the corrected chemical shift of the different compounds would be more negative than the values at equilibrium geometry, and thereby further from the experimental values (mean absolute deviation 529 ppm). 
Percentage wise, the vibrational corrections of the chemical shifts are much larger than for the shielding constants. They range between 7\% for HgI$_2$ and 32\% for MeHgCl. This is of course a combined effect of the vibrational correction to the shielding constant in the respective molecule and the reference molecule.
Moreover, a more accurate comparison with the experimental values\cite{TAYLOR2011_mercury-halides_exp, Jokisaari2002_shieling_Methylmercury_Halides_exp} would require inclusion of the surroundings, given that the experimental data were recorded for solids (mercury halides) or dissolved in liquid crystals (methylmercury halides).

\begin{figure*}[h!]
    \centering
    \includegraphics[width=1.0\textwidth]{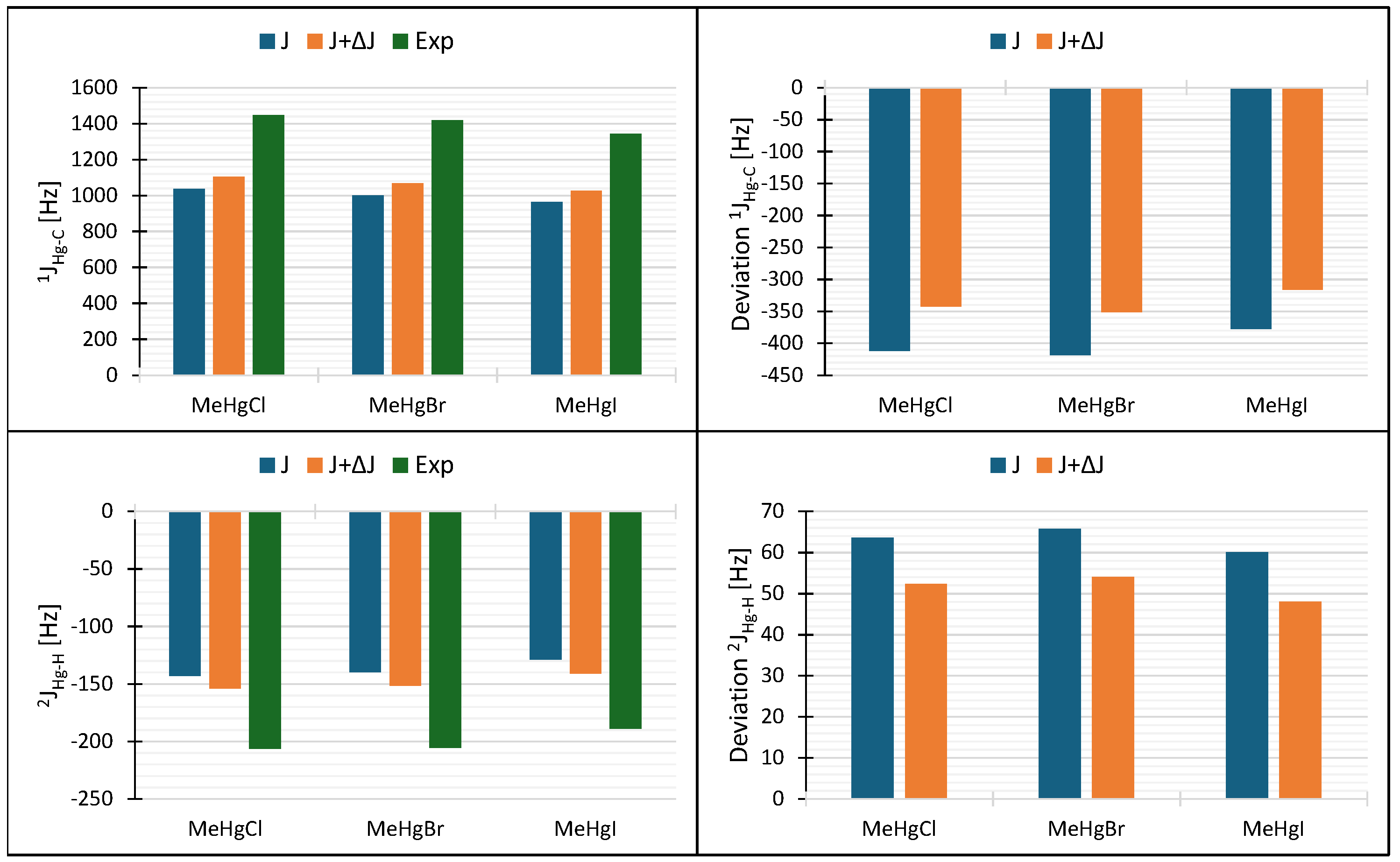}
    \caption{The $^1$J$_{\text{Hg-C}}$ (upper) and  $^2$J$_{\text{Hg-H}}$ (lower) spin-spin coupling constants for the three methylmercury-halides calculated with ZORA(spin-orbit)/PBE0/QZ4P(-J).
    To the left are the values for the SSCC (blue), the SSCC with correction (orange), and the experimental values (green). To the right are the deviations from the experimental values for the equilibrium SSCC (blue) and the corrected SSCC (orange).}
    \label{fig:vib-avg_SSCC_calc_vs_exp}
\end{figure*}

\subsubsection{SSCC}
The vibrational corrections for the methylmercury-halides (MeHgX with X = Cl, Br, I) are between 61 Hz (6.4\%) and 69 Hz (6.7\%) for $^1$J$_{\text{Hg-C}}$, and between -11 Hz (7.9\%) and -12 Hz (9.3\%) for $^2$J$_{\text{Hg-H}}$ (Table \ref{SI-app:vib-avg_calc-vs-exp_sscc}).
This values are very close to the $\sim$53 Hz obtained previously by Autschbach et al.\cite{Autschbach2007_sscc_methyl-mercury-halides_exp} at 300 K and with a different functional and smaller basis set.
The difference between the values and the deviation from the experimental values\cite{Autschbach2007_sscc_methyl-mercury-halides_exp} are illustrated in Figure \ref{fig:vib-avg_SSCC_calc_vs_exp}. 

The SSCCs with vibrational correction agree best with experimental values with a mean absolute deviation of 169 Hz, whereas it is 202 Hz for the SSCCs at equilibrium geometry. 
Very similar differences between the experimental and calculated values were also reported previously.\cite{Autschbach2007_sscc_methyl-mercury-halides_exp} The experimental data were recorded on methylmercury halides dissolved in liquid crystal, and it was already previously discussed that that inclusion of the surroundings in the calculations would improve agreement with the experimental values.\cite{Autschbach2007_sscc_methyl-mercury-halides_exp}

Previously,\cite{spas154, Gleeson_vib_avg_dalton-project} the effect of vibrational averaging of SSCCs had been investigated for small organic molecules at SOPPA and CCSD level, where the agreement with experiment was only improved with corrections at CCSD level due to the errors in the equilibrium geometry values.
However in the present study, the agreement with experiment is already improved at the DFT level.

\section{Conclusion}
The aim of this study was to extend the vibrational averaging module of the Dalton Project to interface also to the ADF program, and to test how important vibrational corrections are for the three molecular properties; electric field gradient, NMR chemical shift and indirect nuclear spin-spin coupling constant of several small mercury(II) compounds. It was further investigated whether inclusion of vibrational corrections improves comparison with experimental values, whether the corrections depend on the inclusion of relativistic effects, and how they change with the basis set employed.

It is found that the values of the properties improved when adding a zero-point vibrational correction. 
Furthemore, including relativistic effects in the calculation of the vibrational corrections has a significant effect, and therefore to obtain the best-described corrections they should be included. Likewise, the magnitude of the corrections differed when changing the basis set. 

It was also observed that in the calculation of the shielding tensor for a mercury complex bound to either sulfur or nitrogen with QZ4P and spin-orbit coupling, there could be a problem with the convergence due to linear dependencies, which was solved by adding the keyword \textit{DEPENDENCY} in the input file. 
Likewise, for the same conditions, the SSCCs sometimes needed more iterations than the default settings to converge.

\section*{DATA AVAILABILITY}
The data that support the findings of this study are available
from the corresponding author upon reasonable request.


\begin{thebibliography}{91}%
\makeatletter
\providecommand \@ifxundefined [1]{%
 \@ifx{#1\undefined}
}%
\providecommand \@ifnum [1]{%
 \ifnum #1\expandafter \@firstoftwo
 \else \expandafter \@secondoftwo
 \fi
}%
\providecommand \@ifx [1]{%
 \ifx #1\expandafter \@firstoftwo
 \else \expandafter \@secondoftwo
 \fi
}%
\providecommand \natexlab [1]{#1}%
\providecommand \enquote  [1]{``#1''}%
\providecommand \bibnamefont  [1]{#1}%
\providecommand \bibfnamefont [1]{#1}%
\providecommand \citenamefont [1]{#1}%
\providecommand \href@noop [0]{\@secondoftwo}%
\providecommand \href [0]{\begingroup \@sanitize@url \@href}%
\providecommand \@href[1]{\@@startlink{#1}\@@href}%
\providecommand \@@href[1]{\endgroup#1\@@endlink}%
\providecommand \@sanitize@url [0]{\catcode `\\12\catcode `\$12\catcode `\&12\catcode `\#12\catcode `\^12\catcode `\_12\catcode `\%12\relax}%
\providecommand \@@startlink[1]{}%
\providecommand \@@endlink[0]{}%
\providecommand \url  [0]{\begingroup\@sanitize@url \@url }%
\providecommand \@url [1]{\endgroup\@href {#1}{\urlprefix }}%
\providecommand \urlprefix  [0]{URL }%
\providecommand \Eprint [0]{\href }%
\providecommand \doibase [0]{http://dx.doi.org/}%
\providecommand \selectlanguage [0]{\@gobble}%
\providecommand \bibinfo  [0]{\@secondoftwo}%
\providecommand \bibfield  [0]{\@secondoftwo}%
\providecommand \translation [1]{[#1]}%
\providecommand \BibitemOpen [0]{}%
\providecommand \bibitemStop [0]{}%
\providecommand \bibitemNoStop [0]{.\EOS\space}%
\providecommand \EOS [0]{\spacefactor3000\relax}%
\providecommand \BibitemShut  [1]{\csname bibitem#1\endcsname}%
\let\auto@bib@innerbib\@empty
\bibitem [{\citenamefont {Schenberg}, \citenamefont {Ducati},\ and\ \citenamefont {Autschbach}(2024)}]{Schenberg2024_NMR}%
  \BibitemOpen
  \bibfield  {author} {\bibinfo {author} {\bibfnamefont {L.~A.}\ \bibnamefont {Schenberg}}, \bibinfo {author} {\bibfnamefont {L.~C.}\ \bibnamefont {Ducati}}, \ and\ \bibinfo {author} {\bibfnamefont {J.}~\bibnamefont {Autschbach}},\ }\bibfield  {title} {\enquote {\bibinfo {title} {{Inquiring $^{199}$Hg NMR Parameters by Combining Ab Initio Molecular Dynamics and Relativistic NMR Calculations}},}\ }\href {\doibase 10.1021/acs.inorgchem.3c03878} {\bibfield  {journal} {\bibinfo  {journal} {Inorg. Chem.}\ }\textbf {\bibinfo {volume} {63}},\ \bibinfo {pages} {2082--2089} (\bibinfo {year} {2024})}\BibitemShut {NoStop}%
\bibitem [{\citenamefont {Erlemeier}\ \emph {et~al.}(2025)\citenamefont {Erlemeier}, \citenamefont {Hertler}, \citenamefont {Wu}, \citenamefont {Xie}, \citenamefont {Autschbach}, \citenamefont {Schenberg}, \citenamefont {Ducati}, \citenamefont {Hayton},\ and\ \citenamefont {von Hänisch}}]{Erlemeier2025}%
  \BibitemOpen
  \bibfield  {author} {\bibinfo {author} {\bibfnamefont {L.}~\bibnamefont {Erlemeier}}, \bibinfo {author} {\bibfnamefont {P.~R.}\ \bibnamefont {Hertler}}, \bibinfo {author} {\bibfnamefont {G.}~\bibnamefont {Wu}}, \bibinfo {author} {\bibfnamefont {X.}~\bibnamefont {Xie}}, \bibinfo {author} {\bibfnamefont {J.}~\bibnamefont {Autschbach}}, \bibinfo {author} {\bibfnamefont {L.~A.}\ \bibnamefont {Schenberg}}, \bibinfo {author} {\bibfnamefont {L.~C.}\ \bibnamefont {Ducati}}, \bibinfo {author} {\bibfnamefont {T.~W.}\ \bibnamefont {Hayton}}, \ and\ \bibinfo {author} {\bibfnamefont {C.}~\bibnamefont {von Hänisch}},\ }\bibfield  {title} {\enquote {\bibinfo {title} {{Synthesis, Structure, and $^{199}$Hg Chemical Shifts of Mercury 1,5,9-trimesityldipyrromethene ($^{\text{Mes}}$DPM) Complexes}},}\ }\href {\doibase https://doi.org/10.1002/chem.202501460} {\bibfield  {journal} {\bibinfo  {journal} {Chem. Eur. J.}\ }\textbf {\bibinfo {volume} {31}},\ \bibinfo {pages} {e202501460} (\bibinfo {year} {2025})}\BibitemShut
  {NoStop}%
\bibitem [{\citenamefont {Visscher}, \citenamefont {Saue},\ and\ \citenamefont {Oddershede}(1997)}]{jod970801vso}%
  \BibitemOpen
  \bibfield  {author} {\bibinfo {author} {\bibfnamefont {L.}~\bibnamefont {Visscher}}, \bibinfo {author} {\bibfnamefont {T.}~\bibnamefont {Saue}}, \ and\ \bibinfo {author} {\bibfnamefont {J.}~\bibnamefont {Oddershede}},\ }\bibfield  {title} {\enquote {\bibinfo {title} {{The 4-component random phase approximation method applied to the calculation of frequency-dependent dipole polarizabilities}},}\ }\href@noop {} {\bibfield  {journal} {\bibinfo  {journal} {Chem. Phys. Lett.}\ }\textbf {\bibinfo {volume} {274}},\ \bibinfo {pages} {181--188} (\bibinfo {year} {1997})}\BibitemShut {NoStop}%
\bibitem [{\citenamefont {Visscher}\ \emph {et~al.}(1999)\citenamefont {Visscher}, \citenamefont {Enevoldsen}, \citenamefont {Saue}, \citenamefont {Jensen},\ and\ \citenamefont {Oddershede}}]{Visscher1999_4comp-relativity}%
  \BibitemOpen
  \bibfield  {author} {\bibinfo {author} {\bibfnamefont {L.}~\bibnamefont {Visscher}}, \bibinfo {author} {\bibfnamefont {T.}~\bibnamefont {Enevoldsen}}, \bibinfo {author} {\bibfnamefont {T.}~\bibnamefont {Saue}}, \bibinfo {author} {\bibfnamefont {H.~J.~A.}\ \bibnamefont {Jensen}}, \ and\ \bibinfo {author} {\bibfnamefont {J.}~\bibnamefont {Oddershede}},\ }\bibfield  {title} {\enquote {\bibinfo {title} {{Full four-component relativistic calculations of NMR shielding and indirect spin–spin coupling tensors in hydrogen halides}},}\ }\href {\doibase https://doi.org/10.1002/(SICI)1096-987X(199909)20:12<1262::AID-JCC6>3.0.CO;2-H} {\bibfield  {journal} {\bibinfo  {journal} {J. Comput. Chem.}\ }\textbf {\bibinfo {volume} {20}},\ \bibinfo {pages} {1262--1273} (\bibinfo {year} {1999})}\BibitemShut {NoStop}%
\bibitem [{\citenamefont {Komorovsk}\ \emph {et~al.}(2010)\citenamefont {Komorovsk}, \citenamefont {Repisky}, \citenamefont {Malkina},\ and\ \citenamefont {Malkin}}]{Komorovsk2010a}%
  \BibitemOpen
  \bibfield  {author} {\bibinfo {author} {\bibfnamefont {S.}~\bibnamefont {Komorovsk}}, \bibinfo {author} {\bibfnamefont {M.}~\bibnamefont {Repisky}}, \bibinfo {author} {\bibfnamefont {O.~L.}\ \bibnamefont {Malkina}}, \ and\ \bibinfo {author} {\bibfnamefont {V.~G.}\ \bibnamefont {Malkin}},\ }\bibfield  {title} {\enquote {\bibinfo {title} {{Fully relativistic calculations of NMR shielding tensors using restricted magnetically balanced basis and gauge including atomic orbitals}},}\ }\href {\doibase 10.1063/1.3359849} {\bibfield  {journal} {\bibinfo  {journal} {J. Chem. Phys.}\ }\textbf {\bibinfo {volume} {132}},\ \bibinfo {pages} {1--8} (\bibinfo {year} {2010})}\BibitemShut {NoStop}%
\bibitem [{\citenamefont {Repisky}\ \emph {et~al.}(2020)\citenamefont {Repisky}, \citenamefont {Komorovsky}, \citenamefont {Kadek}, \citenamefont {Konecny}, \citenamefont {Ekstr{\"{o}}m}, \citenamefont {Malkin}, \citenamefont {Kaupp}, \citenamefont {Ruud}, \citenamefont {Malkina},\ and\ \citenamefont {Malkin}}]{JCP-152-184101-2020-Repisky}%
  \BibitemOpen
  \bibfield  {author} {\bibinfo {author} {\bibfnamefont {M.}~\bibnamefont {Repisky}}, \bibinfo {author} {\bibfnamefont {S.}~\bibnamefont {Komorovsky}}, \bibinfo {author} {\bibfnamefont {M.}~\bibnamefont {Kadek}}, \bibinfo {author} {\bibfnamefont {L.}~\bibnamefont {Konecny}}, \bibinfo {author} {\bibfnamefont {U.}~\bibnamefont {Ekstr{\"{o}}m}}, \bibinfo {author} {\bibfnamefont {E.}~\bibnamefont {Malkin}}, \bibinfo {author} {\bibfnamefont {M.}~\bibnamefont {Kaupp}}, \bibinfo {author} {\bibfnamefont {K.}~\bibnamefont {Ruud}}, \bibinfo {author} {\bibfnamefont {O.~L.}\ \bibnamefont {Malkina}}, \ and\ \bibinfo {author} {\bibfnamefont {V.~G.}\ \bibnamefont {Malkin}},\ }\bibfield  {title} {\enquote {\bibinfo {title} {{ReSpect: Relativistic spectroscopy DFT program package}},}\ }\href {\doibase 10.1063/5.0005094} {\bibfield  {journal} {\bibinfo  {journal} {J. Chem. Phys.}\ }\textbf {\bibinfo {volume} {152}},\ \bibinfo {pages} {184101} (\bibinfo {year} {2020})}\BibitemShut {NoStop}%
\bibitem [{\citenamefont {Komorovsky}\ \emph {et~al.}(2020)\citenamefont {Komorovsky}, \citenamefont {Jakubowska}, \citenamefont {{\'{S}}wider}, \citenamefont {Repisky},\ and\ \citenamefont {Jaszu{\'{n}}ski}}]{JPCA-124-5157-2020-Komorovsky}%
  \BibitemOpen
  \bibfield  {author} {\bibinfo {author} {\bibfnamefont {S.}~\bibnamefont {Komorovsky}}, \bibinfo {author} {\bibfnamefont {K.}~\bibnamefont {Jakubowska}}, \bibinfo {author} {\bibfnamefont {P.}~\bibnamefont {{\'{S}}wider}}, \bibinfo {author} {\bibfnamefont {M.}~\bibnamefont {Repisky}}, \ and\ \bibinfo {author} {\bibfnamefont {M.}~\bibnamefont {Jaszu{\'{n}}ski}},\ }\bibfield  {title} {\enquote {\bibinfo {title} {{NMR Spin-Spin Coupling Constants Derived from Relativistic Four-Component DFT Theory - Analysis and Visualization}},}\ }\href {\doibase 10.1021/acs.jpca.0c02807} {\bibfield  {journal} {\bibinfo  {journal} {J. Phys. Chem. A}\ }\textbf {\bibinfo {volume} {124}},\ \bibinfo {pages} {5157--5169} (\bibinfo {year} {2020})}\BibitemShut {NoStop}%
\bibitem [{\citenamefont {Saue}\ \emph {et~al.}(2020)\citenamefont {Saue}, \citenamefont {Bast}, \citenamefont {Gomes}, \citenamefont {Jensen}, \citenamefont {Visscher}, \citenamefont {Aucar}, \citenamefont {{Di Remigio}}, \citenamefont {Dyall}, \citenamefont {Eliav}, \citenamefont {Fasshauer}, \citenamefont {Fleig}, \citenamefont {Halbert}, \citenamefont {Hedeg{\aa}rd}, \citenamefont {Helmich-Paris}, \citenamefont {Ilia{\v{s}}}, \citenamefont {Jacob}, \citenamefont {Knecht}, \citenamefont {Laerdahl}, \citenamefont {Vidal}, \citenamefont {Nayak}, \citenamefont {Olejniczak}, \citenamefont {Olsen}, \citenamefont {Pernpointner}, \citenamefont {Senjean}, \citenamefont {Shee}, \citenamefont {Sunaga},\ and\ \citenamefont {van Stralen}}]{Saue2020}%
  \BibitemOpen
  \bibfield  {author} {\bibinfo {author} {\bibfnamefont {T.}~\bibnamefont {Saue}}, \bibinfo {author} {\bibfnamefont {R.}~\bibnamefont {Bast}}, \bibinfo {author} {\bibfnamefont {A.~S.~P.}\ \bibnamefont {Gomes}}, \bibinfo {author} {\bibfnamefont {H.~J.~A.}\ \bibnamefont {Jensen}}, \bibinfo {author} {\bibfnamefont {L.}~\bibnamefont {Visscher}}, \bibinfo {author} {\bibfnamefont {I.~A.}\ \bibnamefont {Aucar}}, \bibinfo {author} {\bibfnamefont {R.}~\bibnamefont {{Di Remigio}}}, \bibinfo {author} {\bibfnamefont {K.~G.}\ \bibnamefont {Dyall}}, \bibinfo {author} {\bibfnamefont {E.}~\bibnamefont {Eliav}}, \bibinfo {author} {\bibfnamefont {E.}~\bibnamefont {Fasshauer}}, \bibinfo {author} {\bibfnamefont {T.}~\bibnamefont {Fleig}}, \bibinfo {author} {\bibfnamefont {L.}~\bibnamefont {Halbert}}, \bibinfo {author} {\bibfnamefont {E.~D.}\ \bibnamefont {Hedeg{\aa}rd}}, \bibinfo {author} {\bibfnamefont {B.}~\bibnamefont {Helmich-Paris}}, \bibinfo {author} {\bibfnamefont {M.}~\bibnamefont {Ilia{\v{s}}}}, \bibinfo {author}
  {\bibfnamefont {C.~R.}\ \bibnamefont {Jacob}}, \bibinfo {author} {\bibfnamefont {S.}~\bibnamefont {Knecht}}, \bibinfo {author} {\bibfnamefont {J.~K.}\ \bibnamefont {Laerdahl}}, \bibinfo {author} {\bibfnamefont {M.~L.}\ \bibnamefont {Vidal}}, \bibinfo {author} {\bibfnamefont {M.~K.}\ \bibnamefont {Nayak}}, \bibinfo {author} {\bibfnamefont {M.}~\bibnamefont {Olejniczak}}, \bibinfo {author} {\bibfnamefont {J.~M.~H.}\ \bibnamefont {Olsen}}, \bibinfo {author} {\bibfnamefont {M.}~\bibnamefont {Pernpointner}}, \bibinfo {author} {\bibfnamefont {B.}~\bibnamefont {Senjean}}, \bibinfo {author} {\bibfnamefont {A.}~\bibnamefont {Shee}}, \bibinfo {author} {\bibfnamefont {A.}~\bibnamefont {Sunaga}}, \ and\ \bibinfo {author} {\bibfnamefont {J.~N.}\ \bibnamefont {van Stralen}},\ }\bibfield  {title} {\enquote {\bibinfo {title} {{The DIRAC code for relativistic molecular calculations}},}\ }\href {\doibase 10.1063/5.0004844} {\bibfield  {journal} {\bibinfo  {journal} {J. Chem. Phys.}\ }\textbf {\bibinfo {volume} {152}},\
  \bibinfo {pages} {204104} (\bibinfo {year} {2020})}\BibitemShut {NoStop}%
\bibitem [{\citenamefont {Melo}\ \emph {et~al.}(2003)\citenamefont {Melo}, \citenamefont {Ruiz De~Azua}, \citenamefont {Giribet}, \citenamefont {Aucar},\ and\ \citenamefont {Romero}}]{Melo.LRESC1.JCP.2003}%
  \BibitemOpen
  \bibfield  {author} {\bibinfo {author} {\bibfnamefont {J.~I.}\ \bibnamefont {Melo}}, \bibinfo {author} {\bibfnamefont {M.~C.}\ \bibnamefont {Ruiz De~Azua}}, \bibinfo {author} {\bibfnamefont {C.~G.}\ \bibnamefont {Giribet}}, \bibinfo {author} {\bibfnamefont {G.~A.}\ \bibnamefont {Aucar}}, \ and\ \bibinfo {author} {\bibfnamefont {R.~H.}\ \bibnamefont {Romero}},\ }\bibfield  {title} {\enquote {\bibinfo {title} {Relativistic effects on the nuclear magnetic shielding tensor},}\ }\href {\doibase 10.1063/1.1525808} {\bibfield  {journal} {\bibinfo  {journal} {J. Chem. Phys.}\ }\textbf {\bibinfo {volume} {118}},\ \bibinfo {pages} {471--486} (\bibinfo {year} {2003})}\BibitemShut {NoStop}%
\bibitem [{\citenamefont {Ruiz De~Az\'ua}, \citenamefont {Melo},\ and\ \citenamefont {Giribet}(2003)}]{Melo.LRESC3.MolPhys.2003}%
  \BibitemOpen
  \bibfield  {author} {\bibinfo {author} {\bibfnamefont {M.~C.}\ \bibnamefont {Ruiz De~Az\'ua}}, \bibinfo {author} {\bibfnamefont {J.~I.}\ \bibnamefont {Melo}}, \ and\ \bibinfo {author} {\bibfnamefont {C.~G.}\ \bibnamefont {Giribet}},\ }\bibfield  {title} {\enquote {\bibinfo {title} {Orbital contributions to relativistic corrections of the {NMR} nuclear magnetic shielding tensor originated in scalar field-dependent operators},}\ }\href {\doibase 10.1080/00268970310001617784} {\bibfield  {journal} {\bibinfo  {journal} {Mol. Phys.}\ }\textbf {\bibinfo {volume} {101}},\ \bibinfo {pages} {3103--3109} (\bibinfo {year} {2003})}\BibitemShut {NoStop}%
\bibitem [{\citenamefont {Melo}\ \emph {et~al.}(2004)\citenamefont {Melo}, \citenamefont {Ruiz~de Azua}, \citenamefont {Giribet}, \citenamefont {Aucar},\ and\ \citenamefont {Provasi}}]{Melo.LRESC2.JCP.2004}%
  \BibitemOpen
  \bibfield  {author} {\bibinfo {author} {\bibfnamefont {J.~I.}\ \bibnamefont {Melo}}, \bibinfo {author} {\bibfnamefont {M.~C.}\ \bibnamefont {Ruiz~de Azua}}, \bibinfo {author} {\bibfnamefont {C.~G.}\ \bibnamefont {Giribet}}, \bibinfo {author} {\bibfnamefont {G.~A.}\ \bibnamefont {Aucar}}, \ and\ \bibinfo {author} {\bibfnamefont {P.~F.}\ \bibnamefont {Provasi}},\ }\bibfield  {title} {\enquote {\bibinfo {title} {Relativistic effects on nuclear magnetic shielding constants in {HX} and {CH$_3$X} ({X}= {Br}, {I}) based on the linear response within the elimination of small component approach},}\ }\href {\doibase 10.1063/1.1787495} {\bibfield  {journal} {\bibinfo  {journal} {J. Chem. Phys.}\ }\textbf {\bibinfo {volume} {121}},\ \bibinfo {pages} {6798--6808} (\bibinfo {year} {2004})}\BibitemShut {NoStop}%
\bibitem [{\citenamefont {Melo}\ and\ \citenamefont {Maldonado}(2019)}]{Melo.EFG1.IJQC.2019}%
  \BibitemOpen
  \bibfield  {author} {\bibinfo {author} {\bibfnamefont {J.~I.}\ \bibnamefont {Melo}}\ and\ \bibinfo {author} {\bibfnamefont {A.~F.}\ \bibnamefont {Maldonado}},\ }\bibfield  {title} {\enquote {\bibinfo {title} {Relativistic corrections to the electric field gradient given by linear response elimination of the small component formalism},}\ }\href {\doibase 10.1002/qua.25935} {\bibfield  {journal} {\bibinfo  {journal} {Int. J. Quantum Chem.}\ }\textbf {\bibinfo {volume} {119}},\ \bibinfo {pages} {e25935} (\bibinfo {year} {2019})}\BibitemShut {NoStop}%
\bibitem [{\citenamefont {Aucar}, \citenamefont {Maldonado},\ and\ \citenamefont {Melo}(2021)}]{Melo.EFG2.IJQC.2021}%
  \BibitemOpen
  \bibfield  {author} {\bibinfo {author} {\bibfnamefont {J.~J.}\ \bibnamefont {Aucar}}, \bibinfo {author} {\bibfnamefont {A.~F.}\ \bibnamefont {Maldonado}}, \ and\ \bibinfo {author} {\bibfnamefont {J.~I.}\ \bibnamefont {Melo}},\ }\bibfield  {title} {\enquote {\bibinfo {title} {Relativistic corrections of the electric field gradient in dihalogen molecules \textit{{XY}} ( \textit{{X}} , \textit{{Y}} = {F}, {Cl}, {Br}, {I}, {At}) within the linear response elimination of the small component formalism},}\ }\href {\doibase 10.1002/qua.26769} {\bibfield  {journal} {\bibinfo  {journal} {Int. J. Quantum Chem.}\ }\textbf {\bibinfo {volume} {121}} (\bibinfo {year} {2021}),\ 10.1002/qua.26769}\BibitemShut {NoStop}%
\bibitem [{\citenamefont {Aucar}, \citenamefont {Maldonado},\ and\ \citenamefont {Melo}(2022)}]{Melo.EFG3.JCP.2022}%
  \BibitemOpen
  \bibfield  {author} {\bibinfo {author} {\bibfnamefont {J.~J.}\ \bibnamefont {Aucar}}, \bibinfo {author} {\bibfnamefont {A.~F.}\ \bibnamefont {Maldonado}}, \ and\ \bibinfo {author} {\bibfnamefont {J.~I.}\ \bibnamefont {Melo}},\ }\bibfield  {title} {\enquote {\bibinfo {title} {High order relativistic corrections on the electric field gradient within the {LRESC} formalism},}\ }\href {\doibase 10.1063/5.0124701} {\bibfield  {journal} {\bibinfo  {journal} {J. Chem. Phys.}\ }\textbf {\bibinfo {volume} {157}},\ \bibinfo {pages} {244105} (\bibinfo {year} {2022})}\BibitemShut {NoStop}%
\bibitem [{\citenamefont {Aucar}, \citenamefont {Melo},\ and\ \citenamefont {Maldonado}(2024)}]{Melo.EFG3.JPCA.2024}%
  \BibitemOpen
  \bibfield  {author} {\bibinfo {author} {\bibfnamefont {J.~J.}\ \bibnamefont {Aucar}}, \bibinfo {author} {\bibfnamefont {J.~I.}\ \bibnamefont {Melo}}, \ and\ \bibinfo {author} {\bibfnamefont {A.~F.}\ \bibnamefont {Maldonado}},\ }\bibfield  {title} {\enquote {\bibinfo {title} {Electric field gradient in chiral and tetrahedral molecules within high-order lresc formalism},}\ }\href {\doibase 10.1021/acs.jpca.4c00426} {\bibfield  {journal} {\bibinfo  {journal} {J. Phys. Chem. A}\ }\textbf {\bibinfo {volume} {128}},\ \bibinfo {pages} {5089--5099} (\bibinfo {year} {2024})}\BibitemShut {NoStop}%
\bibitem [{\citenamefont {Chang}, \citenamefont {Pelissier},\ and\ \citenamefont {Durand}(1986)}]{Chang1986}%
  \BibitemOpen
  \bibfield  {author} {\bibinfo {author} {\bibfnamefont {C.}~\bibnamefont {Chang}}, \bibinfo {author} {\bibfnamefont {M.}~\bibnamefont {Pelissier}}, \ and\ \bibinfo {author} {\bibfnamefont {P.}~\bibnamefont {Durand}},\ }\bibfield  {title} {\enquote {\bibinfo {title} {{Regular Two-Component Pauli-Like Effective Hamiltonians in Dirac Theory}},}\ }\href@noop {} {\bibfield  {journal} {\bibinfo  {journal} {Phys. Scr.}\ }\textbf {\bibinfo {volume} {34}},\ \bibinfo {pages} {394--404} (\bibinfo {year} {1986})}\BibitemShut {NoStop}%
\bibitem [{\citenamefont {van Lenthe}, \citenamefont {Baerends},\ and\ \citenamefont {Snijders}(1993)}]{Lenthe1993_ZORA_hamiltonian}%
  \BibitemOpen
  \bibfield  {author} {\bibinfo {author} {\bibfnamefont {E.}~\bibnamefont {van Lenthe}}, \bibinfo {author} {\bibfnamefont {E.~J.}\ \bibnamefont {Baerends}}, \ and\ \bibinfo {author} {\bibfnamefont {J.~G.}\ \bibnamefont {Snijders}},\ }\bibfield  {title} {\enquote {\bibinfo {title} {{Relativistic regular two‐component Hamiltonians}},}\ }\href {\doibase 10.1063/1.466059} {\bibfield  {journal} {\bibinfo  {journal} {J. Chem. Phys.}\ }\textbf {\bibinfo {volume} {99}},\ \bibinfo {pages} {4597--4610} (\bibinfo {year} {1993})}\BibitemShut {NoStop}%
\bibitem [{\citenamefont {van Lenthe}\ \emph {et~al.}(1996)\citenamefont {van Lenthe}, \citenamefont {van Leeuwen}, \citenamefont {Baerends},\ and\ \citenamefont {Snijders}}]{Lenthe1996_ZORA_hamiltonians}%
  \BibitemOpen
  \bibfield  {author} {\bibinfo {author} {\bibfnamefont {E.}~\bibnamefont {van Lenthe}}, \bibinfo {author} {\bibfnamefont {R.}~\bibnamefont {van Leeuwen}}, \bibinfo {author} {\bibfnamefont {E.~J.}\ \bibnamefont {Baerends}}, \ and\ \bibinfo {author} {\bibfnamefont {J.~G.}\ \bibnamefont {Snijders}},\ }\bibfield  {title} {\enquote {\bibinfo {title} {Relativistic regular two-component hamiltonians},}\ }\href {\doibase https://doi.org/10.1002/(SICI)1097-461X(1996)57:3<281::AID-QUA2>3.0.CO;2-U} {\bibfield  {journal} {\bibinfo  {journal} {Int. J. Quantum Chem.}\ }\textbf {\bibinfo {volume} {57}},\ \bibinfo {pages} {281--293} (\bibinfo {year} {1996})}\BibitemShut {NoStop}%
\bibitem [{\citenamefont {van Lenthe}, \citenamefont {Snijders},\ and\ \citenamefont {Baerends}(1996)}]{Lenthe1996_ZORA_effects}%
  \BibitemOpen
  \bibfield  {author} {\bibinfo {author} {\bibfnamefont {E.}~\bibnamefont {van Lenthe}}, \bibinfo {author} {\bibfnamefont {J.~G.}\ \bibnamefont {Snijders}}, \ and\ \bibinfo {author} {\bibfnamefont {E.~J.}\ \bibnamefont {Baerends}},\ }\bibfield  {title} {\enquote {\bibinfo {title} {{The zero‐order regular approximation for relativistic effects: The effect of spin–orbit coupling in closed shell molecules}},}\ }\href {\doibase 10.1063/1.472460} {\bibfield  {journal} {\bibinfo  {journal} {J. Chem. Phys.}\ }\textbf {\bibinfo {volume} {105}},\ \bibinfo {pages} {6505--6516} (\bibinfo {year} {1996})}\BibitemShut {NoStop}%
\bibitem [{\citenamefont {van Lenthe}, \citenamefont {Baerends},\ and\ \citenamefont {Snijders}(1994)}]{Lenthe1994_ZORA_energy}%
  \BibitemOpen
  \bibfield  {author} {\bibinfo {author} {\bibfnamefont {E.}~\bibnamefont {van Lenthe}}, \bibinfo {author} {\bibfnamefont {E.~J.}\ \bibnamefont {Baerends}}, \ and\ \bibinfo {author} {\bibfnamefont {J.~G.}\ \bibnamefont {Snijders}},\ }\bibfield  {title} {\enquote {\bibinfo {title} {{Relativistic total energy using regular approximations}},}\ }\href {\doibase 10.1063/1.467943} {\bibfield  {journal} {\bibinfo  {journal} {J. Chem. Phys.}\ }\textbf {\bibinfo {volume} {101}},\ \bibinfo {pages} {9783--9792} (\bibinfo {year} {1994})}\BibitemShut {NoStop}%
\bibitem [{\citenamefont {van Lenthe}, \citenamefont {Ehlers},\ and\ \citenamefont {Baerends}(1999)}]{Lenthe1999_ZORA_geometry}%
  \BibitemOpen
  \bibfield  {author} {\bibinfo {author} {\bibfnamefont {E.}~\bibnamefont {van Lenthe}}, \bibinfo {author} {\bibfnamefont {A.}~\bibnamefont {Ehlers}}, \ and\ \bibinfo {author} {\bibfnamefont {E.~J.}\ \bibnamefont {Baerends}},\ }\bibfield  {title} {\enquote {\bibinfo {title} {{Geometry optimizations in the zero order regular approximation for relativistic effects}},}\ }\href {\doibase 10.1063/1.478813} {\bibfield  {journal} {\bibinfo  {journal} {J. Chem. Phys.}\ }\textbf {\bibinfo {volume} {110}},\ \bibinfo {pages} {8943--8953} (\bibinfo {year} {1999})}\BibitemShut {NoStop}%
\bibitem [{\citenamefont {Arcisauskaite}\ \emph {et~al.}(2012{\natexlab{a}})\citenamefont {Arcisauskaite}, \citenamefont {Knecht}, \citenamefont {Sauer},\ and\ \citenamefont {Hemmingsen}}]{Vaida2012_EFG}%
  \BibitemOpen
  \bibfield  {author} {\bibinfo {author} {\bibfnamefont {V.}~\bibnamefont {Arcisauskaite}}, \bibinfo {author} {\bibfnamefont {S.}~\bibnamefont {Knecht}}, \bibinfo {author} {\bibfnamefont {S.~P.~A.}\ \bibnamefont {Sauer}}, \ and\ \bibinfo {author} {\bibfnamefont {L.}~\bibnamefont {Hemmingsen}},\ }\bibfield  {title} {\enquote {\bibinfo {title} {{Electric field gradients in Hg compounds: Molecular orbital (MO) analysis and comparison of 4-component and 2-component (ZORA) methods}},}\ }\href {\doibase 10.1039/C2CP42291C} {\bibfield  {journal} {\bibinfo  {journal} {Phys. Chem. Chem. Phys.}\ }\textbf {\bibinfo {volume} {14}},\ \bibinfo {pages} {16070--16079} (\bibinfo {year} {2012}{\natexlab{a}})}\BibitemShut {NoStop}%
\bibitem [{\citenamefont {Arcisauskaite}\ \emph {et~al.}(2012{\natexlab{b}})\citenamefont {Arcisauskaite}, \citenamefont {Knecht}, \citenamefont {Sauer},\ and\ \citenamefont {Hemmingsen}}]{Vaida2012_relativistic}%
  \BibitemOpen
  \bibfield  {author} {\bibinfo {author} {\bibfnamefont {V.}~\bibnamefont {Arcisauskaite}}, \bibinfo {author} {\bibfnamefont {S.}~\bibnamefont {Knecht}}, \bibinfo {author} {\bibfnamefont {S.~P.~A.}\ \bibnamefont {Sauer}}, \ and\ \bibinfo {author} {\bibfnamefont {L.}~\bibnamefont {Hemmingsen}},\ }\bibfield  {title} {\enquote {\bibinfo {title} {{Fully relativistic coupled cluster and DFT study of electric field gradients at Hg in $^{199}$Hg compounds}},}\ }\href {\doibase 10.1039/C2CP23080A} {\bibfield  {journal} {\bibinfo  {journal} {Phys. Chem. Chem. Phys.}\ }\textbf {\bibinfo {volume} {14}},\ \bibinfo {pages} {2651--2657} (\bibinfo {year} {2012}{\natexlab{b}})}\BibitemShut {NoStop}%
\bibitem [{\citenamefont {Arcisauskaite}\ \emph {et~al.}(2011)\citenamefont {Arcisauskaite}, \citenamefont {Melo}, \citenamefont {Hemmingsen},\ and\ \citenamefont {Sauer}}]{Vaida2011_NMR}%
  \BibitemOpen
  \bibfield  {author} {\bibinfo {author} {\bibfnamefont {V.}~\bibnamefont {Arcisauskaite}}, \bibinfo {author} {\bibfnamefont {J.~I.}\ \bibnamefont {Melo}}, \bibinfo {author} {\bibfnamefont {L.}~\bibnamefont {Hemmingsen}}, \ and\ \bibinfo {author} {\bibfnamefont {S.~P.~A.}\ \bibnamefont {Sauer}},\ }\bibfield  {title} {\enquote {\bibinfo {title} {{Nuclear magnetic resonance shielding constants and chemical shifts in linear $^{199}$Hg compounds: A comparison of three relativistic computational methods}},}\ }\href {\doibase 10.1063/1.3608153} {\bibfield  {journal} {\bibinfo  {journal} {J. Chem. Phys.}\ }\textbf {\bibinfo {volume} {135}},\ \bibinfo {pages} {044306} (\bibinfo {year} {2011})}\BibitemShut {NoStop}%
\bibitem [{\citenamefont {Bühl}(2004)}]{Buhl2004_other-ZORA-calc}%
  \BibitemOpen
  \bibfield  {author} {\bibinfo {author} {\bibfnamefont {M.}~\bibnamefont {Bühl}},\ }\enquote {\bibinfo {title} {{NMR of Transition Metal Compounds}},}\ in\ \href {\doibase https://doi.org/10.1002/3527601678.ch26} {\emph {\bibinfo {booktitle} {Calculation of NMR and EPR Parameters}}},\ \bibinfo {editor} {edited by\ \bibinfo {editor} {\bibfnamefont {M.}~\bibnamefont {Kaupp}}, \bibinfo {editor} {\bibfnamefont {M.}~\bibnamefont {Bühl}}, \ and\ \bibinfo {editor} {\bibfnamefont {V.~G.}\ \bibnamefont {Malkin}}}\ (\bibinfo  {publisher} {John Wiley \& Sons, Ltd},\ \bibinfo {year} {2004})\ Chap.~\bibinfo {chapter} {26}, pp.\ \bibinfo {pages} {421--431}\BibitemShut {NoStop}%
\bibitem [{\citenamefont {Kaupp}(2004)}]{Kuapp2004_other-ZORA-calc}%
  \BibitemOpen
  \bibfield  {author} {\bibinfo {author} {\bibfnamefont {M.}~\bibnamefont {Kaupp}},\ }\enquote {\bibinfo {title} {{Interpretation of NMR Chemical Shifts}},}\ in\ \href {\doibase https://doi.org/10.1002/3527601678.ch18} {\emph {\bibinfo {booktitle} {{Calculation of NMR and EPR Parameters}}}},\ \bibinfo {editor} {edited by\ \bibinfo {editor} {\bibfnamefont {M.}~\bibnamefont {Kaupp}}, \bibinfo {editor} {\bibfnamefont {M.}~\bibnamefont {Bühl}}, \ and\ \bibinfo {editor} {\bibfnamefont {V.~G.}\ \bibnamefont {Malkin}}}\ (\bibinfo  {publisher} {John Wiley \& Sons, Ltd},\ \bibinfo {year} {2004})\ Chap.~\bibinfo {chapter} {18}, pp.\ \bibinfo {pages} {293--306}\BibitemShut {NoStop}%
\bibitem [{\citenamefont {Autschbach}\ and\ \citenamefont {Ziegler}(2004)}]{Autshbach2004_other-ZORA-calc}%
  \BibitemOpen
  \bibfield  {author} {\bibinfo {author} {\bibfnamefont {J.}~\bibnamefont {Autschbach}}\ and\ \bibinfo {author} {\bibfnamefont {T.}~\bibnamefont {Ziegler}},\ }\enquote {\bibinfo {title} {{Relativistic Calculations of Spin–Spin Coupling Constants of Heavy Nuclei}},}\ in\ \href {\doibase https://doi.org/10.1002/3527601678.ch15} {\emph {\bibinfo {booktitle} {{Calculation of NMR and EPR Parameters}}}},\ \bibinfo {editor} {edited by\ \bibinfo {editor} {\bibfnamefont {M.}~\bibnamefont {Kaupp}}, \bibinfo {editor} {\bibfnamefont {M.}~\bibnamefont {Bühl}}, \ and\ \bibinfo {editor} {\bibfnamefont {V.~G.}\ \bibnamefont {Malkin}}}\ (\bibinfo  {publisher} {John Wiley \& Sons, Ltd},\ \bibinfo {year} {2004})\ Chap.~\bibinfo {chapter} {15}, pp.\ \bibinfo {pages} {249--264}\BibitemShut {NoStop}%
\bibitem [{\citenamefont {Autschbach}(2004)}]{Autschbach2004_other-ZORA-calc}%
  \BibitemOpen
  \bibfield  {author} {\bibinfo {author} {\bibfnamefont {J.}~\bibnamefont {Autschbach}},\ }\enquote {\bibinfo {title} {{Calculation of Heavy-Nucleus Chemical Shifts. Relativistic All-Electron Methods}},}\ in\ \href {\doibase https://doi.org/10.1002/3527601678.ch14} {\emph {\bibinfo {booktitle} {Calculation of NMR and EPR Parameters}}}\ (\bibinfo  {publisher} {John Wiley \& Sons, Ltd},\ \bibinfo {year} {2004})\ Chap.~\bibinfo {chapter} {14}, pp.\ \bibinfo {pages} {227--247}\BibitemShut {NoStop}%
\bibitem [{\citenamefont {Autschbach}\ and\ \citenamefont {Zheng}(2009)}]{AUTSCHBACH2009_other-ZORA-calc}%
  \BibitemOpen
  \bibfield  {author} {\bibinfo {author} {\bibfnamefont {J.}~\bibnamefont {Autschbach}}\ and\ \bibinfo {author} {\bibfnamefont {S.}~\bibnamefont {Zheng}},\ }\bibfield  {title} {\enquote {\bibinfo {title} {{Chapter 1 Relativistic Computations of NMR Parameters from First Principles: Theory and Applications}},}\ \ }(\bibinfo  {publisher} {Academic Press},\ \bibinfo {year} {2009})\ pp.\ \bibinfo {pages} {1--95}\BibitemShut {NoStop}%
\bibitem [{\citenamefont {Autschbach}(2014)}]{AUTSCHBACH2014_other-ZORA-calc}%
  \BibitemOpen
  \bibfield  {author} {\bibinfo {author} {\bibfnamefont {J.}~\bibnamefont {Autschbach}},\ }\bibfield  {title} {\enquote {\bibinfo {title} {{Relativistic calculations of magnetic resonance parameters: background and some recent developments}},}\ }\href {\doibase https://doi.org/10.1098/rsta.2012.0489} {\bibfield  {journal} {\bibinfo  {journal} {Phil. Trans. R. Soc. A.}\ }\textbf {\bibinfo {volume} {372}},\ \bibinfo {pages} {20120489} (\bibinfo {year} {2014})}\BibitemShut {NoStop}%
\bibitem [{\citenamefont {Repisky}\ \emph {et~al.}(2016)\citenamefont {Repisky}, \citenamefont {Komorovsky}, \citenamefont {Bast},\ and\ \citenamefont {Ruud}}]{Repisky2016_other-ZORA-calc}%
  \BibitemOpen
  \bibfield  {author} {\bibinfo {author} {\bibfnamefont {M.}~\bibnamefont {Repisky}}, \bibinfo {author} {\bibfnamefont {S.}~\bibnamefont {Komorovsky}}, \bibinfo {author} {\bibfnamefont {R.}~\bibnamefont {Bast}}, \ and\ \bibinfo {author} {\bibfnamefont {K.}~\bibnamefont {Ruud}},\ }\bibfield  {title} {\enquote {\bibinfo {title} {{Relativistic Calculations of Nuclear Magnetic Resonance Parameters}},}\ }in\ \href {\doibase 10.1039/9781782623816-00267} {\emph {\bibinfo {booktitle} {{Gas Phase NMR}}}},\ \bibinfo {editor} {edited by\ \bibinfo {editor} {\bibfnamefont {K.}~\bibnamefont {Jackowski}}\ and\ \bibinfo {editor} {\bibfnamefont {M.}~\bibnamefont {Jaszunski}}}\ (\bibinfo  {publisher} {The Royal Society of Chemistry},\ \bibinfo {year} {2016})\BibitemShut {NoStop}%
\bibitem [{\citenamefont {Jankowska}\ \emph {et~al.}(2016)\citenamefont {Jankowska}, \citenamefont {Kupka}, \citenamefont {Stobiński}, \citenamefont {Faber}, \citenamefont {Lacerda~Jr.},\ and\ \citenamefont {Sauer}}]{Jankowska2016_other-ZORA-calc}%
  \BibitemOpen
  \bibfield  {author} {\bibinfo {author} {\bibfnamefont {M.}~\bibnamefont {Jankowska}}, \bibinfo {author} {\bibfnamefont {T.}~\bibnamefont {Kupka}}, \bibinfo {author} {\bibfnamefont {L.}~\bibnamefont {Stobiński}}, \bibinfo {author} {\bibfnamefont {R.}~\bibnamefont {Faber}}, \bibinfo {author} {\bibfnamefont {E.~G.}\ \bibnamefont {Lacerda~Jr.}}, \ and\ \bibinfo {author} {\bibfnamefont {S.~P.~A.}\ \bibnamefont {Sauer}},\ }\bibfield  {title} {\enquote {\bibinfo {title} {{Spin-orbit ZORA and four-component Dirac–Coulomb estimation of relativistic corrections to isotropic nuclear shieldings and chemical shifts of noble gas dimers}},}\ }\href {\doibase https://doi.org/10.1002/jcc.24228} {\bibfield  {journal} {\bibinfo  {journal} {J. Comput. Chem.}\ }\textbf {\bibinfo {volume} {37}},\ \bibinfo {pages} {395--403} (\bibinfo {year} {2016})}\BibitemShut {NoStop}%
\bibitem [{\citenamefont {Lino}, \citenamefont {Sauer},\ and\ \citenamefont {Ramalho}(2020)}]{Lino2020_other-ZORA-calc}%
  \BibitemOpen
  \bibfield  {author} {\bibinfo {author} {\bibfnamefont {J.~B. d.~R.}\ \bibnamefont {Lino}}, \bibinfo {author} {\bibfnamefont {S.~P.~A.}\ \bibnamefont {Sauer}}, \ and\ \bibinfo {author} {\bibfnamefont {T.~C.}\ \bibnamefont {Ramalho}},\ }\bibfield  {title} {\enquote {\bibinfo {title} {{Enhancing NMR Quantum Computation by Exploring Heavy Metal Complexes as Multiqubit Systems: A Theoretical Investigation}},}\ }\href {\doibase 10.1021/acs.jpca.0c01607} {\bibfield  {journal} {\bibinfo  {journal} {J. Phys. Chem. A}\ }\textbf {\bibinfo {volume} {124}},\ \bibinfo {pages} {4946--4955} (\bibinfo {year} {2020})}\BibitemShut {NoStop}%
\bibitem [{\citenamefont {V{\' i}cha}\ \emph {et~al.}(2020)\citenamefont {Vı́cha}, \citenamefont {Novotn{\' y}}, \citenamefont {Komorovsky}, \citenamefont {Straka}, \citenamefont {Kaupp},\ and\ \citenamefont {Marek}}]{Vı́cha2020_other-ZORA-calc}%
  \BibitemOpen
  \bibfield  {author} {\bibinfo {author} {\bibfnamefont {J.}~\bibnamefont {V{\' i}cha}}, \bibinfo {author} {\bibfnamefont {J.}~\bibnamefont {Novotn{\' y}}}, \bibinfo {author} {\bibfnamefont {S.}~\bibnamefont {Komorovsky}}, \bibinfo {author} {\bibfnamefont {M.}~\bibnamefont {Straka}}, \bibinfo {author} {\bibfnamefont {M.}~\bibnamefont {Kaupp}}, \ and\ \bibinfo {author} {\bibfnamefont {R.}~\bibnamefont {Marek}},\ }\bibfield  {title} {\enquote {\bibinfo {title} {{Relativistic Heavy-Neighbor-Atom Effects on NMR Shifts: Concepts and Trends Across the Periodic Table}},}\ }\href {\doibase 10.1021/acs.chemrev.9b00785} {\bibfield  {journal} {\bibinfo  {journal} {Chem. Rev.}\ }\textbf {\bibinfo {volume} {120}},\ \bibinfo {pages} {7065--7103} (\bibinfo {year} {2020})}\BibitemShut {NoStop}%
\bibitem [{\citenamefont {Glent-Madsen}\ \emph {et~al.}(2021)\citenamefont {Glent-Madsen}, \citenamefont {Reinholdt}, \citenamefont {Bendix},\ and\ \citenamefont {Sauer}}]{Glent-Madsen2021_other-ZORA-calc}%
  \BibitemOpen
  \bibfield  {author} {\bibinfo {author} {\bibfnamefont {I.}~\bibnamefont {Glent-Madsen}}, \bibinfo {author} {\bibfnamefont {A.}~\bibnamefont {Reinholdt}}, \bibinfo {author} {\bibfnamefont {J.}~\bibnamefont {Bendix}}, \ and\ \bibinfo {author} {\bibfnamefont {S.~P.~A.}\ \bibnamefont {Sauer}},\ }\bibfield  {title} {\enquote {\bibinfo {title} {{Importance of Relativistic Effects for Carbon as an NMR Reporter Nucleus in Carbide-Bridged [RuCPt] Complexes}},}\ }\href {\doibase 10.1021/acs.organomet.1c00079} {\bibfield  {journal} {\bibinfo  {journal} {Organometallics}\ }\textbf {\bibinfo {volume} {40}},\ \bibinfo {pages} {1443--1453} (\bibinfo {year} {2021})}\BibitemShut {NoStop}%
\bibitem [{\citenamefont {Lino}\ \emph {et~al.}(2022)\citenamefont {Lino}, \citenamefont {Gonçalves}, \citenamefont {Sauer},\ and\ \citenamefont {Ramalho}}]{Lino2022_other-ZORA-calc}%
  \BibitemOpen
  \bibfield  {author} {\bibinfo {author} {\bibfnamefont {J.~B. d.~R.}\ \bibnamefont {Lino}}, \bibinfo {author} {\bibfnamefont {M.~A.}\ \bibnamefont {Gonçalves}}, \bibinfo {author} {\bibfnamefont {S.~P.~A.}\ \bibnamefont {Sauer}}, \ and\ \bibinfo {author} {\bibfnamefont {T.~C.}\ \bibnamefont {Ramalho}},\ }\bibfield  {title} {\enquote {\bibinfo {title} {{Extending NMR Quantum Computation Systems by Employing Compounds with Several Heavy Metals as Qubits}},}\ }\href {\doibase 10.3390/magnetochemistry8050047} {\bibfield  {journal} {\bibinfo  {journal} {Magnetochemistry}\ }\textbf {\bibinfo {volume} {8}},\ \bibinfo {pages} {47} (\bibinfo {year} {2022})}\BibitemShut {NoStop}%
\bibitem [{\citenamefont {Wu}, \citenamefont {Hemmingsen},\ and\ \citenamefont {Sauer}(2024)}]{Wu2023_Hg-geometry}%
  \BibitemOpen
  \bibfield  {author} {\bibinfo {author} {\bibfnamefont {H.}~\bibnamefont {Wu}}, \bibinfo {author} {\bibfnamefont {L.}~\bibnamefont {Hemmingsen}}, \ and\ \bibinfo {author} {\bibfnamefont {S.~P.~A.}\ \bibnamefont {Sauer}},\ }\bibfield  {title} {\enquote {\bibinfo {title} {{On the Geometry Dependence of the NMR Chemical Shift of Mercury in Thiolate Complexes: A Relativistic DFT Study}},}\ }\href {\doibase 10.1002/mrc.5452} {\bibfield  {journal} {\bibinfo  {journal} {Magn. Reson. Chem.}\ }\textbf {\bibinfo {volume} {62}},\ \bibinfo {pages} {648--669} (\bibinfo {year} {2024})}\BibitemShut {NoStop}%
\bibitem [{\citenamefont {Jessen}, \citenamefont {Hemmingsen},\ and\ \citenamefont {Sauer}(2024)}]{Jessen2024}%
  \BibitemOpen
  \bibfield  {author} {\bibinfo {author} {\bibfnamefont {L.~M.}\ \bibnamefont {Jessen}}, \bibinfo {author} {\bibfnamefont {L.}~\bibnamefont {Hemmingsen}}, \ and\ \bibinfo {author} {\bibfnamefont {S.~P.~A.}\ \bibnamefont {Sauer}},\ }\bibfield  {title} {\enquote {\bibinfo {title} {{$^{199}$Hg NMR Shielding and Chemical Shifts of 2-, 3-, and 4-Coordinate Hg(II)-Thiolate Species}},}\ }\href {\doibase 10.1021/acs.inorgchem.4c03518} {\bibfield  {journal} {\bibinfo  {journal} {Inorg. Chem.}\ }\textbf {\bibinfo {volume} {63}},\ \bibinfo {pages} {23614--23619} (\bibinfo {year} {2024})}\BibitemShut {NoStop}%
\bibitem [{\citenamefont {Wrackmeyer}\ and\ \citenamefont {Contreras}(1992)}]{WRACKMEYER1992_Hg-nmr}%
  \BibitemOpen
  \bibfield  {author} {\bibinfo {author} {\bibfnamefont {B.}~\bibnamefont {Wrackmeyer}}\ and\ \bibinfo {author} {\bibfnamefont {R.}~\bibnamefont {Contreras}},\ }\bibfield  {title} {\enquote {\bibinfo {title} {{$^{199}$Hg NMR Parameters}},}\ }in\ \href {\doibase https://doi.org/10.1016/S0066-4103(08)60200-8} {\emph {\bibinfo {booktitle} {Annu. Rep. NMR Spectrosc.}}},\ Vol.~\bibinfo {volume} {24},\ \bibinfo {editor} {edited by\ \bibinfo {editor} {\bibfnamefont {G.}~\bibnamefont {Webb}}}\ (\bibinfo  {publisher} {Academic Press},\ \bibinfo {year} {1992})\ pp.\ \bibinfo {pages} {267--329}\BibitemShut {NoStop}%
\bibitem [{\citenamefont {Utschig}, \citenamefont {Bryson},\ and\ \citenamefont {O'Halloran}(1995)}]{Utschig1995_MerR-NMR}%
  \BibitemOpen
  \bibfield  {author} {\bibinfo {author} {\bibfnamefont {L.~M.}\ \bibnamefont {Utschig}}, \bibinfo {author} {\bibfnamefont {J.~W.}\ \bibnamefont {Bryson}}, \ and\ \bibinfo {author} {\bibfnamefont {T.~V.}\ \bibnamefont {O'Halloran}},\ }\bibfield  {title} {\enquote {\bibinfo {title} {{Mercury-199 NMR of the Metal Receptor Site in MerR and Its Protein-DNA Complex}},}\ }\href@noop {} {\bibfield  {journal} {\bibinfo  {journal} {Science}\ }\textbf {\bibinfo {volume} {268}},\ \bibinfo {pages} {380--385} (\bibinfo {year} {1995})}\BibitemShut {NoStop}%
\bibitem [{\citenamefont {Iranzo}\ \emph {et~al.}(2007)\citenamefont {Iranzo}, \citenamefont {Thulstrup}, \citenamefont {Ryu}, \citenamefont {Hemmingsen},\ and\ \citenamefont {Pecoraro}}]{Iranzo2007_Hg-NMR-first-coordinations-sphere}%
  \BibitemOpen
  \bibfield  {author} {\bibinfo {author} {\bibfnamefont {O.}~\bibnamefont {Iranzo}}, \bibinfo {author} {\bibfnamefont {P.~W.}\ \bibnamefont {Thulstrup}}, \bibinfo {author} {\bibfnamefont {S.-b.}\ \bibnamefont {Ryu}}, \bibinfo {author} {\bibfnamefont {L.}~\bibnamefont {Hemmingsen}}, \ and\ \bibinfo {author} {\bibfnamefont {V.~L.}\ \bibnamefont {Pecoraro}},\ }\bibfield  {title} {\enquote {\bibinfo {title} {{The Application of 199Hg NMR and 199mHg Perturbed Angular Correlation (PAC) Spectroscopy to Define the Biological Chemistry of HgII: A Case Study with Designed Two- and Three-Stranded Coiled Coils}},}\ }\href {\doibase https://doi.org/10.1002/chem.200701208} {\bibfield  {journal} {\bibinfo  {journal} {Chem. Eur. J.}\ }\textbf {\bibinfo {volume} {13}},\ \bibinfo {pages} {9178--9190} (\bibinfo {year} {2007})}\BibitemShut {NoStop}%
\bibitem [{\citenamefont {Łuczkowski}\ \emph {et~al.}(2008)\citenamefont {Łuczkowski}, \citenamefont {Stachura}, \citenamefont {Schirf}, \citenamefont {Demeler}, \citenamefont {Hemmingsen},\ and\ \citenamefont {Pecoraro}}]{Luczkowski2008_Hg-NMR-first_coordinations_sphere}%
  \BibitemOpen
  \bibfield  {author} {\bibinfo {author} {\bibfnamefont {M.}~\bibnamefont {Łuczkowski}}, \bibinfo {author} {\bibfnamefont {M.}~\bibnamefont {Stachura}}, \bibinfo {author} {\bibfnamefont {V.}~\bibnamefont {Schirf}}, \bibinfo {author} {\bibfnamefont {B.}~\bibnamefont {Demeler}}, \bibinfo {author} {\bibfnamefont {L.}~\bibnamefont {Hemmingsen}}, \ and\ \bibinfo {author} {\bibfnamefont {V.~L.}\ \bibnamefont {Pecoraro}},\ }\bibfield  {title} {\enquote {\bibinfo {title} {Design of thiolate rich metal binding sites within a peptidic framework},}\ }\href {\doibase 10.1021/ic8009817} {\bibfield  {journal} {\bibinfo  {journal} {Inorg. Chem.}\ }\textbf {\bibinfo {volume} {47}},\ \bibinfo {pages} {10875--10888} (\bibinfo {year} {2008})}\BibitemShut {NoStop}%
\bibitem [{\citenamefont {Wagner}(1993)}]{Wagner1993_NMR-protein}%
  \BibitemOpen
  \bibfield  {author} {\bibinfo {author} {\bibfnamefont {G.}~\bibnamefont {Wagner}},\ }\bibfield  {title} {\enquote {\bibinfo {title} {{Prospects for NMR of large proteins}},}\ }\href {\doibase 10.1007/BF00176005} {\bibfield  {journal} {\bibinfo  {journal} {J. Biomol. NMR}\ }\textbf {\bibinfo {volume} {3}},\ \bibinfo {pages} {375--385} (\bibinfo {year} {1993})}\BibitemShut {NoStop}%
\bibitem [{\citenamefont {Raman}\ \emph {et~al.}(2010)\citenamefont {Raman}, \citenamefont {Lange}, \citenamefont {Rossi}, \citenamefont {Tyka}, \citenamefont {Wang}, \citenamefont {Aramini}, \citenamefont {Liu}, \citenamefont {Ramelot}, \citenamefont {Eletsky}, \citenamefont {Szyperski}, \citenamefont {Kennedy}, \citenamefont {Prestegard}, \citenamefont {Montelione},\ and\ \citenamefont {Baker}}]{Srivatsan2010_NMR-protein}%
  \BibitemOpen
  \bibfield  {author} {\bibinfo {author} {\bibfnamefont {S.}~\bibnamefont {Raman}}, \bibinfo {author} {\bibfnamefont {O.~F.}\ \bibnamefont {Lange}}, \bibinfo {author} {\bibfnamefont {P.}~\bibnamefont {Rossi}}, \bibinfo {author} {\bibfnamefont {M.}~\bibnamefont {Tyka}}, \bibinfo {author} {\bibfnamefont {X.}~\bibnamefont {Wang}}, \bibinfo {author} {\bibfnamefont {J.}~\bibnamefont {Aramini}}, \bibinfo {author} {\bibfnamefont {G.}~\bibnamefont {Liu}}, \bibinfo {author} {\bibfnamefont {T.~A.}\ \bibnamefont {Ramelot}}, \bibinfo {author} {\bibfnamefont {A.}~\bibnamefont {Eletsky}}, \bibinfo {author} {\bibfnamefont {T.}~\bibnamefont {Szyperski}}, \bibinfo {author} {\bibfnamefont {M.~A.}\ \bibnamefont {Kennedy}}, \bibinfo {author} {\bibfnamefont {J.}~\bibnamefont {Prestegard}}, \bibinfo {author} {\bibfnamefont {G.~T.}\ \bibnamefont {Montelione}}, \ and\ \bibinfo {author} {\bibfnamefont {D.}~\bibnamefont {Baker}},\ }\bibfield  {title} {\enquote {\bibinfo {title} {{NMR Structure Determination for Larger Proteins Using
  Backbone-Only Data}},}\ }\href {\doibase 10.1126/science.1183649} {\bibfield  {journal} {\bibinfo  {journal} {Science}\ }\textbf {\bibinfo {volume} {327}},\ \bibinfo {pages} {1014--1018} (\bibinfo {year} {2010})}\BibitemShut {NoStop}%
\bibitem [{\citenamefont {Danielson}\ and\ \citenamefont {Falke}(1996)}]{Danielson1996_NMR_proteins}%
  \BibitemOpen
  \bibfield  {author} {\bibinfo {author} {\bibfnamefont {M.~A.}\ \bibnamefont {Danielson}}\ and\ \bibinfo {author} {\bibfnamefont {J.~J.}\ \bibnamefont {Falke}},\ }\bibfield  {title} {\enquote {\bibinfo {title} {{Use of 19F NMR to Probe Protein Structure and Conformational Changes}},}\ }\href {\doibase 10.1146/annurev.bb.25.060196.001115} {\bibfield  {journal} {\bibinfo  {journal} {Annu. Rev. Biophys. Biomol. Struct.}\ }\textbf {\bibinfo {volume} {25}},\ \bibinfo {pages} {163--195} (\bibinfo {year} {1996})}\BibitemShut {NoStop}%
\bibitem [{\citenamefont {Duus}, \citenamefont {Gotfredsen},\ and\ \citenamefont {Bock}(2000)}]{Duus2000_NMR-carbohydrates_proteins}%
  \BibitemOpen
  \bibfield  {author} {\bibinfo {author} {\bibfnamefont {J.~\O.}\ \bibnamefont {Duus}}, \bibinfo {author} {\bibfnamefont {C.~H.}\ \bibnamefont {Gotfredsen}}, \ and\ \bibinfo {author} {\bibfnamefont {K.}~\bibnamefont {Bock}},\ }\bibfield  {title} {\enquote {\bibinfo {title} {{Carbohydrate Structural Determination by NMR Spectroscopy: Modern Methods and Limitations}},}\ }\href {\doibase 10.1021/cr990302n} {\bibfield  {journal} {\bibinfo  {journal} {Chem. Rev.}\ }\textbf {\bibinfo {volume} {100}},\ \bibinfo {pages} {4589--4614} (\bibinfo {year} {2000})}\BibitemShut {NoStop}%
\bibitem [{\citenamefont {Tosato}\ \emph {et~al.}(2024)\citenamefont {Tosato}, \citenamefont {P.~Randhawa}, \citenamefont {Hemmingsen}, \citenamefont {O'Shea}, \citenamefont {Thaveenrasingam}, \citenamefont {Sauer}, \citenamefont {Chen}, \citenamefont {Graiff}, \citenamefont {Menegazzo}, \citenamefont {Baron}, \citenamefont {Radchenko}, \citenamefont {Ramogida},\ and\ \citenamefont {Marco}}]{Tosato2024}%
  \BibitemOpen
  \bibfield  {author} {\bibinfo {author} {\bibfnamefont {M.}~\bibnamefont {Tosato}}, \bibinfo {author} {\bibfnamefont {M.~A.}\ \bibnamefont {P.~Randhawa}}, \bibinfo {author} {\bibfnamefont {L.~B.~S.}\ \bibnamefont {Hemmingsen}}, \bibinfo {author} {\bibfnamefont {C.~A.}\ \bibnamefont {O'Shea}}, \bibinfo {author} {\bibfnamefont {P.}~\bibnamefont {Thaveenrasingam}}, \bibinfo {author} {\bibfnamefont {S.~P.~A.}\ \bibnamefont {Sauer}}, \bibinfo {author} {\bibfnamefont {S.}~\bibnamefont {Chen}}, \bibinfo {author} {\bibfnamefont {C.}~\bibnamefont {Graiff}}, \bibinfo {author} {\bibfnamefont {I.}~\bibnamefont {Menegazzo}}, \bibinfo {author} {\bibfnamefont {M.}~\bibnamefont {Baron}}, \bibinfo {author} {\bibfnamefont {V.}~\bibnamefont {Radchenko}}, \bibinfo {author} {\bibfnamefont {C.~F.}\ \bibnamefont {Ramogida}}, \ and\ \bibinfo {author} {\bibfnamefont {V.~D.}\ \bibnamefont {Marco}},\ }\bibfield  {title} {\enquote {\bibinfo {title} {{Capturing Mercury-197m/g for Auger Electron Therapy and Cancer Theranostic with
  Sulfur-Containing Cyclen-Based Macrocycles}},}\ }\href@noop {} {\bibfield  {journal} {\bibinfo  {journal} {Inorg. Chem.}\ }\textbf {\bibinfo {volume} {63}},\ \bibinfo {pages} {14241--14255} (\bibinfo {year} {2024})}\BibitemShut {NoStop}%
\bibitem [{\citenamefont {Olsen}\ \emph {et~al.}(2002)\citenamefont {Olsen}, \citenamefont {Christiansen}, \citenamefont {Hemmingsen}, \citenamefont {Sauer},\ and\ \citenamefont {Mikkelsen}}]{spas051}%
  \BibitemOpen
  \bibfield  {author} {\bibinfo {author} {\bibfnamefont {L.}~\bibnamefont {Olsen}}, \bibinfo {author} {\bibfnamefont {O.}~\bibnamefont {Christiansen}}, \bibinfo {author} {\bibfnamefont {L.}~\bibnamefont {Hemmingsen}}, \bibinfo {author} {\bibfnamefont {S.~P.~A.}\ \bibnamefont {Sauer}}, \ and\ \bibinfo {author} {\bibfnamefont {K.~V.}\ \bibnamefont {Mikkelsen}},\ }\bibfield  {title} {\enquote {\bibinfo {title} {{Electric field gradients of water: A systematic investigation of basis set, electron correlation, and rovibrational effects}},}\ }\href {\doibase 10.1063/1.1428340} {\bibfield  {journal} {\bibinfo  {journal} {J. Chem. Phys.}\ }\textbf {\bibinfo {volume} {116}},\ \bibinfo {pages} {1424--1434} (\bibinfo {year} {2002})}\BibitemShut {NoStop}%
\bibitem [{\citenamefont {O’Shea}\ \emph {et~al.}(2023)\citenamefont {O’Shea}, \citenamefont {Fromsejer}, \citenamefont {Sauer}, \citenamefont {Mikkelsen},\ and\ \citenamefont {Hemmingsen}}]{OShea2023}%
  \BibitemOpen
  \bibfield  {author} {\bibinfo {author} {\bibfnamefont {C.~A.}\ \bibnamefont {O’Shea}}, \bibinfo {author} {\bibfnamefont {R.}~\bibnamefont {Fromsejer}}, \bibinfo {author} {\bibfnamefont {S.~P.~A.}\ \bibnamefont {Sauer}}, \bibinfo {author} {\bibfnamefont {K.~V.}\ \bibnamefont {Mikkelsen}}, \ and\ \bibinfo {author} {\bibfnamefont {L.}~\bibnamefont {Hemmingsen}},\ }\bibfield  {title} {\enquote {\bibinfo {title} {{Calculation of electric field gradients in Cd(II) model complexes of the CueR protein metal site}},}\ }\href {\doibase 10.1039/D2CP05574K} {\bibfield  {journal} {\bibinfo  {journal} {Phys. Chem. Chem. Phys.}\ }\textbf {\bibinfo {volume} {25}},\ \bibinfo {pages} {12277--12283} (\bibinfo {year} {2023})}\BibitemShut {NoStop}%
\bibitem [{\citenamefont {Nagy}\ \emph {et~al.}(2024)\citenamefont {Nagy}, \citenamefont {Reinholdt}, \citenamefont {Jensen}, \citenamefont {Kjellgren}, \citenamefont {Ziems}, \citenamefont {Fitzpatrick}, \citenamefont {Knecht}, \citenamefont {Kongsted}, \citenamefont {Coriani},\ and\ \citenamefont {Sauer}}]{Nagy2024}%
  \BibitemOpen
  \bibfield  {author} {\bibinfo {author} {\bibfnamefont {D.}~\bibnamefont {Nagy}}, \bibinfo {author} {\bibfnamefont {P.}~\bibnamefont {Reinholdt}}, \bibinfo {author} {\bibfnamefont {P.~W.~K.}\ \bibnamefont {Jensen}}, \bibinfo {author} {\bibfnamefont {E.~R.}\ \bibnamefont {Kjellgren}}, \bibinfo {author} {\bibfnamefont {K.~M.}\ \bibnamefont {Ziems}}, \bibinfo {author} {\bibfnamefont {A.}~\bibnamefont {Fitzpatrick}}, \bibinfo {author} {\bibfnamefont {S.}~\bibnamefont {Knecht}}, \bibinfo {author} {\bibfnamefont {J.}~\bibnamefont {Kongsted}}, \bibinfo {author} {\bibfnamefont {S.}~\bibnamefont {Coriani}}, \ and\ \bibinfo {author} {\bibfnamefont {S.~P.~A.}\ \bibnamefont {Sauer}},\ }\bibfield  {title} {\enquote {\bibinfo {title} {{Electric Field Gradient Calculations for Ice VIII and IX using Polarizable Embedding: A Comparative Study on Classical Computers and Quantum Simulators}},}\ }\href@noop {} {\bibfield  {journal} {\bibinfo  {journal} {J. Phys. Chem. A}\ }\textbf {\bibinfo {volume} {128}},\ \bibinfo {pages}
  {6305--6315} (\bibinfo {year} {2024})}\BibitemShut {NoStop}%
\bibitem [{\citenamefont {Hemmingsen}, \citenamefont {Sas},\ and\ \citenamefont {Danielsen}(2004)}]{Hemmingsen2004_PAC}%
  \BibitemOpen
  \bibfield  {author} {\bibinfo {author} {\bibfnamefont {L.}~\bibnamefont {Hemmingsen}}, \bibinfo {author} {\bibfnamefont {K.~N.}\ \bibnamefont {Sas}}, \ and\ \bibinfo {author} {\bibfnamefont {E.}~\bibnamefont {Danielsen}},\ }\bibfield  {title} {\enquote {\bibinfo {title} {{Biological Applications of Perturbed Angular Correlations of $\gamma$-Ray Spectroscopy}},}\ }\href {\doibase 10.1021/cr030030v} {\bibfield  {journal} {\bibinfo  {journal} {Chem. Rev.}\ }\textbf {\bibinfo {volume} {104}},\ \bibinfo {pages} {4027--4062} (\bibinfo {year} {2004})}\BibitemShut {NoStop}%
\bibitem [{\citenamefont {Tröger}(1999)}]{Troger1999_MerR-EFG}%
  \BibitemOpen
  \bibfield  {author} {\bibinfo {author} {\bibfnamefont {W.}~\bibnamefont {Tröger}},\ }\bibfield  {title} {\enquote {\bibinfo {title} {{Nuclear probes in life sciences}},}\ }\href {\doibase 10.1023/A:1017082113854} {\bibfield  {journal} {\bibinfo  {journal} {Hyperfine Interact.}\ }\textbf {\bibinfo {volume} {120}},\ \bibinfo {pages} {117–128} (\bibinfo {year} {1999})}\BibitemShut {NoStop}%
\bibitem [{\citenamefont {Tröger}\ and\ \citenamefont {the ISOLDE~Collaboration}(2001)}]{Troger2001_PAC-biomolecules}%
  \BibitemOpen
  \bibfield  {author} {\bibinfo {author} {\bibfnamefont {W.}~\bibnamefont {Tröger}}\ and\ \bibinfo {author} {\bibnamefont {the ISOLDE~Collaboration}},\ }\bibfield  {title} {\enquote {\bibinfo {title} {{Hg(II) Coordination Studies in Penicillamine Enantiomers by $^{199m}$Hg-TDPAC}},}\ }\href {\doibase https://doi.org/10.1023/A:1020500709562} {\bibfield  {journal} {\bibinfo  {journal} {Hyperfine Interact.}\ }\textbf {\bibinfo {volume} {136}},\ \bibinfo {pages} {673–680} (\bibinfo {year} {2001})}\BibitemShut {NoStop}%
\bibitem [{\citenamefont {Butz}\ \emph {et~al.}(1992)\citenamefont {Butz}, \citenamefont {Tröger}, \citenamefont {Pöhlmann},\ and\ \citenamefont {Nuyken}}]{Butz1992_PAC_biomolecules}%
  \BibitemOpen
  \bibfield  {author} {\bibinfo {author} {\bibfnamefont {T.}~\bibnamefont {Butz}}, \bibinfo {author} {\bibfnamefont {W.}~\bibnamefont {Tröger}}, \bibinfo {author} {\bibfnamefont {T.}~\bibnamefont {Pöhlmann}}, \ and\ \bibinfo {author} {\bibfnamefont {O.}~\bibnamefont {Nuyken}},\ }\bibfield  {title} {\enquote {\bibinfo {title} {{The Nuclear Quadrupole Interaction of $^{199m}$Hg-Cysteine and $^{199m}$Hg-tert-butyl-mercaptide}},}\ }\href {\doibase https://doi.org/10.1515/zna-1992-1-215} {\bibfield  {journal} {\bibinfo  {journal} {Z. Naturforsch. A}\ }\textbf {\bibinfo {volume} {47}},\ \bibinfo {pages} {85--88} (\bibinfo {year} {1992})}\BibitemShut {NoStop}%
\bibitem [{\citenamefont {Ctortecka}\ \emph {et~al.}(1999)\citenamefont {Ctortecka}, \citenamefont {Tröger}, \citenamefont {S.~Mallion}, \citenamefont {Hoffmann},\ and\ \citenamefont {the ISOLDE-Collaboration}}]{Ctortecka1999_PAC_biomolecules}%
  \BibitemOpen
  \bibfield  {author} {\bibinfo {author} {\bibfnamefont {B.}~\bibnamefont {Ctortecka}}, \bibinfo {author} {\bibfnamefont {W.}~\bibnamefont {Tröger}}, \bibinfo {author} {\bibfnamefont {T.~B.}\ \bibnamefont {S.~Mallion}}, \bibinfo {author} {\bibfnamefont {R.}~\bibnamefont {Hoffmann}}, \ and\ \bibinfo {author} {\bibnamefont {the ISOLDE-Collaboration}},\ }\bibfield  {title} {\enquote {\bibinfo {title} {{Hg-coordination studies of oligopeptides containing cysteine, histidine and tyrosine by $^{199m}$Hg-TDPAC.}}}\ }\href {\doibase https://doi.org/10.1023/A:1017008631060} {\bibfield  {journal} {\bibinfo  {journal} {Hyperfine Interact.}\ }\textbf {\bibinfo {volume} {120}},\ \bibinfo {pages} {737–743} (\bibinfo {year} {1999})}\BibitemShut {NoStop}%
\bibitem [{\citenamefont {Faller}\ \emph {et~al.}(2000)\citenamefont {Faller}, \citenamefont {Ctortecka}, \citenamefont {Tröger}, \citenamefont {Butz}, \citenamefont {Collaboration},\ and\ \citenamefont {Va{\v s}{\' a}k}}]{Faller2000_PAC_biomolecules}%
  \BibitemOpen
  \bibfield  {author} {\bibinfo {author} {\bibfnamefont {P.}~\bibnamefont {Faller}}, \bibinfo {author} {\bibfnamefont {B.}~\bibnamefont {Ctortecka}}, \bibinfo {author} {\bibfnamefont {W.}~\bibnamefont {Tröger}}, \bibinfo {author} {\bibfnamefont {T.}~\bibnamefont {Butz}}, \bibinfo {author} {\bibfnamefont {I.}~\bibnamefont {Collaboration}}, \ and\ \bibinfo {author} {\bibfnamefont {M.}~\bibnamefont {Va{\v s}{\' a}k}},\ }\bibfield  {title} {\enquote {\bibinfo {title} {{Optical and TDPAC spectroscopy of Hg(II)-rubredoxin: model for a mononuclear tetrahedral [Hg(CysS)$_4$]$^{2-}$ center}},}\ }\href {\doibase https://doi.org/10.1007/PL00010668} {\bibfield  {journal} {\bibinfo  {journal} {J. Biol. Inorg. Chem.}\ }\textbf {\bibinfo {volume} {5}},\ \bibinfo {pages} {393--401} (\bibinfo {year} {2000})}\BibitemShut {NoStop}%
\bibitem [{\citenamefont {Haas}\ \emph {et~al.}(2017)\citenamefont {Haas}, \citenamefont {Sauer}, \citenamefont {Hemmingsen}, \citenamefont {Kellö},\ and\ \citenamefont {Zhao}}]{Haas_2017}%
  \BibitemOpen
  \bibfield  {author} {\bibinfo {author} {\bibfnamefont {H.}~\bibnamefont {Haas}}, \bibinfo {author} {\bibfnamefont {S.~P.~A.}\ \bibnamefont {Sauer}}, \bibinfo {author} {\bibfnamefont {L.}~\bibnamefont {Hemmingsen}}, \bibinfo {author} {\bibfnamefont {V.}~\bibnamefont {Kellö}}, \ and\ \bibinfo {author} {\bibfnamefont {P.~W.}\ \bibnamefont {Zhao}},\ }\bibfield  {title} {\enquote {\bibinfo {title} {{Quadrupole moments of Cd and Zn nuclei: When solid-state, molecular, atomic, and nuclear theory meet}},}\ }\href {\doibase 10.1209/0295-5075/117/62001} {\bibfield  {journal} {\bibinfo  {journal} {Europhysics Letters}\ }\textbf {\bibinfo {volume} {117}},\ \bibinfo {pages} {62001} (\bibinfo {year} {2017})}\BibitemShut {NoStop}%
\bibitem [{\citenamefont {Kern}\ and\ \citenamefont {Matcha}(1968)}]{Kern1968_vib_avg_original}%
  \BibitemOpen
  \bibfield  {author} {\bibinfo {author} {\bibfnamefont {C.~W.}\ \bibnamefont {Kern}}\ and\ \bibinfo {author} {\bibfnamefont {R.~L.}\ \bibnamefont {Matcha}},\ }\bibfield  {title} {\enquote {\bibinfo {title} {{Nuclear Corrections to Electronic Expectation Values: Zero‐Point Vibrational Effects in the Water Molecule}},}\ }\href {\doibase 10.1063/1.1670369} {\bibfield  {journal} {\bibinfo  {journal} {J. Chem. Phys.}\ }\textbf {\bibinfo {volume} {49}},\ \bibinfo {pages} {2081--2091} (\bibinfo {year} {1968})}\BibitemShut {NoStop}%
\bibitem [{\citenamefont {Gleeson}\ \emph {et~al.}(2024)\citenamefont {Gleeson}, \citenamefont {Aggelund}, \citenamefont {Østergaard}, \citenamefont {Schaltz},\ and\ \citenamefont {Sauer}}]{Gleeson_vib_avg_dalton-project}%
  \BibitemOpen
  \bibfield  {author} {\bibinfo {author} {\bibfnamefont {R.}~\bibnamefont {Gleeson}}, \bibinfo {author} {\bibfnamefont {P.~A.}\ \bibnamefont {Aggelund}}, \bibinfo {author} {\bibfnamefont {F.~C.}\ \bibnamefont {Østergaard}}, \bibinfo {author} {\bibfnamefont {K.~F.}\ \bibnamefont {Schaltz}}, \ and\ \bibinfo {author} {\bibfnamefont {S.~P.~A.}\ \bibnamefont {Sauer}},\ }\bibfield  {title} {\enquote {\bibinfo {title} {{Exploring Alternate Methods for the Calculation of High-Level Vibrational Corrections of NMR Spin–Spin Coupling Constants}},}\ }\href {\doibase 10.1021/acs.jctc.3c01223} {\bibfield  {journal} {\bibinfo  {journal} {J. Chem. Theory Comput.}\ }\textbf {\bibinfo {volume} {20}},\ \bibinfo {pages} {1228--1243} (\bibinfo {year} {2024})}\BibitemShut {NoStop}%
\bibitem [{\citenamefont {Olsen}\ \emph {et~al.}(2020)\citenamefont {Olsen}, \citenamefont {Reine}, \citenamefont {Vahtras}, \citenamefont {Kjellgren}, \citenamefont {Reinholdt}, \citenamefont {{Hjorth Dundas}}, \citenamefont {Li}, \citenamefont {Cukras}, \citenamefont {Ringholm}, \citenamefont {Hedeg{\aa}rd}, \citenamefont {{Di Remigio}}, \citenamefont {List}, \citenamefont {Faber}, \citenamefont {{Cabral Tenorio}}, \citenamefont {Bast}, \citenamefont {Pedersen}, \citenamefont {Rinkevicius}, \citenamefont {Sauer}, \citenamefont {Mikkelsen}, \citenamefont {Kongsted}, \citenamefont {Coriani}, \citenamefont {Ruud}, \citenamefont {Helgaker}, \citenamefont {Jensen},\ and\ \citenamefont {Norman}}]{spas191}%
  \BibitemOpen
  \bibfield  {author} {\bibinfo {author} {\bibfnamefont {J.~M.~H.}\ \bibnamefont {Olsen}}, \bibinfo {author} {\bibfnamefont {S.}~\bibnamefont {Reine}}, \bibinfo {author} {\bibfnamefont {O.}~\bibnamefont {Vahtras}}, \bibinfo {author} {\bibfnamefont {E.}~\bibnamefont {Kjellgren}}, \bibinfo {author} {\bibfnamefont {P.}~\bibnamefont {Reinholdt}}, \bibinfo {author} {\bibfnamefont {K.~O.}\ \bibnamefont {{Hjorth Dundas}}}, \bibinfo {author} {\bibfnamefont {X.}~\bibnamefont {Li}}, \bibinfo {author} {\bibfnamefont {J.}~\bibnamefont {Cukras}}, \bibinfo {author} {\bibfnamefont {M.}~\bibnamefont {Ringholm}}, \bibinfo {author} {\bibfnamefont {E.~D.}\ \bibnamefont {Hedeg{\aa}rd}}, \bibinfo {author} {\bibfnamefont {R.}~\bibnamefont {{Di Remigio}}}, \bibinfo {author} {\bibfnamefont {N.~H.}\ \bibnamefont {List}}, \bibinfo {author} {\bibfnamefont {R.}~\bibnamefont {Faber}}, \bibinfo {author} {\bibfnamefont {B.~N.}\ \bibnamefont {{Cabral Tenorio}}}, \bibinfo {author} {\bibfnamefont {R.}~\bibnamefont {Bast}}, \bibinfo {author}
  {\bibfnamefont {T.~B.}\ \bibnamefont {Pedersen}}, \bibinfo {author} {\bibfnamefont {Z.}~\bibnamefont {Rinkevicius}}, \bibinfo {author} {\bibfnamefont {S.~P.~A.}\ \bibnamefont {Sauer}}, \bibinfo {author} {\bibfnamefont {K.~V.}\ \bibnamefont {Mikkelsen}}, \bibinfo {author} {\bibfnamefont {J.}~\bibnamefont {Kongsted}}, \bibinfo {author} {\bibfnamefont {S.}~\bibnamefont {Coriani}}, \bibinfo {author} {\bibfnamefont {K.}~\bibnamefont {Ruud}}, \bibinfo {author} {\bibfnamefont {T.}~\bibnamefont {Helgaker}}, \bibinfo {author} {\bibfnamefont {H.~J.~A.}\ \bibnamefont {Jensen}}, \ and\ \bibinfo {author} {\bibfnamefont {P.}~\bibnamefont {Norman}},\ }\bibfield  {title} {\enquote {\bibinfo {title} {{Dalton Project: A Python platform for molecular- and electronic-structure simulations of complex systems}},}\ }\href {\doibase 10.1063/1.5144298} {\bibfield  {journal} {\bibinfo  {journal} {J. Chem. Phys.}\ }\textbf {\bibinfo {volume} {152}},\ \bibinfo {pages} {214115} (\bibinfo {year} {2020})}\BibitemShut {NoStop}%
\bibitem [{\citenamefont {Baerends}\ \emph {et~al.}(2023)\citenamefont {Baerends}, \citenamefont {Ziegler}, \citenamefont {Atkins}, \citenamefont {Autschbach}, \citenamefont {Bashford}, \citenamefont {Baseggio}, \citenamefont {B{\'{e}}rces}, \citenamefont {Bickelhaupt}, \citenamefont {Bo}, \citenamefont {Boerritger}, \citenamefont {Cappelli}, \citenamefont {Cavallo}, \citenamefont {Daul}, \citenamefont {Chong}, \citenamefont {Chulhai}, \citenamefont {Deng}, \citenamefont {Dickson}, \citenamefont {Dieterich}, \citenamefont {Egidi}, \citenamefont {Ellis}, \citenamefont {van Faassen}, \citenamefont {Fan}, \citenamefont {Fischer}, \citenamefont {F{\"{o}}rster}, \citenamefont {Fonseca~Guerra}, \citenamefont {Franchini}, \citenamefont {Ghysels}, \citenamefont {Giammona}, \citenamefont {van Gisbergen}, \citenamefont {Goez}, \citenamefont {G{\"{o}}tz}, \citenamefont {Groeneveld}, \citenamefont {Gritsenko}, \citenamefont {Gr{\"{u}}ning}, \citenamefont {Gusarov}, \citenamefont {Harris}, \citenamefont {van~den Hoek},
  \citenamefont {Hu}, \citenamefont {Jacob}, \citenamefont {Jacobsen}, \citenamefont {Jensen}, \citenamefont {Joubert}, \citenamefont {Kaminski}, \citenamefont {van Kessel}, \citenamefont {K{\"{o}}nig}, \citenamefont {Kootstra}, \citenamefont {Kovalenko}, \citenamefont {Krykunov}, \citenamefont {Lafiosca}, \citenamefont {Van~Lenthe}, \citenamefont {McCormack}, \citenamefont {Medves}, \citenamefont {Michalak}, \citenamefont {Mitoraj}, \citenamefont {Morton}, \citenamefont {Neugebauer}, \citenamefont {Nicu}, \citenamefont {Noodleman}, \citenamefont {Osinga}, \citenamefont {Patchkovskii}, \citenamefont {Pavanello}, \citenamefont {Peeples}, \citenamefont {Philipsen}, \citenamefont {Post}, \citenamefont {Pye}, \citenamefont {Ramanantoanina}, \citenamefont {Ramos}, \citenamefont {Ravenek}, \citenamefont {Reimann}, \citenamefont {Rodr{\'{i}}guez}, \citenamefont {Ros}, \citenamefont {R{\"{u}}ger}, \citenamefont {Schipper}, \citenamefont {Schl{\"{u}}ns}, \citenamefont {van Schoot}, \citenamefont {Schreckenbach},
  \citenamefont {Seldenthuis}, \citenamefont {Seth}, \citenamefont {Snijders}, \citenamefont {Sol{\`{a}}}, \citenamefont {Stener}, \citenamefont {Swart}, \citenamefont {Swerhone}, \citenamefont {Tognetti}, \citenamefont {te~Velde}, \citenamefont {Vernooijs}, \citenamefont {Versluis}, \citenamefont {Visscher}, \citenamefont {Visser}, \citenamefont {Wang}, \citenamefont {Wesolowski}, \citenamefont {van Wezenbeek}, \citenamefont {Wiesenekker}, \citenamefont {Wolff}, \citenamefont {Woo},\ and\ \citenamefont {Yakovlev}}]{ADF2023authors}%
  \BibitemOpen
  \bibfield  {author} {\bibinfo {author} {\bibfnamefont {E.~J.}\ \bibnamefont {Baerends}}, \bibinfo {author} {\bibfnamefont {T.}~\bibnamefont {Ziegler}}, \bibinfo {author} {\bibfnamefont {A.~J.}\ \bibnamefont {Atkins}}, \bibinfo {author} {\bibfnamefont {J.}~\bibnamefont {Autschbach}}, \bibinfo {author} {\bibfnamefont {D.}~\bibnamefont {Bashford}}, \bibinfo {author} {\bibfnamefont {O.}~\bibnamefont {Baseggio}}, \bibinfo {author} {\bibfnamefont {A.}~\bibnamefont {B{\'{e}}rces}}, \bibinfo {author} {\bibfnamefont {F.~M.}\ \bibnamefont {Bickelhaupt}}, \bibinfo {author} {\bibfnamefont {C.}~\bibnamefont {Bo}}, \bibinfo {author} {\bibfnamefont {P.~M.}\ \bibnamefont {Boerritger}}, \bibinfo {author} {\bibfnamefont {C.}~\bibnamefont {Cappelli}}, \bibinfo {author} {\bibfnamefont {L.}~\bibnamefont {Cavallo}}, \bibinfo {author} {\bibfnamefont {C.}~\bibnamefont {Daul}}, \bibinfo {author} {\bibfnamefont {D.~P.}\ \bibnamefont {Chong}}, \bibinfo {author} {\bibfnamefont {D.~V.}\ \bibnamefont {Chulhai}}, \bibinfo {author}
  {\bibfnamefont {L.}~\bibnamefont {Deng}}, \bibinfo {author} {\bibfnamefont {R.~M.}\ \bibnamefont {Dickson}}, \bibinfo {author} {\bibfnamefont {J.~M.}\ \bibnamefont {Dieterich}}, \bibinfo {author} {\bibfnamefont {F.}~\bibnamefont {Egidi}}, \bibinfo {author} {\bibfnamefont {D.~E.}\ \bibnamefont {Ellis}}, \bibinfo {author} {\bibfnamefont {M.}~\bibnamefont {van Faassen}}, \bibinfo {author} {\bibfnamefont {L.}~\bibnamefont {Fan}}, \bibinfo {author} {\bibfnamefont {T.~H.}\ \bibnamefont {Fischer}}, \bibinfo {author} {\bibfnamefont {A.}~\bibnamefont {F{\"{o}}rster}}, \bibinfo {author} {\bibfnamefont {C.}~\bibnamefont {Fonseca~Guerra}}, \bibinfo {author} {\bibfnamefont {M.}~\bibnamefont {Franchini}}, \bibinfo {author} {\bibfnamefont {A.}~\bibnamefont {Ghysels}}, \bibinfo {author} {\bibfnamefont {A.}~\bibnamefont {Giammona}}, \bibinfo {author} {\bibfnamefont {S.~J.~A.}\ \bibnamefont {van Gisbergen}}, \bibinfo {author} {\bibfnamefont {A.}~\bibnamefont {Goez}}, \bibinfo {author} {\bibfnamefont {A.~W.}\ \bibnamefont
  {G{\"{o}}tz}}, \bibinfo {author} {\bibfnamefont {J.~A.}\ \bibnamefont {Groeneveld}}, \bibinfo {author} {\bibfnamefont {O.~V.}\ \bibnamefont {Gritsenko}}, \bibinfo {author} {\bibfnamefont {M.}~\bibnamefont {Gr{\"{u}}ning}}, \bibinfo {author} {\bibfnamefont {S.}~\bibnamefont {Gusarov}}, \bibinfo {author} {\bibfnamefont {F.~E.}\ \bibnamefont {Harris}}, \bibinfo {author} {\bibfnamefont {P.}~\bibnamefont {van~den Hoek}}, \bibinfo {author} {\bibfnamefont {Z.}~\bibnamefont {Hu}}, \bibinfo {author} {\bibfnamefont {C.~R.}\ \bibnamefont {Jacob}}, \bibinfo {author} {\bibfnamefont {H.}~\bibnamefont {Jacobsen}}, \bibinfo {author} {\bibfnamefont {L.}~\bibnamefont {Jensen}}, \bibinfo {author} {\bibfnamefont {L.}~\bibnamefont {Joubert}}, \bibinfo {author} {\bibfnamefont {J.~W.}\ \bibnamefont {Kaminski}}, \bibinfo {author} {\bibfnamefont {G.}~\bibnamefont {van Kessel}}, \bibinfo {author} {\bibfnamefont {C.}~\bibnamefont {K{\"{o}}nig}}, \bibinfo {author} {\bibfnamefont {F.}~\bibnamefont {Kootstra}}, \bibinfo {author}
  {\bibfnamefont {A.}~\bibnamefont {Kovalenko}}, \bibinfo {author} {\bibfnamefont {M.~V.}\ \bibnamefont {Krykunov}}, \bibinfo {author} {\bibfnamefont {P.}~\bibnamefont {Lafiosca}}, \bibinfo {author} {\bibfnamefont {E.}~\bibnamefont {Van~Lenthe}}, \bibinfo {author} {\bibfnamefont {D.~A.}\ \bibnamefont {McCormack}}, \bibinfo {author} {\bibfnamefont {M.}~\bibnamefont {Medves}}, \bibinfo {author} {\bibfnamefont {A.}~\bibnamefont {Michalak}}, \bibinfo {author} {\bibfnamefont {M.}~\bibnamefont {Mitoraj}}, \bibinfo {author} {\bibfnamefont {S.~M.}\ \bibnamefont {Morton}}, \bibinfo {author} {\bibfnamefont {J.}~\bibnamefont {Neugebauer}}, \bibinfo {author} {\bibfnamefont {V.~P.}\ \bibnamefont {Nicu}}, \bibinfo {author} {\bibfnamefont {L.}~\bibnamefont {Noodleman}}, \bibinfo {author} {\bibfnamefont {V.~P.}\ \bibnamefont {Osinga}}, \bibinfo {author} {\bibfnamefont {S.}~\bibnamefont {Patchkovskii}}, \bibinfo {author} {\bibfnamefont {M.}~\bibnamefont {Pavanello}}, \bibinfo {author} {\bibfnamefont {C.~A.}\ \bibnamefont
  {Peeples}}, \bibinfo {author} {\bibfnamefont {P.~H.~T.}\ \bibnamefont {Philipsen}}, \bibinfo {author} {\bibfnamefont {D.}~\bibnamefont {Post}}, \bibinfo {author} {\bibfnamefont {C.~C.}\ \bibnamefont {Pye}}, \bibinfo {author} {\bibfnamefont {H.}~\bibnamefont {Ramanantoanina}}, \bibinfo {author} {\bibfnamefont {P.}~\bibnamefont {Ramos}}, \bibinfo {author} {\bibfnamefont {W.}~\bibnamefont {Ravenek}}, \bibinfo {author} {\bibfnamefont {M.}~\bibnamefont {Reimann}}, \bibinfo {author} {\bibfnamefont {J.~I.}\ \bibnamefont {Rodr{\'{i}}guez}}, \bibinfo {author} {\bibfnamefont {P.}~\bibnamefont {Ros}}, \bibinfo {author} {\bibfnamefont {R.}~\bibnamefont {R{\"{u}}ger}}, \bibinfo {author} {\bibfnamefont {P.~R.~T.}\ \bibnamefont {Schipper}}, \bibinfo {author} {\bibfnamefont {D.}~\bibnamefont {Schl{\"{u}}ns}}, \bibinfo {author} {\bibfnamefont {H.}~\bibnamefont {van Schoot}}, \bibinfo {author} {\bibfnamefont {G.}~\bibnamefont {Schreckenbach}}, \bibinfo {author} {\bibfnamefont {J.~S.}\ \bibnamefont {Seldenthuis}}, \bibinfo
  {author} {\bibfnamefont {M.}~\bibnamefont {Seth}}, \bibinfo {author} {\bibfnamefont {J.~G.}\ \bibnamefont {Snijders}}, \bibinfo {author} {\bibfnamefont {M.}~\bibnamefont {Sol{\`{a}}}}, \bibinfo {author} {\bibfnamefont {M.}~\bibnamefont {Stener}}, \bibinfo {author} {\bibfnamefont {M.}~\bibnamefont {Swart}}, \bibinfo {author} {\bibfnamefont {D.}~\bibnamefont {Swerhone}}, \bibinfo {author} {\bibfnamefont {V.}~\bibnamefont {Tognetti}}, \bibinfo {author} {\bibfnamefont {G.}~\bibnamefont {te~Velde}}, \bibinfo {author} {\bibfnamefont {P.}~\bibnamefont {Vernooijs}}, \bibinfo {author} {\bibfnamefont {L.}~\bibnamefont {Versluis}}, \bibinfo {author} {\bibfnamefont {L.}~\bibnamefont {Visscher}}, \bibinfo {author} {\bibfnamefont {O.}~\bibnamefont {Visser}}, \bibinfo {author} {\bibfnamefont {F.}~\bibnamefont {Wang}}, \bibinfo {author} {\bibfnamefont {T.~A.}\ \bibnamefont {Wesolowski}}, \bibinfo {author} {\bibfnamefont {E.~M.}\ \bibnamefont {van Wezenbeek}}, \bibinfo {author} {\bibfnamefont {G.}~\bibnamefont
  {Wiesenekker}}, \bibinfo {author} {\bibfnamefont {S.~K.}\ \bibnamefont {Wolff}}, \bibinfo {author} {\bibfnamefont {T.~K.}\ \bibnamefont {Woo}}, \ and\ \bibinfo {author} {\bibfnamefont {A.~L.}\ \bibnamefont {Yakovlev}},\ }\href@noop {} {\enquote {\bibinfo {title} {{ADF2023.1, SCM, Theoretical Chemistry, Vrije Universiteit, Amsterdam, The Netherlands, https://www.scm.com}},}\ } (\bibinfo {year} {2023})\BibitemShut {NoStop}%
\bibitem [{\citenamefont {te~Velde}\ \emph {et~al.}(2001)\citenamefont {te~Velde}, \citenamefont {Bickelhaupt}, \citenamefont {Baerends}, \citenamefont {Fonseca~Guerra}, \citenamefont {van Gisbergen}, \citenamefont {Snijders},\ and\ \citenamefont {Ziegler}}]{Velde2001_ADF}%
  \BibitemOpen
  \bibfield  {author} {\bibinfo {author} {\bibfnamefont {G.}~\bibnamefont {te~Velde}}, \bibinfo {author} {\bibfnamefont {F.~M.}\ \bibnamefont {Bickelhaupt}}, \bibinfo {author} {\bibfnamefont {E.~J.}\ \bibnamefont {Baerends}}, \bibinfo {author} {\bibfnamefont {C.}~\bibnamefont {Fonseca~Guerra}}, \bibinfo {author} {\bibfnamefont {S.~J.~A.}\ \bibnamefont {van Gisbergen}}, \bibinfo {author} {\bibfnamefont {J.~G.}\ \bibnamefont {Snijders}}, \ and\ \bibinfo {author} {\bibfnamefont {T.}~\bibnamefont {Ziegler}},\ }\bibfield  {title} {\enquote {\bibinfo {title} {{Chemistry with ADF}},}\ }\href {\doibase https://doi.org/10.1002/jcc.1056} {\bibfield  {journal} {\bibinfo  {journal} {J. Comput. Chem.}\ }\textbf {\bibinfo {volume} {22}},\ \bibinfo {pages} {931--967} (\bibinfo {year} {2001})}\BibitemShut {NoStop}%
\bibitem [{\citenamefont {Fonseca~Guerra}\ \emph {et~al.}(1998)\citenamefont {Fonseca~Guerra}, \citenamefont {Snijders}, \citenamefont {te~Velde},\ and\ \citenamefont {Baerends}}]{Guerra1998-ADF}%
  \BibitemOpen
  \bibfield  {author} {\bibinfo {author} {\bibfnamefont {C.}~\bibnamefont {Fonseca~Guerra}}, \bibinfo {author} {\bibfnamefont {J.~G.}\ \bibnamefont {Snijders}}, \bibinfo {author} {\bibfnamefont {G.}~\bibnamefont {te~Velde}}, \ and\ \bibinfo {author} {\bibfnamefont {E.~J.}\ \bibnamefont {Baerends}},\ }\bibfield  {title} {\enquote {\bibinfo {title} {{Towards an order-N DFT method}},}\ }\href {\doibase https://doi.org/10.1007/s002140050353} {\bibfield  {journal} {\bibinfo  {journal} {Theor. Chem. Acc.}\ }\textbf {\bibinfo {volume} {99}},\ \bibinfo {pages} {391–403} (\bibinfo {year} {1998})}\BibitemShut {NoStop}%
\bibitem [{\citenamefont {Jakubowska}, \citenamefont {Pecul},\ and\ \citenamefont {Ruud}(2022)}]{Jakubowska2022_vib-avg_relativity}%
  \BibitemOpen
  \bibfield  {author} {\bibinfo {author} {\bibfnamefont {K.}~\bibnamefont {Jakubowska}}, \bibinfo {author} {\bibfnamefont {M.}~\bibnamefont {Pecul}}, \ and\ \bibinfo {author} {\bibfnamefont {K.}~\bibnamefont {Ruud}},\ }\bibfield  {title} {\enquote {\bibinfo {title} {{Vibrational Corrections to NMR Spin–Spin Coupling Constants from Relativistic Four-Component DFT Calculations}},}\ }\href {\doibase 10.1021/acs.jpca.2c05019} {\bibfield  {journal} {\bibinfo  {journal} {J. Phys. Chem. A}\ }\textbf {\bibinfo {volume} {126}},\ \bibinfo {pages} {7013--7020} (\bibinfo {year} {2022})}\BibitemShut {NoStop}%
\bibitem [{\citenamefont {Faber}, \citenamefont {Kaminsky},\ and\ \citenamefont {Sauer}(2016)}]{Faber2016_vib-avg_book}%
  \BibitemOpen
  \bibfield  {author} {\bibinfo {author} {\bibfnamefont {R.}~\bibnamefont {Faber}}, \bibinfo {author} {\bibfnamefont {J.}~\bibnamefont {Kaminsky}}, \ and\ \bibinfo {author} {\bibfnamefont {S.~P.~A.}\ \bibnamefont {Sauer}},\ }\bibfield  {title} {\enquote {\bibinfo {title} {{Rovibrational and Temperature Effects in Theoretical Studies of NMR Parameters}},}\ }in\ \href {\doibase 10.1039/9781782623816-00218} {\emph {\bibinfo {booktitle} {{Gas Phase NMR}}}},\ \bibinfo {editor} {edited by\ \bibinfo {editor} {\bibfnamefont {K.}~\bibnamefont {Jackowski}}\ and\ \bibinfo {editor} {\bibfnamefont {M.}~\bibnamefont {Jaszunski}}}\ (\bibinfo  {publisher} {The Royal Society of Chemistry},\ \bibinfo {year} {2016})\ Chap.~\bibinfo {chapter} {7}, pp.\ \bibinfo {pages} {218--266}\BibitemShut {NoStop}%
\bibitem [{\citenamefont {Ruud}, \citenamefont {{\AA}strand},\ and\ \citenamefont {Taylor}(2000)}]{pro000208rat}%
  \BibitemOpen
  \bibfield  {author} {\bibinfo {author} {\bibfnamefont {K.}~\bibnamefont {Ruud}}, \bibinfo {author} {\bibfnamefont {P.-O.}\ \bibnamefont {{\AA}strand}}, \ and\ \bibinfo {author} {\bibfnamefont {P.~R.}\ \bibnamefont {Taylor}},\ }\bibfield  {title} {\enquote {\bibinfo {title} {{An efficient approach for calculating vibrational wave functions and zero-point vibrational corrections to molecular properties of polyatomic molecules}},}\ }\href@noop {} {\bibfield  {journal} {\bibinfo  {journal} {J. Chem. Phys.}\ }\textbf {\bibinfo {volume} {112}},\ \bibinfo {pages} {2668--2683} (\bibinfo {year} {2000})}\BibitemShut {NoStop}%
\bibitem [{\citenamefont {Schneider}\ and\ \citenamefont {Thiel}(1989)}]{schneider1989anharmonic}%
  \BibitemOpen
  \bibfield  {author} {\bibinfo {author} {\bibfnamefont {W.}~\bibnamefont {Schneider}}\ and\ \bibinfo {author} {\bibfnamefont {W.}~\bibnamefont {Thiel}},\ }\bibfield  {title} {\enquote {\bibinfo {title} {Anharmonic force fields from analytic second derivatives: Method and application to methyl bromide},}\ }\href@noop {} {\bibfield  {journal} {\bibinfo  {journal} {Chem. Phys. Lett.}\ }\textbf {\bibinfo {volume} {157}},\ \bibinfo {pages} {367--373} (\bibinfo {year} {1989})}\BibitemShut {NoStop}%
\bibitem [{\citenamefont {Barone}(2005)}]{barone2005anharmonic}%
  \BibitemOpen
  \bibfield  {author} {\bibinfo {author} {\bibfnamefont {V.}~\bibnamefont {Barone}},\ }\bibfield  {title} {\enquote {\bibinfo {title} {{Anharmonic vibrational properties by a fully automated second-order perturbative approach}},}\ }\href@noop {} {\bibfield  {journal} {\bibinfo  {journal} {J. Phys. Chem.}\ }\textbf {\bibinfo {volume} {122}},\ \bibinfo {pages} {14108} (\bibinfo {year} {2005})}\BibitemShut {NoStop}%
\bibitem [{\citenamefont {Aidas}\ \emph {et~al.}(2014)\citenamefont {Aidas}, \citenamefont {Angeli}, \citenamefont {Bak}, \citenamefont {Bakken}, \citenamefont {Bast}, \citenamefont {Boman}, \citenamefont {Christiansen}, \citenamefont {Cimiraglia}, \citenamefont {Coriani}, \citenamefont {Dahle}, \citenamefont {Dalskov}, \citenamefont {Ekstr{\"{o}}m}, \citenamefont {Enevoldsen}, \citenamefont {Eriksen}, \citenamefont {Ettenhuber}, \citenamefont {Fern{\'{a}}ndez}, \citenamefont {Ferrighi}, \citenamefont {Fliegl}, \citenamefont {Frediani}, \citenamefont {Hald}, \citenamefont {Halkier}, \citenamefont {H{\"{a}}ttig}, \citenamefont {Heiberg}, \citenamefont {Helgaker}, \citenamefont {Hennum}, \citenamefont {Hettema}, \citenamefont {Hjerten{\ae}s}, \citenamefont {H{\o}st}, \citenamefont {H{\o}yvik}, \citenamefont {Iozzi}, \citenamefont {Jans{\'{i}}k}, \citenamefont {Jensen}, \citenamefont {Jonsson}, \citenamefont {J{\o}rgensen}, \citenamefont {Kauczor}, \citenamefont {Kirpekar}, \citenamefont {Kj{\ae}rgaard},
  \citenamefont {Klopper}, \citenamefont {Knecht}, \citenamefont {Kobayashi}, \citenamefont {Koch}, \citenamefont {Kongsted}, \citenamefont {Krapp}, \citenamefont {Kristensen}, \citenamefont {Ligabue}, \citenamefont {Lutn{\ae}s}, \citenamefont {Melo}, \citenamefont {Mikkelsen}, \citenamefont {Myhre}, \citenamefont {Neiss}, \citenamefont {Nielsen}, \citenamefont {Norman}, \citenamefont {Olsen}, \citenamefont {Olsen}, \citenamefont {Osted}, \citenamefont {Packer}, \citenamefont {Pawlowski}, \citenamefont {Pedersen}, \citenamefont {Provasi}, \citenamefont {Reine}, \citenamefont {Rinkevicius}, \citenamefont {Ruden}, \citenamefont {Ruud}, \citenamefont {Rybkin}, \citenamefont {Sa{\l}ek}, \citenamefont {Samson}, \citenamefont {de~Mer{\'{a}}s}, \citenamefont {Saue}, \citenamefont {Sauer}, \citenamefont {Schimmelpfennig}, \citenamefont {Sneskov}, \citenamefont {Steindal}, \citenamefont {Sylvester-Hvid}, \citenamefont {Taylor}, \citenamefont {Teale}, \citenamefont {Tellgren}, \citenamefont {Tew}, \citenamefont
  {Thorvaldsen}, \citenamefont {Th{\o}gersen}, \citenamefont {Vahtras}, \citenamefont {Watson}, \citenamefont {Wilson}, \citenamefont {Ziolkowski},\ and\ \citenamefont {{\AA}gren}}]{Dalton2014}%
  \BibitemOpen
  \bibfield  {author} {\bibinfo {author} {\bibfnamefont {K.}~\bibnamefont {Aidas}}, \bibinfo {author} {\bibfnamefont {C.}~\bibnamefont {Angeli}}, \bibinfo {author} {\bibfnamefont {K.~L.}\ \bibnamefont {Bak}}, \bibinfo {author} {\bibfnamefont {V.}~\bibnamefont {Bakken}}, \bibinfo {author} {\bibfnamefont {R.}~\bibnamefont {Bast}}, \bibinfo {author} {\bibfnamefont {L.}~\bibnamefont {Boman}}, \bibinfo {author} {\bibfnamefont {O.}~\bibnamefont {Christiansen}}, \bibinfo {author} {\bibfnamefont {R.}~\bibnamefont {Cimiraglia}}, \bibinfo {author} {\bibfnamefont {S.}~\bibnamefont {Coriani}}, \bibinfo {author} {\bibfnamefont {P.}~\bibnamefont {Dahle}}, \bibinfo {author} {\bibfnamefont {E.~K.}\ \bibnamefont {Dalskov}}, \bibinfo {author} {\bibfnamefont {U.}~\bibnamefont {Ekstr{\"{o}}m}}, \bibinfo {author} {\bibfnamefont {T.}~\bibnamefont {Enevoldsen}}, \bibinfo {author} {\bibfnamefont {J.~J.}\ \bibnamefont {Eriksen}}, \bibinfo {author} {\bibfnamefont {P.}~\bibnamefont {Ettenhuber}}, \bibinfo {author} {\bibfnamefont
  {B.}~\bibnamefont {Fern{\'{a}}ndez}}, \bibinfo {author} {\bibfnamefont {L.}~\bibnamefont {Ferrighi}}, \bibinfo {author} {\bibfnamefont {H.}~\bibnamefont {Fliegl}}, \bibinfo {author} {\bibfnamefont {L.}~\bibnamefont {Frediani}}, \bibinfo {author} {\bibfnamefont {K.}~\bibnamefont {Hald}}, \bibinfo {author} {\bibfnamefont {A.}~\bibnamefont {Halkier}}, \bibinfo {author} {\bibfnamefont {C.}~\bibnamefont {H{\"{a}}ttig}}, \bibinfo {author} {\bibfnamefont {H.}~\bibnamefont {Heiberg}}, \bibinfo {author} {\bibfnamefont {T.}~\bibnamefont {Helgaker}}, \bibinfo {author} {\bibfnamefont {A.~C.}\ \bibnamefont {Hennum}}, \bibinfo {author} {\bibfnamefont {H.}~\bibnamefont {Hettema}}, \bibinfo {author} {\bibfnamefont {E.}~\bibnamefont {Hjerten{\ae}s}}, \bibinfo {author} {\bibfnamefont {S.}~\bibnamefont {H{\o}st}}, \bibinfo {author} {\bibfnamefont {I.~M.}\ \bibnamefont {H{\o}yvik}}, \bibinfo {author} {\bibfnamefont {M.~F.}\ \bibnamefont {Iozzi}}, \bibinfo {author} {\bibfnamefont {B.}~\bibnamefont {Jans{\'{i}}k}}, \bibinfo
  {author} {\bibfnamefont {H.~J.~A.}\ \bibnamefont {Jensen}}, \bibinfo {author} {\bibfnamefont {D.}~\bibnamefont {Jonsson}}, \bibinfo {author} {\bibfnamefont {P.}~\bibnamefont {J{\o}rgensen}}, \bibinfo {author} {\bibfnamefont {J.}~\bibnamefont {Kauczor}}, \bibinfo {author} {\bibfnamefont {S.}~\bibnamefont {Kirpekar}}, \bibinfo {author} {\bibfnamefont {T.}~\bibnamefont {Kj{\ae}rgaard}}, \bibinfo {author} {\bibfnamefont {W.}~\bibnamefont {Klopper}}, \bibinfo {author} {\bibfnamefont {S.}~\bibnamefont {Knecht}}, \bibinfo {author} {\bibfnamefont {R.}~\bibnamefont {Kobayashi}}, \bibinfo {author} {\bibfnamefont {H.}~\bibnamefont {Koch}}, \bibinfo {author} {\bibfnamefont {J.}~\bibnamefont {Kongsted}}, \bibinfo {author} {\bibfnamefont {A.}~\bibnamefont {Krapp}}, \bibinfo {author} {\bibfnamefont {K.}~\bibnamefont {Kristensen}}, \bibinfo {author} {\bibfnamefont {A.}~\bibnamefont {Ligabue}}, \bibinfo {author} {\bibfnamefont {O.~B.}\ \bibnamefont {Lutn{\ae}s}}, \bibinfo {author} {\bibfnamefont {J.~I.}\ \bibnamefont
  {Melo}}, \bibinfo {author} {\bibfnamefont {K.~V.}\ \bibnamefont {Mikkelsen}}, \bibinfo {author} {\bibfnamefont {R.~H.}\ \bibnamefont {Myhre}}, \bibinfo {author} {\bibfnamefont {C.}~\bibnamefont {Neiss}}, \bibinfo {author} {\bibfnamefont {C.~B.}\ \bibnamefont {Nielsen}}, \bibinfo {author} {\bibfnamefont {P.}~\bibnamefont {Norman}}, \bibinfo {author} {\bibfnamefont {J.}~\bibnamefont {Olsen}}, \bibinfo {author} {\bibfnamefont {J.~M.~H.}\ \bibnamefont {Olsen}}, \bibinfo {author} {\bibfnamefont {A.}~\bibnamefont {Osted}}, \bibinfo {author} {\bibfnamefont {M.~J.}\ \bibnamefont {Packer}}, \bibinfo {author} {\bibfnamefont {F.}~\bibnamefont {Pawlowski}}, \bibinfo {author} {\bibfnamefont {T.~B.}\ \bibnamefont {Pedersen}}, \bibinfo {author} {\bibfnamefont {P.~F.}\ \bibnamefont {Provasi}}, \bibinfo {author} {\bibfnamefont {S.}~\bibnamefont {Reine}}, \bibinfo {author} {\bibfnamefont {Z.}~\bibnamefont {Rinkevicius}}, \bibinfo {author} {\bibfnamefont {T.~A.}\ \bibnamefont {Ruden}}, \bibinfo {author} {\bibfnamefont
  {K.}~\bibnamefont {Ruud}}, \bibinfo {author} {\bibfnamefont {V.~V.}\ \bibnamefont {Rybkin}}, \bibinfo {author} {\bibfnamefont {P.}~\bibnamefont {Sa{\l}ek}}, \bibinfo {author} {\bibfnamefont {C.~C.~M.}\ \bibnamefont {Samson}}, \bibinfo {author} {\bibfnamefont {A.~S.}\ \bibnamefont {de~Mer{\'{a}}s}}, \bibinfo {author} {\bibfnamefont {T.}~\bibnamefont {Saue}}, \bibinfo {author} {\bibfnamefont {S.~P.~A.}\ \bibnamefont {Sauer}}, \bibinfo {author} {\bibfnamefont {B.}~\bibnamefont {Schimmelpfennig}}, \bibinfo {author} {\bibfnamefont {K.}~\bibnamefont {Sneskov}}, \bibinfo {author} {\bibfnamefont {A.~H.}\ \bibnamefont {Steindal}}, \bibinfo {author} {\bibfnamefont {K.~O.}\ \bibnamefont {Sylvester-Hvid}}, \bibinfo {author} {\bibfnamefont {P.~R.}\ \bibnamefont {Taylor}}, \bibinfo {author} {\bibfnamefont {A.~M.}\ \bibnamefont {Teale}}, \bibinfo {author} {\bibfnamefont {E.~I.}\ \bibnamefont {Tellgren}}, \bibinfo {author} {\bibfnamefont {D.~P.}\ \bibnamefont {Tew}}, \bibinfo {author} {\bibfnamefont {A.~J.}\ \bibnamefont
  {Thorvaldsen}}, \bibinfo {author} {\bibfnamefont {L.}~\bibnamefont {Th{\o}gersen}}, \bibinfo {author} {\bibfnamefont {O.}~\bibnamefont {Vahtras}}, \bibinfo {author} {\bibfnamefont {M.~A.}\ \bibnamefont {Watson}}, \bibinfo {author} {\bibfnamefont {D.~J.~D.}\ \bibnamefont {Wilson}}, \bibinfo {author} {\bibfnamefont {M.}~\bibnamefont {Ziolkowski}}, \ and\ \bibinfo {author} {\bibfnamefont {H.}~\bibnamefont {{\AA}gren}},\ }\bibfield  {title} {\enquote {\bibinfo {title} {{The Dalton quantum chemistry program system}},}\ }\href {\doibase 10.1002/WCMS.1172} {\bibfield  {journal} {\bibinfo  {journal} {Wiley Interdiscip. Rev. Comput. Mol. Sci.}\ }\textbf {\bibinfo {volume} {4}},\ \bibinfo {pages} {269--284} (\bibinfo {year} {2014})}\BibitemShut {NoStop}%
\bibitem [{\citenamefont {Becke}(1993)}]{Becke_BHandH_BHandHLYP}%
  \BibitemOpen
  \bibfield  {author} {\bibinfo {author} {\bibfnamefont {A.~D.}\ \bibnamefont {Becke}},\ }\bibfield  {title} {\enquote {\bibinfo {title} {A new mixing of hartree–fock and local density‐functional theories},}\ }\href {\doibase 10.1063/1.464304} {\bibfield  {journal} {\bibinfo  {journal} {J. Chem. Phys.}\ }\textbf {\bibinfo {volume} {98}},\ \bibinfo {pages} {1372--1377} (\bibinfo {year} {1993})}\BibitemShut {NoStop}%
\bibitem [{\citenamefont {Bryce}\ and\ \citenamefont {Autschbach}(2009)}]{Bryce2009_basis-set_TZ2P-J-QZ4P-J}%
  \BibitemOpen
  \bibfield  {author} {\bibinfo {author} {\bibfnamefont {D.~L.}\ \bibnamefont {Bryce}}\ and\ \bibinfo {author} {\bibfnamefont {J.}~\bibnamefont {Autschbach}},\ }\bibfield  {title} {\enquote {\bibinfo {title} {{Relativistic hybrid density functional calculations of indirect nuclear spin–spin coupling tensors — Comparison with experiment for diatomic alkali metal halides}},}\ }\href {\doibase 10.1139/V09-040} {\bibfield  {journal} {\bibinfo  {journal} {Can. J. Chem.}\ }\textbf {\bibinfo {volume} {87}},\ \bibinfo {pages} {927--941} (\bibinfo {year} {2009})}\BibitemShut {NoStop}%
\bibitem [{\citenamefont {Ernzerhof}\ and\ \citenamefont {Scuseria}(1999)}]{Ernzerhof1999_PBE0}%
  \BibitemOpen
  \bibfield  {author} {\bibinfo {author} {\bibfnamefont {M.}~\bibnamefont {Ernzerhof}}\ and\ \bibinfo {author} {\bibfnamefont {G.~E.}\ \bibnamefont {Scuseria}},\ }\bibfield  {title} {\enquote {\bibinfo {title} {{Assessment of the Perdew–Burke–Ernzerhof exchange-correlation functional}},}\ }\href {\doibase 10.1063/1.478401} {\bibfield  {journal} {\bibinfo  {journal} {J. Chem. Phys.}\ }\textbf {\bibinfo {volume} {110}},\ \bibinfo {pages} {5029--5036} (\bibinfo {year} {1999})}\BibitemShut {NoStop}%
\bibitem [{\citenamefont {Faber}\ and\ \citenamefont {Sauer}(2012)}]{Faber2012_vib_avg_SSCC}%
  \BibitemOpen
  \bibfield  {author} {\bibinfo {author} {\bibfnamefont {R.}~\bibnamefont {Faber}}\ and\ \bibinfo {author} {\bibfnamefont {S.~P.~A.}\ \bibnamefont {Sauer}},\ }\bibfield  {title} {\enquote {\bibinfo {title} {{On the discrepancy between theory and experiment for the F–F spin–spin coupling constant of difluoroethyne}},}\ }\href {\doibase 10.1039/C2CP42198D} {\bibfield  {journal} {\bibinfo  {journal} {Phys. Chem. Chem. Phys.}\ }\textbf {\bibinfo {volume} {14}},\ \bibinfo {pages} {16440--16447} (\bibinfo {year} {2012})}\BibitemShut {NoStop}%
\bibitem [{\citenamefont {Faber}\ and\ \citenamefont {Sauer}(2015)}]{spas154}%
  \BibitemOpen
  \bibfield  {author} {\bibinfo {author} {\bibfnamefont {R.}~\bibnamefont {Faber}}\ and\ \bibinfo {author} {\bibfnamefont {S.~P.~A.}\ \bibnamefont {Sauer}},\ }\bibfield  {title} {\enquote {\bibinfo {title} {{SOPPA and CCSD vibrational corrections to NMR indirect spin-spin coupling constants of small hydrocarbons}},}\ }in\ \href {\doibase 10.1063/1.4938843} {\emph {\bibinfo {booktitle} {AIP Conf. Proc.}}},\ Vol.\ \bibinfo {volume} {1702}\ (\bibinfo {year} {2015})\ p.\ \bibinfo {pages} {090035}\BibitemShut {NoStop}%
\bibitem [{\citenamefont {Faber}\ \emph {et~al.}(2017)\citenamefont {Faber}, \citenamefont {Buczek}, \citenamefont {Kupka},\ and\ \citenamefont {Sauer}}]{spas160}%
  \BibitemOpen
  \bibfield  {author} {\bibinfo {author} {\bibfnamefont {R.}~\bibnamefont {Faber}}, \bibinfo {author} {\bibfnamefont {A.}~\bibnamefont {Buczek}}, \bibinfo {author} {\bibfnamefont {T.}~\bibnamefont {Kupka}}, \ and\ \bibinfo {author} {\bibfnamefont {S.~P.~A.}\ \bibnamefont {Sauer}},\ }\bibfield  {title} {\enquote {\bibinfo {title} {{On the convergence of zero-point vibrational corrections to nuclear shieldings and shielding anisotropies towards the complete basis set limit in water}},}\ }\href {\doibase 10.1080/00268976.2016.1210831} {\bibfield  {journal} {\bibinfo  {journal} {Mol. Phys.}\ }\textbf {\bibinfo {volume} {115}},\ \bibinfo {pages} {144--160} (\bibinfo {year} {2017})}\BibitemShut {NoStop}%
\bibitem [{\citenamefont {Sauer}, \citenamefont {{\v{S}}pirko},\ and\ \citenamefont {Oddershede}(1991)}]{spas002}%
  \BibitemOpen
  \bibfield  {author} {\bibinfo {author} {\bibfnamefont {S.~P.~A.}\ \bibnamefont {Sauer}}, \bibinfo {author} {\bibfnamefont {V.}~\bibnamefont {{\v{S}}pirko}}, \ and\ \bibinfo {author} {\bibfnamefont {J.}~\bibnamefont {Oddershede}},\ }\bibfield  {title} {\enquote {\bibinfo {title} {{The magnetizability and g-factor surfaces of ammonia}},}\ }\href@noop {} {\bibfield  {journal} {\bibinfo  {journal} {Chem. Phys.}\ }\textbf {\bibinfo {volume} {153}},\ \bibinfo {pages} {189--200} (\bibinfo {year} {1991})}\BibitemShut {NoStop}%
\bibitem [{\citenamefont {Sauer}\ \emph {et~al.}(1994)\citenamefont {Sauer}, \citenamefont {{\v{S}}pirko}, \citenamefont {Paidarov{\'{a}}},\ and\ \citenamefont {Oddershede}}]{spas011}%
  \BibitemOpen
  \bibfield  {author} {\bibinfo {author} {\bibfnamefont {S.~P.~A.}\ \bibnamefont {Sauer}}, \bibinfo {author} {\bibfnamefont {V.}~\bibnamefont {{\v{S}}pirko}}, \bibinfo {author} {\bibfnamefont {I.}~\bibnamefont {Paidarov{\'{a}}}}, \ and\ \bibinfo {author} {\bibfnamefont {J.}~\bibnamefont {Oddershede}},\ }\bibfield  {title} {\enquote {\bibinfo {title} {{The vibrational and temperature dependence of the magnetic properties of the oxonium ion (H3O$^+$)}},}\ }\href {\doibase 10.1016/0301-0104(94)00080-8} {\bibfield  {journal} {\bibinfo  {journal} {Chem. Phys.}\ }\textbf {\bibinfo {volume} {184}},\ \bibinfo {pages} {1--11} (\bibinfo {year} {1994})}\BibitemShut {NoStop}%
\bibitem [{\citenamefont {Sauer}\ and\ \citenamefont {Paidarov{\'{a}}}(1995)}]{spas018}%
  \BibitemOpen
  \bibfield  {author} {\bibinfo {author} {\bibfnamefont {S.~P.~A.}\ \bibnamefont {Sauer}}\ and\ \bibinfo {author} {\bibfnamefont {I.}~\bibnamefont {Paidarov{\'{a}}}},\ }\bibfield  {title} {\enquote {\bibinfo {title} {{Calculations of magnetic hyperfine structure constants for the low-lying rovibrational levels of LiH, HF, CH$^+$, and BH}},}\ }\href@noop {} {\bibfield  {journal} {\bibinfo  {journal} {Chem. Phys.}\ }\textbf {\bibinfo {volume} {201}},\ \bibinfo {pages} {405--425} (\bibinfo {year} {1995})}\BibitemShut {NoStop}%
\bibitem [{\citenamefont {Sauer}\ \emph {et~al.}(1997)\citenamefont {Sauer}, \citenamefont {{\v{S}}pirko}, \citenamefont {Paidarov{\'{a}}},\ and\ \citenamefont {Kraemer}}]{spas020}%
  \BibitemOpen
  \bibfield  {author} {\bibinfo {author} {\bibfnamefont {S.~P.~A.}\ \bibnamefont {Sauer}}, \bibinfo {author} {\bibfnamefont {V.}~\bibnamefont {{\v{S}}pirko}}, \bibinfo {author} {\bibfnamefont {I.}~\bibnamefont {Paidarov{\'{a}}}}, \ and\ \bibinfo {author} {\bibfnamefont {W.~P.}\ \bibnamefont {Kraemer}},\ }\bibfield  {title} {\enquote {\bibinfo {title} {{The vibrational dependence of the hydrogen and oxygen nuclear magnetic shielding constants in OH$^-$ and OH$^-$ {\textperiodcentered} H2O}},}\ }\href {\doibase 10.1016/S0301-0104(96)00308-4} {\bibfield  {journal} {\bibinfo  {journal} {Chem. Phys.}\ }\textbf {\bibinfo {volume} {214}},\ \bibinfo {pages} {91--101} (\bibinfo {year} {1997})}\BibitemShut {NoStop}%
\bibitem [{\citenamefont {Wigglesworth}\ \emph {et~al.}(1997)\citenamefont {Wigglesworth}, \citenamefont {Raynes}, \citenamefont {Sauer},\ and\ \citenamefont {Oddershede}}]{spas022}%
  \BibitemOpen
  \bibfield  {author} {\bibinfo {author} {\bibfnamefont {R.~D.}\ \bibnamefont {Wigglesworth}}, \bibinfo {author} {\bibfnamefont {W.~T.}\ \bibnamefont {Raynes}}, \bibinfo {author} {\bibfnamefont {S.~P.~A.}\ \bibnamefont {Sauer}}, \ and\ \bibinfo {author} {\bibfnamefont {J.}~\bibnamefont {Oddershede}},\ }\bibfield  {title} {\enquote {\bibinfo {title} {{The calculation and analysis of isotope effects on the nuclear spin-spin coupling constants of methane at various temperatures}},}\ }\href {\doibase 10.1080/002689797170635} {\bibfield  {journal} {\bibinfo  {journal} {Mol. Phys.}\ }\textbf {\bibinfo {volume} {92}},\ \bibinfo {pages} {77--88} (\bibinfo {year} {1997})}\BibitemShut {NoStop}%
\bibitem [{\citenamefont {Wigglesworth}\ \emph {et~al.}(1998)\citenamefont {Wigglesworth}, \citenamefont {Raynes}, \citenamefont {Sauer},\ and\ \citenamefont {Oddershede}}]{spas026}%
  \BibitemOpen
  \bibfield  {author} {\bibinfo {author} {\bibfnamefont {R.~D.}\ \bibnamefont {Wigglesworth}}, \bibinfo {author} {\bibfnamefont {W.~T.}\ \bibnamefont {Raynes}}, \bibinfo {author} {\bibfnamefont {S.~P.~A.}\ \bibnamefont {Sauer}}, \ and\ \bibinfo {author} {\bibfnamefont {J.}~\bibnamefont {Oddershede}},\ }\bibfield  {title} {\enquote {\bibinfo {title} {{Calculated spin-spin coupling surfaces in the water molecule; prediction and analysis of J(O, H), J(O, D) and J(H, D) in water isotopomers}},}\ }\href {\doibase 10.1080/00268979809482379} {\bibfield  {journal} {\bibinfo  {journal} {Mol. Phys.}\ }\textbf {\bibinfo {volume} {94}},\ \bibinfo {pages} {851--862} (\bibinfo {year} {1998})}\BibitemShut {NoStop}%
\bibitem [{\citenamefont {Sauer}\ \emph {et~al.}(1998)\citenamefont {Sauer}, \citenamefont {M{\o}ller}, \citenamefont {Koch}, \citenamefont {Paidarov{\'{a}}},\ and\ \citenamefont {{\v{S}}pirko}}]{spas028}%
  \BibitemOpen
  \bibfield  {author} {\bibinfo {author} {\bibfnamefont {S.~P.~A.}\ \bibnamefont {Sauer}}, \bibinfo {author} {\bibfnamefont {C.~K.}\ \bibnamefont {M{\o}ller}}, \bibinfo {author} {\bibfnamefont {H.}~\bibnamefont {Koch}}, \bibinfo {author} {\bibfnamefont {I.}~\bibnamefont {Paidarov{\'{a}}}}, \ and\ \bibinfo {author} {\bibfnamefont {V.}~\bibnamefont {{\v{S}}pirko}},\ }\bibfield  {title} {\enquote {\bibinfo {title} {{The vibrational and temperature dependence of the indirect nuclear spin–spin coupling constants of the oxonium (H3O$^+$) and hydroxyl (OH$^-$) ions}},}\ }\href {\doibase 10.1016/S0301-0104(98)00329-2} {\bibfield  {journal} {\bibinfo  {journal} {Chem. Phys.}\ }\textbf {\bibinfo {volume} {238}},\ \bibinfo {pages} {385--399} (\bibinfo {year} {1998})}\BibitemShut {NoStop}%
\bibitem [{\citenamefont {Wigglesworth}\ \emph {et~al.}(1999)\citenamefont {Wigglesworth}, \citenamefont {Raynes}, \citenamefont {Sauer},\ and\ \citenamefont {Oddershede}}]{spas032}%
  \BibitemOpen
  \bibfield  {author} {\bibinfo {author} {\bibfnamefont {R.~D.}\ \bibnamefont {Wigglesworth}}, \bibinfo {author} {\bibfnamefont {W.~T.}\ \bibnamefont {Raynes}}, \bibinfo {author} {\bibfnamefont {S.~P.~A.}\ \bibnamefont {Sauer}}, \ and\ \bibinfo {author} {\bibfnamefont {J.}~\bibnamefont {Oddershede}},\ }\bibfield  {title} {\enquote {\bibinfo {title} {{Calculated nuclear shielding surfaces in the water molecule; prediction and analysis of sigma(O), sigma(H) and sigma(D) in water isotopomers}},}\ }\href {\doibase 10.1080/002689799164289} {\bibfield  {journal} {\bibinfo  {journal} {Mol. Phys.}\ }\textbf {\bibinfo {volume} {96}},\ \bibinfo {pages} {1595--1607} (\bibinfo {year} {1999})}\BibitemShut {NoStop}%
\bibitem [{\citenamefont {Wigglesworth}\ \emph {et~al.}(2000{\natexlab{a}})\citenamefont {Wigglesworth}, \citenamefont {Raynes}, \citenamefont {Kirpekar}, \citenamefont {Oddershede},\ and\ \citenamefont {Sauer}}]{spas034}%
  \BibitemOpen
  \bibfield  {author} {\bibinfo {author} {\bibfnamefont {R.~D.}\ \bibnamefont {Wigglesworth}}, \bibinfo {author} {\bibfnamefont {W.~T.}\ \bibnamefont {Raynes}}, \bibinfo {author} {\bibfnamefont {S.}~\bibnamefont {Kirpekar}}, \bibinfo {author} {\bibfnamefont {J.}~\bibnamefont {Oddershede}}, \ and\ \bibinfo {author} {\bibfnamefont {S.~P.~A.}\ \bibnamefont {Sauer}},\ }\bibfield  {title} {\enquote {\bibinfo {title} {{Nuclear magnetic shielding in the acetylene isotopomers calculated from correlated shielding surfaces}},}\ }\href {\doibase 10.1063/1.480697} {\bibfield  {journal} {\bibinfo  {journal} {J. Chem. Phys.}\ }\textbf {\bibinfo {volume} {112}},\ \bibinfo {pages} {736--746} (\bibinfo {year} {2000}{\natexlab{a}})}\BibitemShut {NoStop}%
\bibitem [{\citenamefont {Wigglesworth}\ \emph {et~al.}(2000{\natexlab{b}})\citenamefont {Wigglesworth}, \citenamefont {Raynes}, \citenamefont {Kirpekar}, \citenamefont {Oddershede},\ and\ \citenamefont {Sauer}}]{spas036}%
  \BibitemOpen
  \bibfield  {author} {\bibinfo {author} {\bibfnamefont {R.~D.}\ \bibnamefont {Wigglesworth}}, \bibinfo {author} {\bibfnamefont {W.~T.}\ \bibnamefont {Raynes}}, \bibinfo {author} {\bibfnamefont {S.}~\bibnamefont {Kirpekar}}, \bibinfo {author} {\bibfnamefont {J.}~\bibnamefont {Oddershede}}, \ and\ \bibinfo {author} {\bibfnamefont {S.~P.~A.}\ \bibnamefont {Sauer}},\ }\bibfield  {title} {\enquote {\bibinfo {title} {{Nuclear spin–spin coupling in the acetylene isotopomers calculated from ab initio correlated surfaces for $^1$J(C,H), $^1$J(C,C), $^2$J(C,H), and $^3$J(H,H)}},}\ }\href {\doibase 10.1063/1.480525} {\bibfield  {journal} {\bibinfo  {journal} {J. Chem. Phys.}\ }\textbf {\bibinfo {volume} {112}},\ \bibinfo {pages} {3735--3746} (\bibinfo {year} {2000}{\natexlab{b}})}\BibitemShut {NoStop}%
\bibitem [{\citenamefont {Sauer}, \citenamefont {Raynes},\ and\ \citenamefont {Nicholls}(2001)}]{spas049}%
  \BibitemOpen
  \bibfield  {author} {\bibinfo {author} {\bibfnamefont {S.~P.~A.}\ \bibnamefont {Sauer}}, \bibinfo {author} {\bibfnamefont {W.~T.}\ \bibnamefont {Raynes}}, \ and\ \bibinfo {author} {\bibfnamefont {R.~A.}\ \bibnamefont {Nicholls}},\ }\bibfield  {title} {\enquote {\bibinfo {title} {{Nuclear spin-spin coupling in silane and its isotopomers: ab initio calculation and experimental investigation}},}\ }\href@noop {} {\bibfield  {journal} {\bibinfo  {journal} {J. Chem. Phys.}\ }\textbf {\bibinfo {volume} {115}},\ \bibinfo {pages} {5994--6006} (\bibinfo {year} {2001})}\BibitemShut {NoStop}%
\bibitem [{\citenamefont {Yachmenev}\ \emph {et~al.}(2010)\citenamefont {Yachmenev}, \citenamefont {Yurchenko}, \citenamefont {Paidarov{\'{a}}}, \citenamefont {Jensen}, \citenamefont {Thiel},\ and\ \citenamefont {Sauer}}]{spas093}%
  \BibitemOpen
  \bibfield  {author} {\bibinfo {author} {\bibfnamefont {A.}~\bibnamefont {Yachmenev}}, \bibinfo {author} {\bibfnamefont {S.~N.}\ \bibnamefont {Yurchenko}}, \bibinfo {author} {\bibfnamefont {I.}~\bibnamefont {Paidarov{\'{a}}}}, \bibinfo {author} {\bibfnamefont {P.}~\bibnamefont {Jensen}}, \bibinfo {author} {\bibfnamefont {W.}~\bibnamefont {Thiel}}, \ and\ \bibinfo {author} {\bibfnamefont {S.~P.~A.}\ \bibnamefont {Sauer}},\ }\bibfield  {title} {\enquote {\bibinfo {title} {{Thermal averaging of the indirect nuclear spin-spin coupling constants of ammonia: The importance of the large amplitude inversion mode}},}\ }\href@noop {} {\bibfield  {journal} {\bibinfo  {journal} {J. Chem. Phys.}\ }\textbf {\bibinfo {volume} {132}},\ \bibinfo {pages} {114305} (\bibinfo {year} {2010})}\BibitemShut {NoStop}%
\bibitem [{\citenamefont {Haas}\ \emph {et~al.}(2021)\citenamefont {Haas}, \citenamefont {R\"oder}, \citenamefont {Correia}, \citenamefont {Schell}, \citenamefont {Fenta}, \citenamefont {Vianden}, \citenamefont {Larsen}, \citenamefont {Aggelund}, \citenamefont {Fromsejer}, \citenamefont {Hemmingsen}, \citenamefont {Sauer}, \citenamefont {Lupascu},\ and\ \citenamefont {Amaral}}]{Haas2021_EFG_exp-vibavg_Q}%
  \BibitemOpen
  \bibfield  {author} {\bibinfo {author} {\bibfnamefont {H.}~\bibnamefont {Haas}}, \bibinfo {author} {\bibfnamefont {J.}~\bibnamefont {R\"oder}}, \bibinfo {author} {\bibfnamefont {J.~G.}\ \bibnamefont {Correia}}, \bibinfo {author} {\bibfnamefont {J.}~\bibnamefont {Schell}}, \bibinfo {author} {\bibfnamefont {A.~S.}\ \bibnamefont {Fenta}}, \bibinfo {author} {\bibfnamefont {R.}~\bibnamefont {Vianden}}, \bibinfo {author} {\bibfnamefont {E.~M.~H.}\ \bibnamefont {Larsen}}, \bibinfo {author} {\bibfnamefont {P.~A.}\ \bibnamefont {Aggelund}}, \bibinfo {author} {\bibfnamefont {R.}~\bibnamefont {Fromsejer}}, \bibinfo {author} {\bibfnamefont {L.~B.~S.}\ \bibnamefont {Hemmingsen}}, \bibinfo {author} {\bibfnamefont {S.~P.~A.}\ \bibnamefont {Sauer}}, \bibinfo {author} {\bibfnamefont {D.~C.}\ \bibnamefont {Lupascu}}, \ and\ \bibinfo {author} {\bibfnamefont {V.~S.}\ \bibnamefont {Amaral}},\ }\bibfield  {title} {\enquote {\bibinfo {title} {{Free Molecule Studies by Perturbed
  $\ensuremath{\gamma}\text{\ensuremath{-}}\ensuremath{\gamma}$ Angular Correlation: A New Path to Accurate Nuclear Quadrupole Moments}},}\ }\href {\doibase 10.1103/PhysRevLett.126.103001} {\bibfield  {journal} {\bibinfo  {journal} {Phys. Rev. Lett.}\ }\textbf {\bibinfo {volume} {126}},\ \bibinfo {pages} {103001} (\bibinfo {year} {2021})}\BibitemShut {NoStop}%
\bibitem [{\citenamefont {Taylor}, \citenamefont {Bai},\ and\ \citenamefont {Dybowski}(2011)}]{TAYLOR2011_mercury-halides_exp}%
  \BibitemOpen
  \bibfield  {author} {\bibinfo {author} {\bibfnamefont {R.}~\bibnamefont {Taylor}}, \bibinfo {author} {\bibfnamefont {S.}~\bibnamefont {Bai}}, \ and\ \bibinfo {author} {\bibfnamefont {C.}~\bibnamefont {Dybowski}},\ }\bibfield  {title} {\enquote {\bibinfo {title} {{A solid-state $^{199}$Hg NMR study of mercury halides}},}\ }\href {\doibase https://doi.org/10.1016/j.molstruc.2010.12.013} {\bibfield  {journal} {\bibinfo  {journal} {J. Mol. Struct.}\ }\textbf {\bibinfo {volume} {987}},\ \bibinfo {pages} {193--198} (\bibinfo {year} {2011})}\BibitemShut {NoStop}%
\bibitem [{\citenamefont {Jokisaari}\ \emph {et~al.}(2002)\citenamefont {Jokisaari}, \citenamefont {Järvinen}, \citenamefont {Autschbach},\ and\ \citenamefont {Ziegler}}]{Jokisaari2002_shieling_Methylmercury_Halides_exp}%
  \BibitemOpen
  \bibfield  {author} {\bibinfo {author} {\bibfnamefont {J.}~\bibnamefont {Jokisaari}}, \bibinfo {author} {\bibfnamefont {S.}~\bibnamefont {Järvinen}}, \bibinfo {author} {\bibfnamefont {J.}~\bibnamefont {Autschbach}}, \ and\ \bibinfo {author} {\bibfnamefont {T.}~\bibnamefont {Ziegler}},\ }\bibfield  {title} {\enquote {\bibinfo {title} {{$^{199}$Hg Shielding Tensor in Methylmercury Halides: NMR Experiments and ZORA DFT Calculations}},}\ }\href {\doibase 10.1021/jp025797q} {\bibfield  {journal} {\bibinfo  {journal} {J. Phys. Chem. A}\ }\textbf {\bibinfo {volume} {106}},\ \bibinfo {pages} {9313--9318} (\bibinfo {year} {2002})}\BibitemShut {NoStop}%
\bibitem [{\citenamefont {Autschbach}, \citenamefont {Kantola},\ and\ \citenamefont {Jokisaari}(2007)}]{Autschbach2007_sscc_methyl-mercury-halides_exp}%
  \BibitemOpen
  \bibfield  {author} {\bibinfo {author} {\bibfnamefont {J.}~\bibnamefont {Autschbach}}, \bibinfo {author} {\bibfnamefont {A.~M.}\ \bibnamefont {Kantola}}, \ and\ \bibinfo {author} {\bibfnamefont {J.}~\bibnamefont {Jokisaari}},\ }\bibfield  {title} {\enquote {\bibinfo {title} {{NMR Measurements and Density Functional Calculations of the $^{199}$Hg-$^{13}$C Spin-Spin Coupling Tensor in Methylmercury Halides}},}\ }\href {\doibase 10.1021/jp0713817} {\bibfield  {journal} {\bibinfo  {journal} {J. Phys. Chem. A}\ }\textbf {\bibinfo {volume} {111}},\ \bibinfo {pages} {5343--5348} (\bibinfo {year} {2007})}\BibitemShut {NoStop}%
\end{thebibliography}
%

\end{document}